\documentclass[12pt]{iopart}
\usepackage{color}
\usepackage{yfonts}
\usepackage{enumitem}

\usepackage{graphicx}
\usepackage{dcolumn}
\usepackage{bm}
\usepackage{iopams}

\newcommand{\beqar}{\begin{eqnarray}}
\newcommand{\eeqar}{\end{eqnarray}}

\newcommand{\bcen}{\begin{center}}
\newcommand{\ecen}{\end{center}}

\newcommand{\eps}{\varepsilon}

\newcommand{\f}[2]{\frac{#1}{#2}}
\renewcommand{\b}[1]{\left({#1}\right)}
\renewcommand{\v}[1]{\vec{#1}}

\renewcommand{\sb}[1]{\left[{#1}\right]}
\newcommand{\mean}[1]{\langle {#1} \rangle}

\newcommand{\ra}{\rightarrow}

\begin{document}

\title[Quantum finite-time thermodynamics: 
insight from a single qubit engine]{Quantum finite-time thermodynamics: 
insight from a single qubit engine}

\author{Roie Dann\footnote{Equal contribution of all authors.}}
\address{The Institute of Chemistry, The Hebrew University of Jerusalem, Jerusalem 9190401, Israel}
\ead{roie.dann@mail.huji.ac.il}

\author{Ronnie Kosloff}
\address{The Institute of Chemistry, The Hebrew University of Jerusalem, Jerusalem 9190401, Israel}
\ead{ronnie@fh.huji.ac.il}

\author{Peter Salamon}
\address{Department of Mathematics and Statistics, San Diego State University, 5500 Campanile Drive, San Diego CA 92182-7720}
\ead{salamon@sdsu.edu}
\vspace{10pt}





\begin{abstract}
Incorporating time into thermodynamics
allows addressing the tradeoff between efficiency and power. A qubit engine serves as a toy model to study  this tradeoff from first principles, based on the quantum theory of open systems.
We study the quantum origin of irreversibility, originating from heat transport, quantum friction and thermalization in the presence of external driving. 
We construct various finite-time engine cycles based on the Otto and Carnot templates. Our analysis highlights the role of coherence and the quantum origin of entropy production.
\end{abstract}

%
%
%
%
%

\section{Introduction}

The tradeoff between power and efficiency is well embedded in our everyday experience.
It is witnessed in the performance of any realistic engine or refrigerator, from operation of large nuclear plants, through the internal combustion engines of our automobiles, and all the way to microscopic biological engines and the quantum regime. Despite the intuitive notion, a theoretical analysis is quite involved as it requires a theoretical construction which encompasses both thermodynamics and transient dynamics.

The limiting case was first treated by Carnot, who linked an engine's maximum attainable work production to reversible thermodynamic transformations, thereby obtaining the thermodynamic temperature scale and the universal optimal efficiency that depends only on the hot and cold bath temperatures \cite{carnot1824reflexions}. Unlike efficiency, power requires knowledge of the transient dynamics, which is outside the realm of classical thermodynamics.
Finite-time thermodynamics (FFT) was developed to 
include the  limitations the process duration places on the performance of an engine.
\cite{salamon2001principles,andresen1984thermodynamics,andresen1977thermodynamics,hoffmann1997endoreversible,andresen2011Angewandte}. 
Originally, the pioneers of FTT incorporated empirical kinetic laws to introduce an intrinsic timescale in the analysis of engine cycles  \cite{curzon1975efficiency,salamon1981finite}. Some results from these efforts are recapped in section \ref{FTT}. In this paper, we address the need for kinetic laws by following a different approach: building upon a complete quantum description of the engine and baths.

Such complete quantum description however is not as straightforward as it sounds. Quantum mechanics is a dynamical theory which can supply
equations of motion for thermodynamic processes. The well established portion of this theory has predominantly dealt with \emph{closed} systems which conserve entropy and thus cannot deal with dissipation phenomena at the heart of thermodynamic analyses.
This forces us to turn to open quantum systems, whose description from first principles relies on a reduction from a closed composite system. The reduced description is achieved by tracing over the degrees of freedom of the surroundings, interacting with our system of interest. This description does not conserve entropy and allows exploration of thermodynamic processes in the quantum regime. Notably, the approach is based on the completely positive trace preserving (CPTP) dynamical map \cite{kraus1971general} and the Markovian Gorini-Kossakowski-Lindblad-Sudarshan (GKLS)   master equation \cite{lindblad1976generators,gorini1976completely}. A 
thermodynamically consistent dynamical framework \cite{alicki2018} is obtained by a first principle
derivation based on weak interaction of an open system with a heat bath. The derivation is commonly termed the `Davies construction' \cite{davies1974markovian}.

The quantum open system approach resulted in a number of surprises, initially reported as claimed contradictions to the second law \cite{levy2014local}. These include reported claims of breakdown of the Carnot bound in such engines \cite{allahverdyan2000extraction,boukobza2013breaking}. In turn,
these results led to resolutions, explained by unexpected work available from quantum resources including coherence \cite{Nieuwenhuizen2000BadBath}, squeezed bath
\cite{rossnagel2014nanoscale}, entanglement 
\cite{Alicki2013entanglement}, and information \cite{Mandal2012information,vidrighin2016photonic}. 

Another set of surprises came from attempts to use a naive GKLS formalism with a time-dependent driving, which possibly violates the second law \cite{geva1995relaxation,geva1996quantum,correa2013performance,correa2014quantum}. This led to the realization that the inconsistency arises from the derivation of the reduced dynamics of the system.  For periodic driving a thermodynamically consistent GKLS equation was derived in Ref. \cite{alicki2012periodically}.
In the case of a general (non-periodic) driving, only the adiabatic
master equation was available \cite{albash2012quantum}.
This restricted quantum heat engine analysis to Otto-type cycles whose strokes include either heat exchange or work exchange but never both at the same time. Only recently has a derivation of the GKLS master equation for simultaneous heat and work exchange been found, and it is this discovery whose implications we explore in the present paper \cite{dann2018time}. 

We adopt the dogma that thermodynamics and quantum mechanics address the same subject matter, therefore have to be consistent \cite{kosloff2013quantum}. In this framework quantum mechanics provides the tools to describe the dynamics, while the strict laws of thermodynamics must be obeyed. 
In addition, recent progress in the theory of quantum speed limits can illuminate fundamental bounds on the
process timescale \cite{uzdin2016speed,funo2019speed}.

Engines have been an intrinsic part in the development of classical thermodynamics. Their analysis still serves as an integral part of current research in finite-time and quantum thermodynamics.   These theories allow describing engines more realistically including non-ideal performance. It has been realized that any practical engine operates in a non-ideal irreversible mode.
Typically, there are four sources of irreversible phenomena in engines:
\begin{enumerate}
    \item { Finite heat transport.}
    \item{ Friction.}
    \item{ Heat leaks.}
    \item{ Cost of switching contacts between subsystems.}
\end{enumerate}

Following the thermodynamic tradition of learning from example, we  employ the most elementary working medium, 
a spin one half system to explore
a quantum version of finite-time thermodynamics. A decade ago such an example would have been criticized as a theoretician's toy with no connection to the world of real engines. The finite-time Otto type cycle, which our cars operate by, do not seem related to a single spin quantum engine. Nevertheless, recent experimental progress in miniaturization has enabled a
realization of an Otto cycle engine constructed from a single spin of an atom in an ion trap \cite{von2019spin}, or a single qubit in an
impurity electron spin \cite{ono2020analog}.

The unfortunate collision of the different usages of the word adiabatic in thermodynamics and quantum mechanics have been sidestepped by using the term "unitary dynamics" for dynamics along what thermodynamics would call an adiabat leaving the use of adiabatic for the quantum meaning. 

The present paper begins by laying the quantum thermodynamic foundations for the qubit, stating the quantum definitions for energy, work, heat, entropy and temperature, Sec. \ref{sec:quantum_engine_model}.  We continue by discussing sources of reversibility: heat transport, Sec. \ref{sec:non_vanishing_heat}, the quantum origin of friction, Sec. \ref{sec:fric}, and thermalization processes which combine heat transport and external work, Sec. \ref{sec:thermalization}.
The quantum version of finite-time thermodynamics is studied by constructing two  basic engine platforms: Carnot and Otto.
These models illuminate different aspects of the tradeoff between power and efficiency and the role of coherence on the engines performance, Sec. \ref{sec:local}, \ref{sec:global} and \ref{sec:Q_signiture_global_coherence}. 

\section{Some Preliminaries}

\subsection{Classical engines operating in finite-time}
\label{FTT}

Classical textbook treatments of heat engines define various kinds of engine cycles. These cycles are mostly four-stroke and consist of two unitary strokes and two open strokes in contact with a heat bath -- one hot and one cold. Finite-time thermodynamic analyses of these cycles has given us the simplifying model of endoreversible processes -- processes in which the participating systems are at each instant in equilibrium states and all irreversibility resides in the interactions between such systems. Endoreversible cycles play an important role by edging closer towards real cycles, being relatively easy to analyse and providing checks along the way for more ambitious treatments. They also provide an accurate picture of reality when the slow timescale is the interaction. The simplifying condition of instantaneous lossless adiabatic jumps, made possible for quantum systems using shortcuts to adiabaticity (cf. section \ref{subsec:STA}), is a hallmark simplifying feature that we inherit from these studies.

Also important for these analyses is a much older result known as the Gouy-Stodola theorem \cite{bejan1983entropy}
which established a connection between dissipated work and entropy production, cf.
\begin{equation}
    \Delta {\cal A}^U = - T_0 \Delta {\cal S}^U
    \label{Gouy}
\end{equation}
where the superscript $U$ refers to the universe (all participating systems), ${\cal S}$ is the entropy, ${\cal A}$ is the available work, and $T_0$ is the temperature at which heat is freely available which means it carries no available work. The environment temperature $T_0$ is also used in the availability (also called exergy) state function ${\cal A} = {\cal E}-T_0{\cal S}$, where $\cal E$ is the internal energy. As a consequence Eq. (\ref{Gouy}) valid with any temperature choice for $T_0$. In engineering treatments it is always the atmospheric temperature, but any $T$ will do. In the physics literature $T_0$ is almost always taken to be the system's temperature making ${\cal A}={\cal F}$, where $\cal F$ is the Helmholtz free energy. 

For our purposes, the importance of Eq. (\ref{Gouy}) arises from the fact that it shows that dissipation can equivalently be measured in energetic or entropic terms, even when the system does not have a temperature or when this temperature is changing during the process of interest.  

\subsection{Qubit engine model}
\label{sec:quantum_engine_model}
The engine model is constructed from a hot and cold bath
and a controllable two-level-system shuttling between them.
The  Hamiltonian of the working medium,  a qubit, is
\begin{equation}
    \hat H = \omega(t) \hat S_z +\epsilon(t) \hat S_x
    \label{eq:hamil}
\end{equation}
where $\hat S_j$ 
are the spin operators with the commutation relation of the $SU(2)$ algebra $ [\hat S_i,\hat S_j] =i \hbar \epsilon_{ijk} \hat S_k$, see Appendix \ref{appendixA}. 
The time-dependent driving parameters $\omega(t)$ and $\epsilon(t)$,  define a typical energy scale
\begin{equation}
\hbar \Omega (t)=\hbar \sqrt{\omega^2+\epsilon^2}~~~,
\label{eq:rabi}
\end{equation}
where $\Omega$ is the Rabi frequency.

\begin{figure}[htb!]
\centering
\includegraphics[width=7 cm]{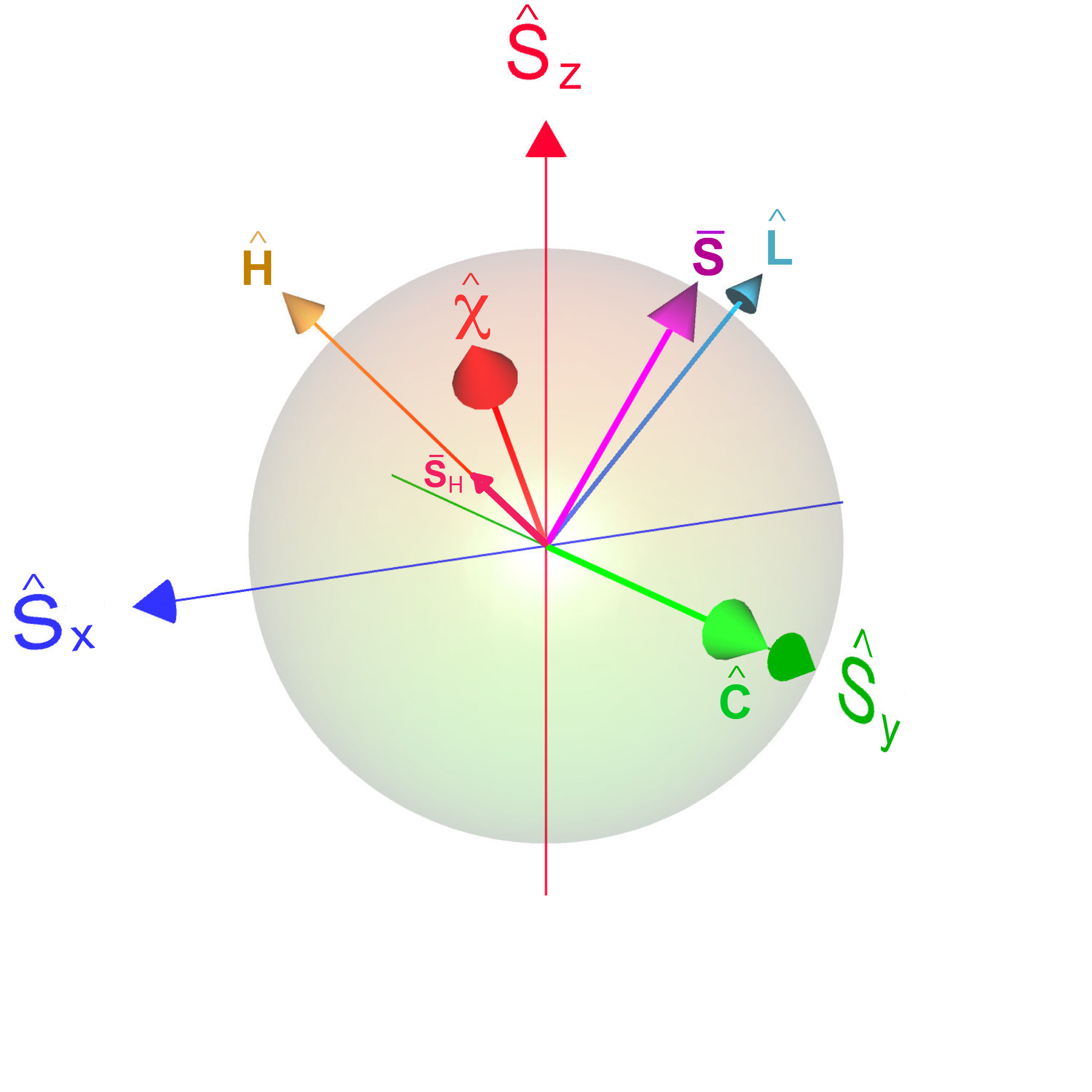}
\caption{The state of the system, Eq. (\ref{eq:statew}), is represented by the polarization vector $\bar S$ in the Bloch sphere (purple). Alternatively, the state can be represented in a rotated frame Eq. (\ref{eq:hlc-state}),
defined by the  set of coordinates $\mean{\hat{H}}, \mean{\hat{L}},\mean{\hat{C}}$ in Eq. (\ref{eq:hlc}). These coordinates are rotated about
the $\mean{\hat{S}_y}$ axis relative  to the static direction. 
The projection of the polarization on the energy direction
$\bar S_H$, Eq. (\ref{eq:denergy}), is shown in light red.
The invariant of the free propagator $\mean{\hat{\chi}}$    
in Eq. (\ref{eq:chi-sigma}) is shown in red. The direction of $\mean{\hat{\chi}}$   is rotated  around the $\mean{\hat{L}}$ axis with respect to the $\mean{\hat{H}}$ direction (cf. Appendix \ref{appendixA}).
The Bloch sphere representation can represent either the expectation values of the operators or the operators themselves. The latter constitute orthogonal vectors in
Liouville space. }
\label{fig:1}
\end{figure}

The state of the qubit  working medium $\hat \rho$, can be expanded using any orthonormal set of operators satisfying $\rm{tr}\{\hat A_i^\dagger\hat A_j \}=\delta_{ij}$.
Choosing the polarizations $\hat S_j$ as basis operators, the state $\hat \rho$
is completely determined by the expectation value of the three polarizations
\begin{equation}
    \hat \rho = \frac{1}{2}\hat I + \frac{2}{ \hbar^2}\left(
    \langle \hat S_x \rangle \hat S_x
    +\langle \hat S_y \rangle \hat S_y+
    \langle \hat S_z \rangle \hat S_z
    \right)~~.
    \label{eq:statew}
\end{equation}
It elucidates the analysis to represent the polarization vector as a geometric object $\v{S}=\{\mean{\hat{S}_x},\mean{\hat{S}_y},\mean{{\hat{S}}_z}\}^T$ which resides inside the Bloch sphere (see Fig. \ref{fig:1} and Appendix \ref{appendixA}). 
The polarization value is defined as
\begin{equation}
\bar S\equiv-|\v{S}|=-\sqrt{\langle \hat{S}_x \rangle^2 +\langle \hat{S}_y \rangle^2+ \langle \hat{S}_z \rangle^2}~~.
\label{eq:polar2}
\end{equation}
It is related to the purity of the state, 
where $0 \geq \bar S \geq -\hbar/2$ where
$|\bar S|=\hbar/2$ for a pure state. The sign convention of the polarization is motivated by the fact that we consider only positive temperatures, see Eq. (\ref{eq:polar}). The polarization value is invariant
under unitary transformations generated by the $SU(2)$ group, which represents rotations of the polarization vector.
It is related to the expectation value of the energy by
${\cal E}=\langle  \hat H \rangle=\hbar \Omega \bar S_H$, where $\bar S_H$ is the projection of the polarization vector on the direction representing the Hamiltonian. In thermal 
equilibrium at temperature $T$, the polarization becomes 
\begin{equation}
\mean{ \v S} =\bar S_H=\bar S_{eq} =-\frac{\hbar}{2} \tanh \left(\frac{\hbar \Omega}{2 k_B T} \right)~~,
\label{eq:polar}
\end{equation}
where $k_B$ is the Boltzmann constant and $T$ is the bath temperature.

\begin{figure}[htb!]
\centering
\includegraphics[width=10 cm]{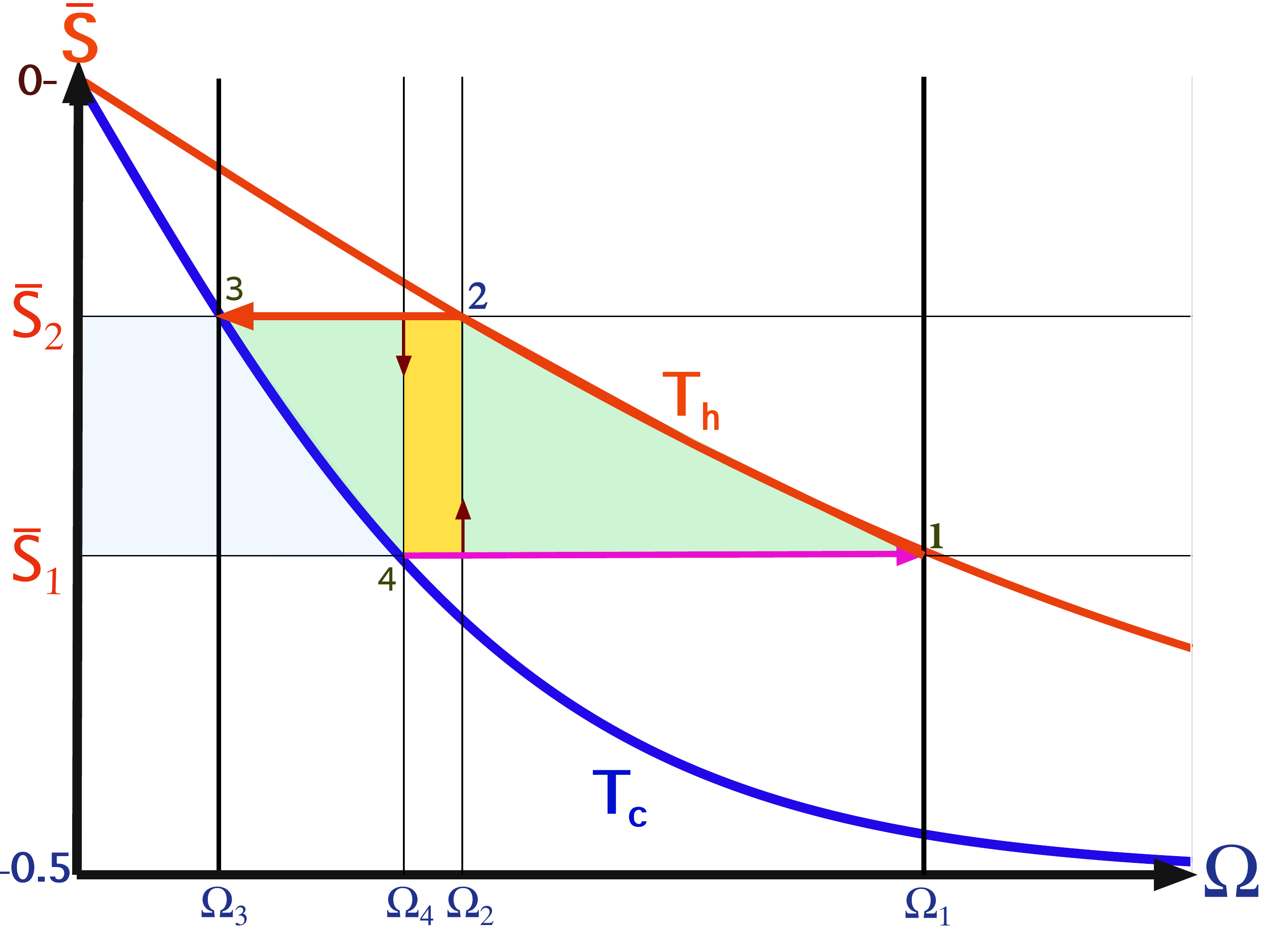}
\caption{Carnot and Otto cycles: Polarization $\bar S$ as a function of frequency $\Omega$ is shown along the hot (red) and cold (blue) isotherms.  The four switching points between the strokes are indicated by numbers 1-4.
 The various strokes are represented by lines. Carnot cycle: Hot isotherm $\b{\Omega_1,\bar S_1} \rightarrow \b{\Omega_2,\bar S_2}$, unitary expansion $\b{\Omega_2,\bar S_2} \rightarrow \b{\Omega_3,\bar S_2}$, cold isotherm 
$\b{\Omega_3,\bar S_2} \rightarrow \b{\Omega_4,\bar S_1}$  and unitary compression  $\b{\Omega_4,\bar S_1} \rightarrow \b{\Omega_1,\bar S_1}$.
The filled area in light green represents the work output, while the area in light blue represents the heat absorbed by the cold bath. The sum of the two areas equals the heat exchange with the hot bath. This implies a nice geometric representation of the efficiency, as the ratio between the light green area and the combined area. Otto cycle embedded in the Carnot cycle (orange area); hot isochore $\b{\Omega_2,\bar S_1}\ra\b{\Omega_2,\bar S_2}$, unitary expansion $\b{\Omega_2,\bar S_2} \rightarrow \b{\Omega_4,\bar S_2}$, cold isochore $\b{\Omega_4,\bar S_2} \rightarrow \b{\Omega_4,\bar S_1}$, unitary compression $\b{\Omega_4,\bar S_1} \rightarrow \b{\Omega_2,\bar S_1}$. The short arrows designate the thermalization isochores.}
\label{fig:1b}
\end{figure}

The engines to be analysed are discrete four stroke cycle models.
Specifically, we will compare the Carnot cycle with the Otto cycle. 
Both cycles are constructed from the following sequence of strokes:
\begin{enumerate}[label=(\Alph*) ]
\item{$1 \rightarrow 2 $ Hot bath thermalization.}
\item {$2 \rightarrow 3 $ Unitary expansion from hot to cold}
    \item{$3 \rightarrow 4 $ Cold bath thermalization}
    \item {$4 \rightarrow 1 $ Unitary compression from cold to hot}
\end{enumerate}
The two cycles differ by the nature of the thermalization strokes, $1\rightarrow 2$  and 
$3 \rightarrow 4$. The reversible Carnot cycle includes isothermal strokes during the thermalization processes, while the Otto cycle utilizes isochores, see Fig. \ref{fig:1b}. In the following study, we sometimes refer to the thermalization strokes as open-strokes, alluding to the fact that the working medium constitutes an open quantum system during these strokes. 
The four stroke cycles can be described 
by the corresponding cycle propagator $\Lambda_{cyc}$, which is a product of individual stroke propagators:
\begin{equation}
\Lambda_{cyc}=\Lambda_{c \rightarrow h}\Lambda_{c}\Lambda_{h\rightarrow c}\Lambda_{h}
\label{eq:propagator}
\end{equation}
These propagators are  completely positive trace-preserving (CPTP) maps on the space of qubit states \cite{kraus1971general}.
The properties of the engine are extracted from the fixed point of the 
cycle map, $\rho_{fp}$, which represents the limit cycle and satisfies
$\Lambda_{cyc} \hat \rho_{fp}=\hat \rho_{fp}$
\cite{feldmann2004characteristics}. The fixed point $\hat{\rho}_{fp}$ along with the stroke propagators fully determine the qubit state throughout the fixed cycle. The existence of a single invariant
of CPTP map guarantees monotonic convergence to the fixed point \cite{lindblad1974expectations,feldmann2004characteristics}. The value of $\hat{\rho}_{fp}$ at the four switching points of the cycle
allow the evaluation of thermodynamic variables, such as work, heat, and entropy.

Two important quantities for the finite-time thermodynamic analysis of these cycles  are 
the von-Neumann and the energy entropies of the qubit
\begin{equation}
    {\cal S}_{v.n} \equiv -\rm{tr}\sb{\hat{\rho}\rm{ln}\hat \rho}=
-\b{\f 12-\f{\bar S}{\hbar}}\log\left(\frac{1}{2}-\f {\bar S}{\hbar}\right)-\b{\f 12+\f{ \bar S}{\hbar}}\log\left(\f{ \bar S}{\hbar}+\frac{1}{2}\right)~~,
\end{equation}
and
\begin{equation}
    {\cal S}_H = -p_H \rm{ln}p_H -(1-p_H)\rm{ln}\b{1-p_H}~~,
\end{equation}
where $p_H=\b{\f 12-\f{\bar S_H}{\hbar}}$.
Generally, we have ${\cal S}_H\geq{\cal S}_{v.n}$ with equality when the state is diagonal in the energy representation.  At equilibrium both entropies  reduce to ${\cal S}_{v.n}\b{\bar{S}_{eq}}$, where $\bar S_{eq}$ is given in Eq. (\ref{eq:polar}). The difference between
the energy entropy and the von-Neumann entropy
is a quantifier \cite{feldmann2003quantum,baumgratz2014quantifying,PhysRevLett.113.170401,feldmann2016transitions} of coherence. It is commonly known as the divergence \cite{lin1991divergence}
\begin{equation}
{\cal D}(\hat \rho| \hat \rho_d)=\rm{tr}\{ \hat \rho \ln \hat \rho - \hat \rho \ln \hat \rho_d \}=
{\cal S}_H-{\cal S}_{v.n}~~,
\label{eq:divergence}
\end{equation}
where $\hat \rho_d $ is diagonal in the energy representation and defined in Eq. (\ref{eq:denergy}).
During the cycle's operation, the unitary strokes $\Lambda_{c \rightarrow h}$ and $\Lambda_{h\rightarrow c}$ maintain a constant von-Neumann entropy, while the energy entropy may increase with the generation of coherence.


\section{Frictionless Engines}
\label{sec:non_vanishing_heat}

A non-vanishing heat transport rate is a prime source of irreversibility. Such heat transfer occurs when there exists a temperature gap on the interface between the engine and the baths. The influence of a realistic heat transport on cycle performance was first addressed by the classical endoreversible model  \cite{curzon1975efficiency}. Such a cycle assumes an empirical Newtonian heat transport law to describe the heat rate. For the qubit engine, we can replace the empirical Newtonian heat transport law with a quantum first principle derivation.
The starting point is the composite Hamiltonian:
\begin{equation}
    \hat H_{tot} = \hat H(t)+ \hat H_{h/c} +  
    \hat H_{s-h/c}~~,
\label{eq:htotal} 
\end{equation}
where $\hat H_{h/c}$ are the hot and cold bath Hamiltonians and $ \hat H_{s-h/c}$ represent the system-bath interaction, correspondingly. Reduced equations of motion for the system are obtained 
in the framework of the theory of open quantum systems \cite{breuer2002theory}. This theory constitutes a general setting from which the dynamics can be derived from first principles, by employing a number of idealization. 
The main assumptions, included in the derivation, are weak system-bath coupling and a separation of timescales between a fast bath and a sluggish system \cite{davies1974markovian}. These assumptions are justified on the basis of physical reasoning and the fact that the obtained dynamical equations are indisputably consistent with the laws of thermodynamics 
\cite{kosloff2013quantum,alicki2018}.

To concentrate only on heat transport we can assume $\epsilon(t)=0$, therefore $\Omega(t)=\omega(t) $, which leads to
\begin{equation}
     \hat{H}^{elem}\b t =\hat H\b t_{\epsilon=0} = \Omega(t) \hat S_z= \omega(t) \hat S_z~~.
    \label{eq:elem_Hamil}
\end{equation}
We refer to Eq. (\ref{eq:elem_Hamil}) as the {\emph{elementary Hamiltonian}}.
For such a case, the Hamiltonian satisfies
$[\hat  H^{elem}(t),\hat H^{elem}(t')]=0$, which decouples the dynamics of the populations and the coherence. This means that when the qubit is initialized in a diagonal state in the energy basis, the dynamics generated by the elementary Hamiltonian remain on the energy shell and are equivalent to frictionless solutions (Cf. section \ref{sec:fric}) of stochastic dynamics. For such instant, the analysis has common features with quantum adiabatic dynamics \cite{messiah1962quantum}.
The state of the system then becomes:
\begin{equation}
\hat \rho_d= \frac{1}{2}\hat I + \frac{2}{(\hbar \Omega)^2}\langle \hat H^{elem} \rangle \hat H^{elem} ~~,
\label{eq:denergy}
\end{equation}
which is diagonal in the energy representation, implying that
$[\hat \rho_d, \hat H^{elem}]=0$, $\bar S = -|\langle \hat S_z \rangle| $ and ${\cal S}_{v.n}={\cal S}_E$. When the initial state exhibits quantum coherence, Eq. (\ref{eq:statew}),
under these operating conditions and after a sufficient time, any initial coherence decays to zero.
We will refer to this model as the {\emph{elementary}} qubit engine.

In this framework, the reduced dissipative dynamics of the qubit is of the following structure  \cite{dann2018time}
\begin{equation}
    \frac{d}{dt} \hat \rho =-
    \frac{i}{\hbar}[\hat H^{elem} (t) , \hat \rho]+
    {\cal L}_D(\hat \rho)
    \label{eq:lind}
\end{equation}
where the dissipator ${\cal L}_D$ has a Gorini-Kossakowski-Lindblad-Sudarshan form (GKLS) \cite{gorini1976completely,lindblad1976generators}
\begin{equation}
    {\cal L}_D( \rho) =\f{4}{\hbar^2}\sb{ k_\uparrow(t)\left( \hat S_+ \hat \rho \hat S_- -\frac{1}{2}\{\hat S_- \hat S_+ ,\hat \rho \} \right)
    + 
   k_\downarrow(t)\left( \hat S_- \hat \rho \hat S_+ -\frac{1}{2}\{\hat S_+ \hat S_- ,\hat \rho \} \right)}
   \label{eq:lind1}
\end{equation}
where $\hat{S}_{\pm}= \hat S_x \pm i \hat S_y$  and $k_\uparrow$ and $k_\downarrow$
obey instantaneous detailed balance:
\begin{equation}
    \frac{k_\uparrow(t)}{k_\downarrow (t)}=e^{ -\frac{ \hbar \Omega(t)}{k_B T}}~~.
    \label{eq:detailed balance}
\end{equation}
The kinetic coefficients typically have
a power dependence on $\Omega$: 
$k_\downarrow (t) \propto \Omega(t)^n$, where $n\in \mathbb{R}$ depends on the spectral properties of the bath \cite{breuer2002theory}.
An alternative representation of the dynamics utilizes the Heisenberg picture in which the equation of motion are of the form
\begin{equation}
    \frac{d}{dt} \hat X = 
    \frac{i}{\hbar}[\hat H^{elem} (t) , \hat X]+
    {\cal L}_D^*(\hat X) + \frac{\partial}{\partial t} \hat X~~;
    \label{eq:lind-h}
\end{equation}
where ${\cal L}_D^*(\bullet)$ is the adjoint generator.
The relation to thermodynamics is achieved by setting $\hat X =\hat H$ and identifying the rate of change of the average energy as the quantum dynamical version of the first law
of thermodynamics \cite{spohn1978irreversible,alicki1979quantum}
\begin{equation}
    \frac{d}{dt}E = {\cal P}+{\dot{\cal Q}}~~,
\end{equation}
where: ${\cal P} =\langle \frac{\partial}{\partial t} \hat H \rangle$ is the power and $\dot {\cal Q}= \langle {\cal L}_D^* (\hat H) \rangle $ is the heat flux. Power is associated with the unitary part of the dynamics, for which the von-Neumann entropy remains constant, and heat flux is identified  as the average energy transfer that induces entropy production.
For the elementary qubit system, the power becomes 
\begin{equation}
{\cal P} = \bar S_H(t) \frac{\partial \omega(t)}{\partial t}  
\label{eq:power}
\end{equation}
This result is analogous to the classical definition of power, where $\partial \omega/\partial t$ takes the role of the generalized force and the polarization is its conjugate variable.
The expression for the heat flux reads 
\begin{equation}
\dot {\cal Q}=- \Gamma (t) \left(  \langle \hat H^{elem}(t) \rangle -  \langle \hat H_{eq}(\Omega(t),T \rangle \right)   ~~,  
\end{equation}
where $\Gamma =  k_\uparrow+k_\downarrow$. $\mean{\hat{H}_{eq}\b{\Omega\b t,T}}=\Omega\b t \bar{S}_{eq}\b{\Omega\b t}$ and
$\bar S_{eq}(\Omega(t)) $ is the instantaneous attractor which is defined by the changing frequency $\Omega(t)$, Eq. \ref{eq:polar}. As expected, the heat flux is proportional to the deviation from equilibrium and the relaxation rate.

The equilibration of energy is accompanied by decay of  coherence. The coherence dynamics are obtained by substituting $\hat S_x$ or $\hat S_y$ for $\hat X$ in Eq. (\ref{eq:lind-h}), leading to
\begin{equation}
\f d{dt}\sb{\begin{array}{c}
\hat{S}_{x}\\
\hat{S}_{y}
\end{array}}=\sb{\begin{array}{cc}
- \frac{1}{2}\Gamma\b t & -\Omega\b t\\
\Omega\b t & -\frac{1}{2} \Gamma\b t
\end{array}}\sb{\begin{array}{c}
\hat{S}_{x}\\
\hat{S}_{y}
\end{array}}~~.
\label{eq:decaycoher}
\end{equation}
This set of equations reflects the separation of  the coherence dynamics from the population dynamics \cite{alicki2018}.
It implies that any initial coherence will decay to zero once the limit cycle is reached. This mode of operation is equivalent to stochastic thermodynamics where all thermodynamic observables are obtained in terms of the populations of the energy levels \cite{seifert2012stochastic,sekimoto2010stochastic,feldmann1996heat}.

\subsection{Elementary cycles}

Utilizing the quantum  description of heat transport, introduced above, we can assemble a finite-time model
of a heat engine. We construct a  Carnot-type cycle and an Otto cycle whose working fluids are governed by elementary Hamiltonians, Eq. (\ref{eq:elem_Hamil}
and compare their finite-time thermodynamic performance.
We will compare the work produced per cycle $-{\cal W}$ and heat ${\cal Q}_{h/c}$
which define the efficiency: $\eta=-\f{{\cal W}}{{\cal Q}_h}$.

\subsection{Elementary Carnot-type cycle}
\label{subsec:endo_reversible}
\begin{figure}[htb!]
\centering
\includegraphics[width=10 cm]{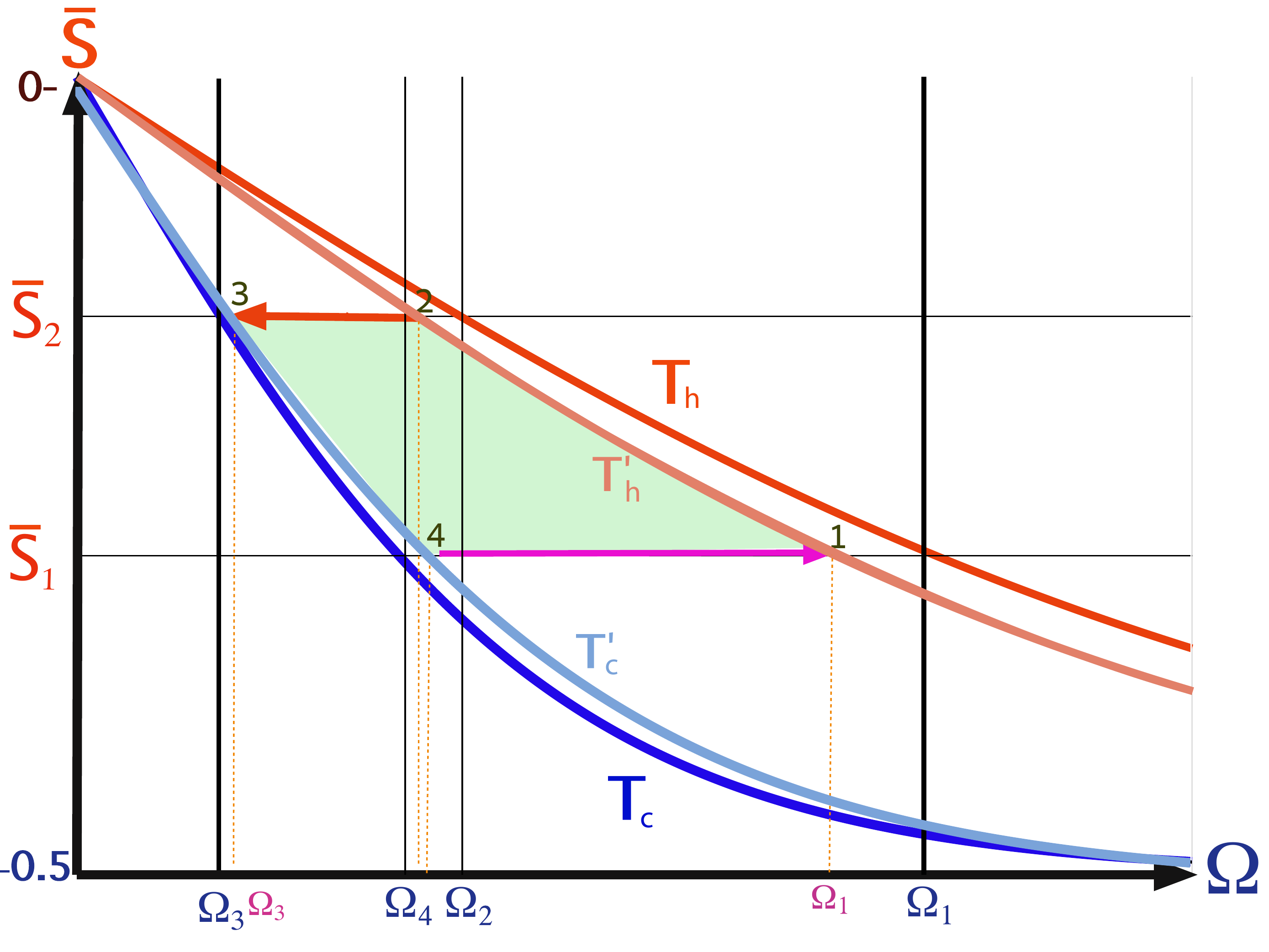}
\caption{Endoreversible Carnot cycle: Polarization $\bar S$ as a function of frequency $\Omega$. The hot and cold isotherms are designated in red and blue, correspondingly. Blue frequencies designate the corners of the reversible Carnot cycle and the small purple frequencies correspond to the corners of the endoreversible cycle. The endoreversible frequencies depend on the cycle time, in the quasi-static limit they converge to the frequencies of the reversible Carnot cycle.
The finite-time engine follows internal
hot $1 \rightarrow 2$ and cold $ 3 \rightarrow 4$ isotherms (light red and blue curved lines) with associated temperatures $T'_h$ and $T'_c$, allowing finite heat transport.
The area in green equals the total work output. }
\label{fig:2}
\end{figure}

Consider a quantum version of a finite-time Carnot-type cycle shown in Fig. \ref{fig:2}. When in contact with the heat bath, the qubit of the endoreversible engine maintains a constant internal temperature
$T'$, generating a temperature gap with the bath. In this scenario, one can optimize the power by varying the temperature gap \cite{geva1992quantum}.
The efficiency then shows a monotonic decrease with the deviation from the ideal Carnot cycle: $\eta_C \ge \eta \ge 0$.
At the high temperature limit the performance is very similar to the Curzon-Ahlborn empirical model \cite{novikov1958efficiency,curzon1975efficiency} or low dissipation limit \cite{esposito2010efficiency}, where the heat conductance was modelled by the Newtonian heat transfer law. In this limit, the efficiency at maximum power converges to
\begin{equation}
    \eta_{CA} = 1 - \sqrt{\frac{T_c}{T_h}}~~,
    \label{eq:eta_ca}
\end{equation}
and the work per cycle becomes half the reversible work, Eq. (\ref{eq:wcarnot}), ${\cal W}_{CA}=\f{1}{2}{\cal W}_C$. 
The optimal power at high temperature can be approximated as \cite{geva1992quantum}
\begin{equation}
    {\cal P}_{Endo}= \Gamma k_B\b{\sqrt{T}_h-\sqrt{T}_c}^2\f{2}{\hbar ^2}  \left( \bar S_2^2-\bar S_1^2\right)\frac{1}{\ln(\bar S_2/\bar S_1)}~~.
    \label{eq:endop}
\end{equation}
This expression is reminiscent of the ideal work at the high temperature limit Eq. (\ref{eq:ideal_work_high_temp}), with a modified temperature gap. 
In this temperature regime, the optimum entropy production average rate per cycle  obtains a similar form 
\begin{equation}
   \f{{\sigma}^u_{cyc}}{\tau_{cyc}} = \Gamma \b{ \f{T_h- T_c}{\sqrt{{T_c}{T_h}}}}  \f{2}{\hbar ^2}  \left( \bar S_2^2-\bar S_1^2\right)\frac{1}{\ln(\bar S_2/\bar S_1)}~~.
    \label{eq:entropy_production_endo_rev}
\end{equation}
A similar structure to
Eq. (\ref{eq:endop}) has been  derived recently \cite{abiuso2020optimal,abiuso2019non} based on a low dissipation limit.

The qubit Carnot-type cycle posses finite power and approaches the reversible limit when the cycle time tends to infinity. 
In the limit of infinite cycle time, the cycle operates reversibly to obtain the Carnot efficiency
\begin{equation}
\eta_C=1-\frac{T_c}{T_h}~~.
\label{eq:carnot}
\end{equation}
The work per reversible cycle then becomes 
\begin{equation}
{\cal W}_{C}=k_B \Delta T \Delta{\cal S}_{v.n}~~,
\label{eq:wcarnot}
\end{equation}
where $\Delta T=T_h-T_c$ is temperature gap and $ \Delta{\cal S}_{v.n}$ is the change of the qubit's von-Neumann entropy 
on the cold or hot isotherms. For the ideal cycle, the von-Neumann and energy entropy ${S}_E$ coincide.
Another important characteristic of the engine is the compression ratio ${\cal C} = \Omega_{max}/\Omega_{min}$, for the ideal Carnot engine
${\cal C}_{Carnot} = \Omega_{1}/\Omega_{3}$, see Fig. \ref{fig:2}.
The entropy of the qubit is bounded by $\rm{ln}2$, giving a maximum  possible work of
$\rm{max}\b{{\cal W}_C}= k_B(T_h-T_c) \rm{ln}(2)$.

At the high temperature limit 
$\hbar \Omega \ll k_B T$ the energy entropy can be approximated as
 ${\cal S}_{v.n} \simeq -\rm{ln}2+2\b{\f{\bar{S}}{\hbar}}^2$ and the work becomes
\begin{equation}
 {\cal W}_C\simeq k_B\Delta T \f{2}{\hbar^2}(\bar S_2^2 -\bar S_1^2 )~~. 
 \label{eq:ideal_work_high_temp}
\end{equation}
This typical dependence is a general feature of any entropy dependent variable. The characteristic quadratic functionality of the polarization stems from the vicinity to the maximum entropy point.

\subsection{Elementary Otto cycle}
\label{subsec:elemOtto}

\begin{figure}[htb!]
\centering
\includegraphics[width=10 cm]{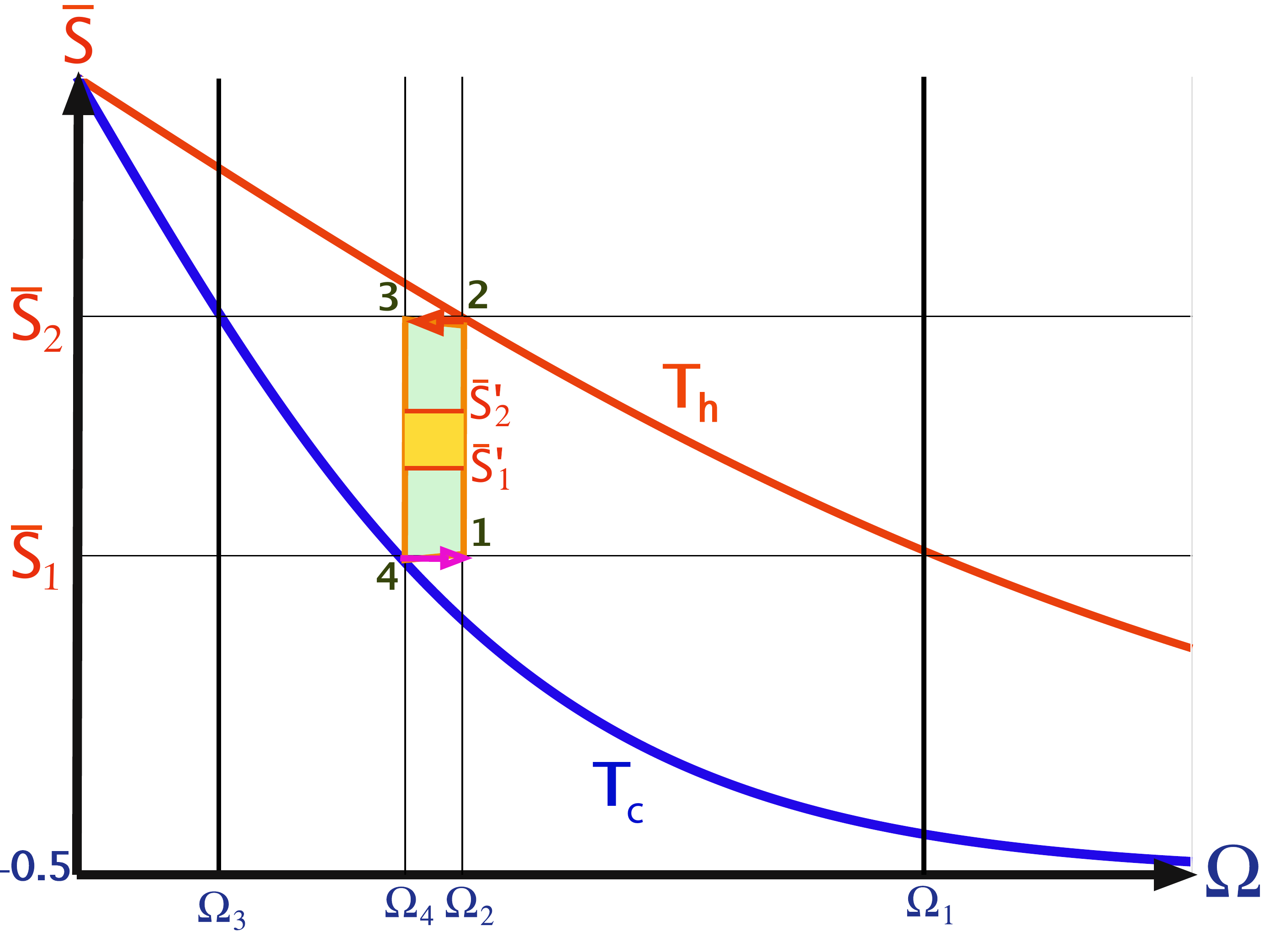}
\caption{Otto cycle embedded within the Carnot cycle: Polarization $\bar S$ as a function of frequency $\Omega$. The same extreme polarizations $\bar{S}_1$ and $\bar{S}_2$ are used. The area in light green is the work output. The area in orange represents a finite-time Oto engine operating between $\bar S_2'$ and $\bar S_1'$. The compression ratio ${\cal C}_{Otto}$ 
is reduced relative to ${\cal C}_{Carnot}$ as ${\Omega_2 } < \Omega_1$ and
${\Omega_3 } < \Omega_4$.}
\label{fig:4}
\end{figure}

We consider an Otto cycle which is embedded within the same isotherms and frequency range of the elementary Carnot cycle and is limited by the polarizations $\bar S_1$ and $\bar S_2$, see Fig. \ref{fig:4}  \cite{feldmann1996heat}. For an engine operation mode the compression ratio of the Otto cycle is constrained by  ${\cal C}_{Otto}=\Omega_c/\Omega_h\leq {\cal C}_{Carnot}$.
The engine's work obtains the simple form
\begin{equation}
    {\cal W}_{Otto} = \Delta \Omega \Delta \bar S~~,
    \label{eq:wotto}
\end{equation}
where $\Delta \Omega = \Omega_h -\Omega_c=\Omega_2 -\Omega_4$ and $\Delta S= \bar S_2-\bar S_4$. It is represented geometrically by the confined area between the frequencies and polarizations, colored as light green in Fig. \ref{fig:4}. 
Such an engine is characterized by a  constant efficiency
\begin{equation}
\eta_{Otto}= 1- \frac{\Omega_c}{\Omega_h}~~,   \label{eq:e-otto} 
\end{equation}
which leads to
$\eta_{Otto}= 1- \frac{\bar S_2 T_c}{\bar S_1 T_h} \le \eta_C$ for the analysed cycle.
When $\bar S_2 \rightarrow \bar S_1$, the cycle operation becomes reversible and the Carnot bound is recovered $\eta_{Otto} \rightarrow \eta_C$. This limiting case is the transition point between the engine and refrigerator operation mode (${\cal C}_{Otto}={\cal{C}}_{Carnot}$).

The heat dissipated during the cycle operation leads to rise in entropy. The entropy production per cycle obtains the form
\begin{equation}
    \sigma^u_{cyc}=\f{1}{k_B}\b{\f{\Omega_h}{T_h}-\f{\Omega_c}{T_c}}\Delta \bar S~~.
    \label{eq:otto_entropy_production}
\end{equation}
We obtain a linear dependence on the polarization difference, which contrasts with the endoreversible result at high temperature, Eq. (\ref{eq:entropy_production_endo_rev}), characterized by a quadratic difference dependence. 

We can compare between the geometric interpretation of the work output of the elementary Otto and Carnot cycles, Eq. (\ref{eq:wotto} and Eq.  (\ref{eq:wcarnot}). In the Otto cycle work is represented by the area enclosed by the cycle in the  $\b{\Omega,\bar S}$ plane, and in the 
$\b{T,{\cal S}_{v.n}}$ plane for the Carnot cycle, see description in the caption of  Fig. \ref{fig:2}.

\subsection{Optimization of the elementary Otto cycle} 
\label{subsec:otto-opt}

A modification of the present cycle includes optimizing the work per cycle with respect to the frequency $\Omega_2$. 
At the high temperature limit $\hbar \Omega \ll k_B T$, this optimization procedure leads to
${\cal W}_{max}=\frac{\hbar^2 \Omega_4^2}{k_B T_h}\left(1- (\f{T_c}{T_h})^2 \right)$
with efficiency $\eta_{Otto}=1-\frac{2T_c}{T_h+T_c}$. Such optimization is equivalent to a maximization of the area of a rectangular region embedded within the Carnot cycle.

Finite power is obtained when the working medium does not completely relax to thermal equilibrium during the open strokes. Power optimization is carried out with respect to the thermalization time. Surprisingly, the optimal cycle was found to be of the bang-bang type, with a vanishing cycle time.
The optimal power becomes \cite{feldmann1996heat}
\begin{equation}
    {\cal P}_{Otto}^{B.B}=\frac{1}{4}\Gamma \Delta  \Omega \Delta \bar S
    \label{eq:opower}
    \end{equation}
where  $\Gamma_h=\Gamma_c =\Gamma$. This gives a simple relation to the maximum work ${\cal P}_{Otto}=\f{\Gamma}{2}{\cal W}_{Otto}$, Eq. (\ref{eq:wotto}). Such an engine operates at
the polarization $ \bar S=\frac{1}{2}(\bar S_2+\bar S_1)$. We refer to cycles with vanishing cycle times as {\emph{sudden cycles}}. 
A generalization of Eq. (\ref{eq:opower}) where the relaxation rates $\Gamma$ are difference on the hot and cold side can be found in \cite{erdman2019maximum}.
A sudden type qubit refrigerator Otto cycle has also been investigated with similar conclusions \cite{karimi2016otto,pekola2019supremacy}. 

The optimal power of the endoreversible cycle Eq. (\ref{eq:endop}) 
can be compared with the Otto cycle Eq. (\ref{eq:opower}) at the high temperature limit. The comparison shows that the optimal power of the Otto cycle may exceed the power of endoreversible Carnot cycle with the same polarization and bath temperatures. 
The counter intuitive result where the optimum power 
for the Otto cycle is obtained for
vanishing cycle time forces one to reexamine the model. We expect that 
engines, as physical objects, do not display singular dependence on
operation parameters. In addition such a cycle can produce finite power
with Carnot efficiency \cite{erdman2019maximum}.
If we put a restriction on the minimum time allocation of the unitary strokes, the power becomes optimal for a finite cycle time \cite{feldmann1996heat}. This was the original motivation for adding friction phenomenologically \cite{feldmann2000performance}. An important outcome of the incorporation of phenomenological friction within the
model is a minimum cycle time for an engine operation. The phenomenological friction is taken to be inversely proportional to the stroke time. Thus, rapid driving leads to enhanced friction, reducing the power. Below  a minimum cycle time the output power vanishes and the cycle operates as an accelerator or a dissipator,  converting useful work to heat. In a dissipator, work is consumed (${\cal{W}}>0$) while heat is dissipated to both the hot and cold baths (${\cal Q}_h,{\cal Q}_c<0$). Using the efficiency definition, $\eta\equiv-{\cal W}/{\cal Q}_{h}$, the "efficiency" exceeds one. An accelerator operation mode includes positive work, accelerating the transfer of heat from the hot to the cold bath (${\cal Q}_h>0$ and ${\cal Q}_c<0$). This leads to  negative values of $\eta$.

\section{The quantum origin of friction}
\label{sec:fric}

Quantum friction is associated with the consumption of energy in the generation of coherence, which thereafter dissipates to the bath. In a sense, coherence constitutes potential work \cite{scully2003extracting}, and the process of coherence generation can be viewed as a temporary storage of energy in the coherence degree(s) of freedom. When these modes decay, the associated potential work is lost. In terms of the work consumption, the dissipation of coherence is equivalent to dissipation of work and amounts to an additional cost.  While such dissipation generally degrades the engines performance \cite{francica2019role,dann2020quantum,feldmann2003quantum}, it also speeds it up.

As in classical engines, quantum friction emerges naturally under rapid external driving. The driving generates coherence, which in turn, leads to a higher work cost and friction but a higher speed of operation. Generation of coherence is closely related to non-adiabatic quantum  dynamics, which occurs whenever the system Hamiltonian does not self-commute at different times $[\hat H(t),\hat H(t')] \ne 0$ \cite{kosloff2002discrete}.

We employ the quantum qubit model to study the influence of quantum friction on the cycle performance. This model is simple enough to allow an explicit solution and includes the sufficient condition for observing quantum friction. That is, 
the qubit working medium does not self commute if $\epsilon(t)$ and $\omega(t)$, in Eq. (\ref{eq:hamil}), are not proportionate to each other.  
A natural time-dependent framework to describe the dynamics of the working medium employs the
set of time-dependent quantum operators
\begin{eqnarray}
\begin{array}{l}
\hat H = \omega(t) \hat S_z +\epsilon(t) \hat S_x \\
\hat L = \epsilon(t) \hat S_z -\omega(t) \hat S_x\\
\hat C = \Omega(t) \hat S_y~~.
\end{array}
\label{eq:hlc}
\end{eqnarray}
This operator basis set $\v v = \{\hat H,\hat L, \hat C\}^T$,
completely defines the state
of the working medium (Cf. Appendix \ref{appendixA})
\begin{equation}
\hat \rho = \frac{1}{2}\hat I +
\frac{2}{(\hbar \Omega)^2}\left(
    \langle \hat H \rangle \hat H
    +\langle \hat L \rangle \hat L+
    \langle \hat C \rangle \hat C
    \right)~~,
    \label{eq:hlc-state}
\end{equation}
which rotates with respect to the static polarization basis set
Eq. (\ref{eq:statew}).
The advantage of such a representation is the straightforward thermodynamic interpretation, where the energy, ${\cal E}=\mean{\hat{H}}$, and coherence of the qubit 
\begin{equation}
    {\textfrak C} = \frac{1}{\hbar \Omega}
    \sqrt{ \langle \hat L \rangle^2 + \langle \hat C \rangle^2}~~,
    \label{eq:coherence}
\end{equation}
have a simple geometric interpretation in the parameter space $\{\hat{H},\hat{L},\hat{C}   \}$.
The coherence measure ${\textfrak C}$ can be viewed as the distance of the state from a dephased state,
diagonal in the energy representation.
${\textfrak C}$ serves as a quantifier of coherence similar to the divergence introduced in
Eq. (\ref{eq:divergence})   \cite{feldmann2016transitions}.

The cost of generating coherence can be evaluated by recalling two invariants
of the unitary dynamics: the Casimir and the Casimir companion \cite{boldt2013casimir}. For the $SU(2)$ algebra in the $\v v$ basis, the Casimir Companion obtains the simple form
\begin{equation}
\bar X =\frac{1}{\b{\hbar \Omega}^2}\left( \langle \hat H \rangle^2
    +\langle \hat L \rangle^2 +
    \langle \hat C \rangle^2 \right)~~,
    \label{eq:casimir}
\end{equation}
and the Casimir is obtained by replacing the squares of expectation values by the expectation values of the squares, i.e., $\mean{\hat{H}}^2\ra \mean{\hat{H}^2}, \mean{\hat{L}}^2\ra \mean{\hat{L}^2}$, and $\mean{\hat{C}}^2\ra \mean{\hat{C}^2}$. 
One consequence of the invariance of $\bar{X}$, is the conservation of the polarization amplitude along
a unitary (isolated) stroke. Thus, starting from an initial equilibrium state with polarization $\bar S_i$,
the Casimir companion throughout the stroke becomes $\bar X = \bar S_i$
and the initial energy is $\langle \hat H \rangle_i= \Omega_i \bar S_i$. This implies that the final energy of the unitary stroke is of the form 
\begin{equation}
\langle   \hat H \rangle_f =
\sqrt{\b{\f{\Omega_f}{\Omega_i}}^2\mean{\hat{H}_i}^2- \b{\hbar\Omega_f{\textfrak C_f}}^2}\approx  \f{\Omega_f}{\Omega_i}\mean{\hat{H}_i}-\f{\hbar^2\Omega_i\Omega_f}{2\mean{\hat{H}_i}} {\textfrak C}_f^2~~,
\label{eq:trade}
\end{equation}
where the RHS is obtained in the limit of small coherence.
This relation allows identifying the quantum adiabatic energy   (first term on the RHS) corresponding to the optimal process, and an additional coherence ${\cal W}_{fric} \equiv|
{\cal W}-{\cal W}_{ideal}|\approx \f{\hbar^2\Omega_i\Omega_f}{2\mean{\hat{H}_i}} {\textfrak C}_f^2$, which arises from the non-adiabatic dynamics. ${\cal W}_{fric}$ equals the extra work required to generate coherence.

 We will now demonstrate that rapid unitary strokes lead to generation of coherence.
Employing equation (\ref{eq:hamil}) we can obtain the Heisenberg equation of motion
for the unitary strokes:
\begin{eqnarray}
\frac{1}{\Omega} \frac{d}{dt}
\left(
\begin{array}{c}
\hat H(t)  \\
\hat L(t)\\
\hat C(t)
\end{array}
\right)
=
\left(\left(
\begin{array}{ccc}
0 & \mu & 0\\
-\mu & 0 & 1\\
0 & -1 & 0
\end{array} \right)
+ \frac{\dot \Omega}{\Omega^2} \hat I\right)
\left(
\begin{array}{c}
\hat H(t)\\
\hat L(t)\\
\hat C(t)
\end{array}
\right)
~~,
\label{eq:TLS propagator}
\end{eqnarray}
where
\begin{equation}
{\mu}=\frac{\dot{\omega}\epsilon -\omega \dot{\epsilon }}{{\Omega}^{3}}   \label{eq:mu} 
\end{equation}
is the adiabatic parameter. 

Finite-time processes require $\mu\neq0$, and  in the limit $\mu \rightarrow 0$ we recover the adiabatic solutions. The exact relation between the stroke duration $\tau_{adi}$ and $\mu$ depends on the protocol. Generally, it can be expressed as
\begin{equation}
    \mu =\frac{K}{\tau_{adi}}
\end{equation}
where  $K=\b{\frac {d\omega}{ds}\epsilon-\omega \frac{d\epsilon}{ds}}/{\Omega^3}$, with $s=t/\tau_{adi}$. For constant $\epsilon$, 
$K$ simplifies to
$K =\frac{1}{\epsilon}(\frac{\omega_i}{\Omega_i}-\frac{\omega_f}{\Omega_f})$. 

For protocols that keep $\mu=\rm{constant}$, Eq. (\ref{eq:TLS propagator}) can be integrated to obtain the dynamical propagator $\Lambda_{adi}$. In general, a driven system's propagator depends explicitly on two reference times, $t_{initial}$ and $t_{final}$. We assume that $t_{initial}=0$, and therefore index the propagator only in terms of the final time.  The propagator of the unitary stroke of a product form:
$\Lambda_{adi}\b{t} ={\cal U}_1\b{t} {\cal U}_2\b{t}$, where
${\cal U}_1 \b t$ is a scaling by the compression ratio
\begin{equation}
    {\cal U}_1(t) =  {\cal C} {\hat I} = \frac{\Omega(t)}{\Omega(0)}\hat I
    \label{eq:u1}
\end{equation}
and ${\cal U}_2\b t$ represents the dynamical map of the polarization. In the $\{\hat{H},\hat{L},\hat{C}\}$ operator basis
\begin{eqnarray}
{\cal U}_2 \b t=  \frac{1}{\kappa^2}\left(
\begin{array}{ccc}
1+\mu^2c&\kappa \mu s&\mu(1-c)\\
-\kappa \mu s&\kappa^2 c&\kappa s\\
\mu(1-c)&-\kappa s&\mu^2+c\\
\end{array}
\right)~~,
\label{eq:prop}
\end{eqnarray}
where $\kappa=\sqrt{1+\mu^2}$ and
$s=\sin(\kappa \theta)$, $c=\cos(\kappa \theta)$ and
$\theta(t) = \int_0^t \Omega(t') dt'$.

Accelerating the driving increases $\mu$, which in turn increases the coupling of the Hamiltonian $\hat{H}$ and coherence related operators ${\hat{L}}$ and $\hat{C}$. Therefore, rapid driving transforms energy to coherence. The constancy of the Casimir companion Eq. (\ref{eq:casimir}) implies that when the final state exhibits coherence, the work extraction relative to the equivalent adiabatic procedure is degraded. 

\subsection{Slow driving regime}
We can use Eq. (\ref{eq:prop}) to  estimate the additional fraction of work 
during the unitary strokes
due to the finite-time operation. Assuming slow driving ($\mu\ll1$ or long stroke duration), we expand ${\cal U}_2$ up to second order in the adiabatic parameter $\mu$ to obtain
\begin{equation}
\frac{{\cal W}_{fric}}{\cal W} \approx \mu^2
\label{eq:wextra}~~.
\end{equation}
This expression relates the ratio of the additional work that is consumed due to  friction, Eq. (\ref{eq:trade}), and the total work $\cal W$,  to the adiabatic parameter. Hence, in the slow driving regime, speeding up the stroke requires additional work. The corresponding work cost for  coherence generation is in accordance
with the notion of geometric thermodynamic distance and the low dissipation limit \cite{brandner2020thermodynamic,Abiuso2020geometric}. When the dissipation becomes significant, the power loss can exceed the gain, which imposes a minimum stroke duration for engine operation.

\subsection{Sudden limit}
In the opposite driving regime, including a sudden modulation of the driving parameters, the dynamical propagator is obtained by employing the sudden approximation \cite{messiah1962quantum}. The propagator in the sudden limit, $\tau_{adi} \rightarrow 0$, is given by
\begin{eqnarray}
\Lambda^{sudd}_{i\ra f}=
\frac{\Omega_f}{\Omega_i}
\left(
\begin{array}{ccc}
\cos(\Phi)&\sin(\Phi)&0\\
\sin(\Phi)&-\cos(\Phi)&0\\
0&0&1\\
\end{array}
\right)~~,
\label{eq:sudden_adiabats_prop}
\end{eqnarray}
where $\Phi =\phi_f-\phi_i$ is the angle of rotation 
between the initial and final polarizations and $\phi=\arccos(\omega/\Omega)$.
In the sudden limit, the work becomes
\begin{equation}
{\cal W} =\mean{\hat{H}\b 0}\b{\f{\Omega_f}{\Omega_i}\cos \b{\Phi}-1}~~,
\label{eq:sudword}
\end{equation}
and the ratio between the frictional and total work is
\begin{equation}
\Bigg |\frac{{\cal W}_{fric}}{\cal W}\Bigg | =\f{1-\cos \b{\Phi}}{|\cos \b{\Phi}-\b{\Omega_i/\Omega_f}|}~~.
\label{eq:sudw}
\end{equation}
The frictional work dissipates during the open stroke (isotherm or isochore) which takes place after the unitary stroke. 
For a compression protocol ($\Omega_f>\Omega_i$) the ratio can diverge since the work may vanish when $\cos \b{\Phi}=\Omega_i/\Omega_f$. In contrast, during an expansion process the ratio is bounded by $\f{2\Omega_{f}}{\Omega_{i}+\Omega_{f}}$.

\subsection{Shortcuts to adiabaticity}
\label{subsec:STA}

The argument that fast dynamics on the adiabats generates coherence and leads to friction like phenomena \cite{kosloff2002discrete,feldmann2003quantum} has a loophole.
The unitary dynamics on the adiabats is in principle reversible. Since the dissipation
of coherence, which seals the loss, does not take place until the thermalization stroke that follows the adiabatic stroke, protocols that null the coherence at the end of the adiabat  will be frictionless.

\begin{figure}[htb!]
\centering
\includegraphics[width=7. cm]{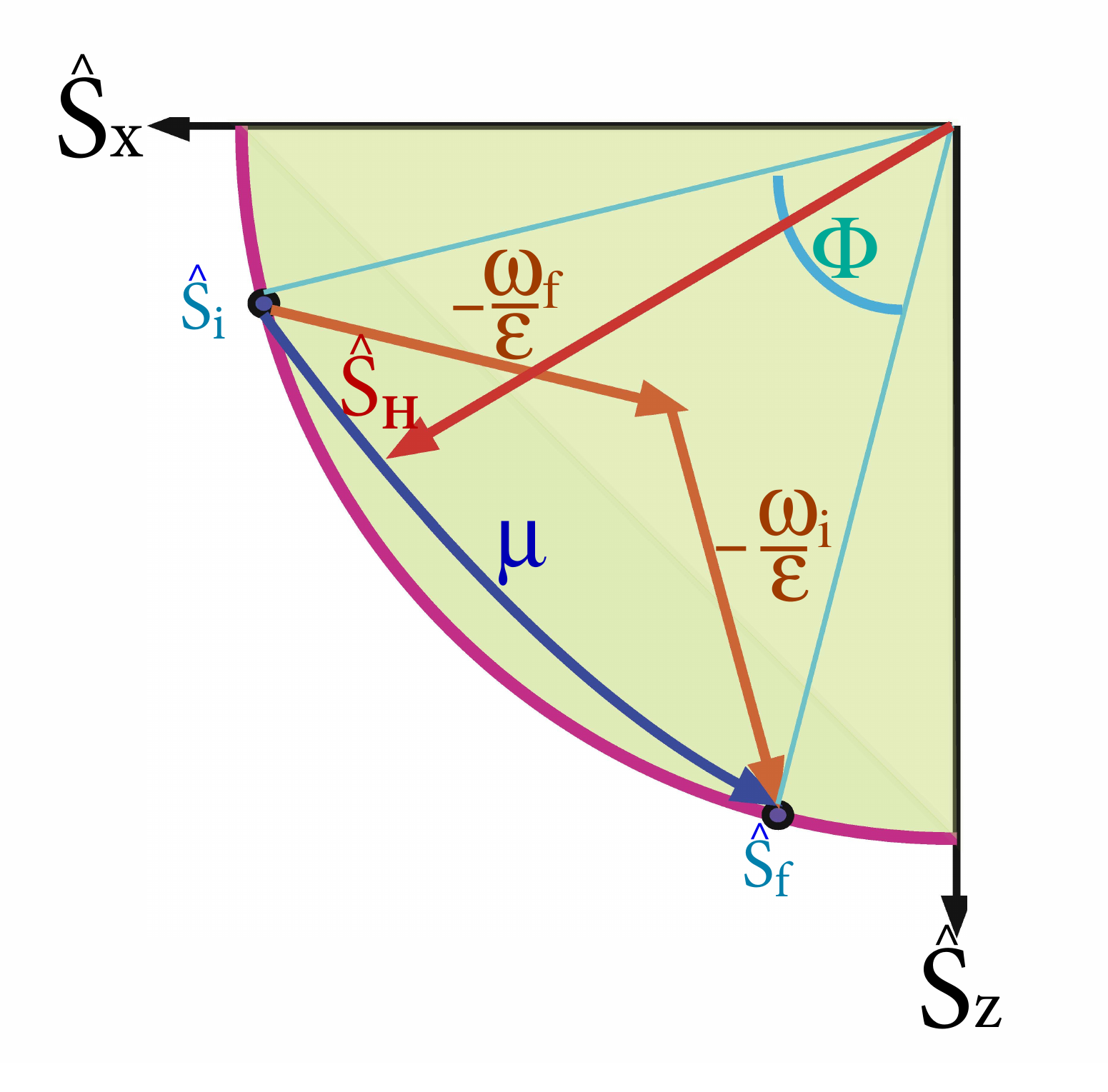}
\caption{Optimal frictionless trajectories for the unitary
stroke displayed on the polarization axis 
of $\mean{\hat{S}_x}$ and $\mean{\hat{S}_z}$. The radius of the semicircle is equivalent to the initial polarization $\bar S_i= \bar S_{H_i}$ which is on the energy axis. 
Since the unitary dynamics preserves the polarization and the chosen energy direction is
on the $x,z$ plane, the semicircle shows all
the possible states on the energy axis. Any interior point possesses coherence. In orange is the optimal bang-bang protocol composed of two segments
the first with a slope of $-\omega_f/\epsilon$ and the second
$-\omega_i/\epsilon$. The blue arrow represents the
constant $\mu$ protocol. 
}
\label{fig:6}
\end{figure}

Examining Eq. (\ref{eq:prop}), we  find that solutions for which $\cos(\theta_f)=1$ are frictionless. These solutions impose a quantization rule on $\mu$:
\begin{equation}
    \mu_l = \frac{1}{\sqrt{\left(\frac{ 2 \pi l}{\Phi } \right)^2-1}}~~~~~~~~,~~ l=1,2,...
    \label{eq:quantum}
\end{equation}
where we used the identity $\theta=-\Phi/\mu$ \footnote{The identity is derived by substituting Eq. (\ref{eq:control_protocols}) into the expression for $\mu$, arranging the equation and integrating.}.
The relation between $\mu $ and the stroke duration leads to the  minimum constant  stroke duration
\begin{equation}
    \tau_{adi}(l=1)= K \sqrt{\left(\frac{ 2 \pi}{\Phi} \right)^2-1}~~.
    \label{eq:tmin}
\end{equation}
Adiabatic trajectories that begin and end
with no coherence are frictionless (cf. Fig. \ref{fig:6}). 
In the limit of small and constant $\epsilon$ we get
$\tau_{adi}(l=1) \propto \epsilon ( \frac{1}{\omega_f^2}-\frac{1}{\omega_i^2} )$.
These frictionless protocols are termed shortcuts to adiabaticity (STA) \cite{chen2010fast,guery2019shortcuts}. 
At intermediate times coherence is generated requiring extra work, but if there is no dissipation in the drive this coherence is converted back by the working medium, arriving at the final target with no coherence. The associated speedup  \cite{chen2010fast,guery2019shortcuts} may come with an accompanying cost if the control is prone to additional dissipation.
Here we consider the ideal case, assuming no dissipation and view the temporary investment of energy as a catalytic process since this energy can in principle be recouped \cite{torrontegui2017energy}. An opposite viewpoint
considers the average energy, stored during the shortcut, as wasted work  
\cite{ccakmak2019spin}.

Can the protocol duration  be shortened further? This is a problem in the framework of quantum control, a field which governs tasks related to manipulation of quantum systems by external fields under defined restrictions. The present control task is to transfer an 
initial thermal state $\hat \rho_i =\frac{1}{Z} e^{ - \hat H_i/k_B T}$ to a final thermal state
$\hat \rho_f =\frac{1}{Z} e^{ - \hat H_f/k_B T'}$
as fast as possible on the unitary strokes of the cycle. Optimal control theory has been applied to address this task
\cite{boldt2012time}, obtaining a the minimum time solution, a so called Fastest Effectively Adiabatic Transition (FEAT) \cite{boldt2016fastest}.
The task of minimizing the time can be reduced to minimizing $\int \hat{S}_z d\hat{S}_x$ while following the dynamics generated by Eq. (\ref{eq:hamil}), which here gives
\begin{equation}
    \frac{d \hat S_z }{d \hat S_x} = -\frac{\omega}{\epsilon}~~.
\end{equation}
For fixed $\epsilon$ and for $\omega$ in the range $\omega_i < \omega(t) < \omega_f$, the geometric solution is to keep the curve as close to the $\hat{S}_x$ axis as possible until the last moment to reach the final state, at which time the solution switches to the steepest curve possible \cite{boldt2012time}.  The solution is thus of the bang-bang type, switching from the initial
$\omega = \omega_i$ to the final $\omega = \omega_f$ to get the process started, keeping $\omega(t) = \omega_f$ for time $0 \leq t < \tau_1$, switching back to 
$\omega(t) = \omega_i$ for a time $\tau_1 \leq t < \tau_1+\tau_2$ and finally switching to $\omega(\tau_1+\tau_2) = \omega_f$ to reach the final state. The resulting two line segment trajectory is shown in Fig. \ref{fig:6}.
The coherence generated during this protocol can be seen as the distance from the purple quarter-circle  of zero coherence. The total FEAT time reads
\begin{equation}
    \tau_{adi}(opt)= \tau_1 + \tau_2 =
    \left(\frac{1}{2 \Omega_i } +\frac{1}{2 \Omega_f }\right) \arccos(\zeta)
\end{equation}
where $\zeta = \frac{\Omega_i \Omega_f (\epsilon^2 + \omega_i \omega_f)-(\epsilon^2 + \omega_i \omega_f)^2}
{\epsilon^2(\omega_i-\omega_f)^2}$. While the FEAT time is much shorter than the time for a constant $\mu$ protocol,
$\tau_{adi}(opt) < \tau_{adi}(l=1)$, the FEAT solution does pay a significantly higher price in intermediate coherence. 

Additional insight can be obtained by adding another control operator; a counter-diabatic term \cite{ccakmak2019spin,funo2019speeding} with the control function $\upsilon(t)$ to the Hamiltonian Eq. (\ref{eq:hamil})
\begin{equation}
    \hat H_{CA}= \upsilon(t) \hat S_y~~.
    \label{eq:ca}
\end{equation}
This term generates a rotation around the $y$ axis in the $z,x$
plane which can rotate the initial to the final Hamiltonian 
in a rate depending on the frequency $\upsilon (t)$. If $\upsilon(0)= \upsilon(t_f)=0$
energy is only stored temporarily in the counter-diabatic drive, which classifies it as a catalyst.

To summarize, fast frictionless protocols for the unitary strokes are possible provided
the coherence is not transferred to the thermalization strokes. The price for acceleration of the stroke is generation of intermediate coherence which requires a temporary investment of power.
If no restrictions are imposed on the power invested, or analogously on the range of $\omega\b t$, the time period $\tau_{adi}$ can be shrunk to zero. Another scheme to achieve a vanishing time period includes adding an unrestricted
counter-diabatic term, Eq. (\ref{eq:ca}), to the Hamiltonian \cite{ccakmak2019spin}. For a more realistic description of the storage device, restriction on the control are introduced. One possible restriction is to limit the averaged 
stored energy. Another possibility is to restrict coherence.
In principle, all the temporary power can be retrieved in the external 
controller when there the drive is completely isolated. However, in practice this is an idealization, and any real storage device is sure to have some dissipation. Thus, one expects some dissipation from the controller \cite{tobalina2019vanishing}.

\section{Thermalization}
\label{sec:thermalization}

Thermalization is the process of relaxing the system toward equilibrium
with an external heat reservoir, e.g., the hot or cold baths. The relaxation is mediated by the system-bath interaction term $\hat H_{s-h/c}$ (Eq. \ref{eq:htotal}), which generally depends on both the system and bath operators as well as the coupling strength $g$. The question arises, can we actively influence the thermalization process?
Three options for control are possible. The first, which we consider here, is to control the system Hamiltonian \cite{dann2018shortcut,dann2020fast}, the second is to vary the coupling strength \cite{pancotti2020speed,villazon2019swift}, and the third is controlling the temperature of the bath.

\subsection{Isochoric thermalization}

Compared to quantum Carnot-type cycles the analysis of the quantum Otto cycle is simplified, since the thermalization strokes are carried out at constant frequency $\Omega$. For this reason, it is has been thoroughly studied, and originally constituted the main platform to investigate thermodynamics at the quantum level \cite{henrich2007quantum,kosloff2017quantum}. 

By definition during an isochore the Hamiltonian is static. As a result, the only adjustable control parameter of the thermalization is the contact time with the bath. This is equivalent to adjusting the system bath coupling. The dynamics along the stroke
is described by Eqs. (\ref{eq:lind}) and (\ref{eq:lind1})
with a constant Hamiltonian. These lead to transfer of energy and exponential decay of coherence,
Eq. (\ref{eq:decaycoher}), until the system reaches equilibrium. 

\subsection{Isothermal thermalization}
\label{subsec:isothermal}

We here concentrate on the finite-time thermalization strokes, which transfer heat to and from the engine. Within the limits imposed by the isotherms of the working medium, we can find various choices for cycles with finite power \cite{wu2000finite,yin2017optimal,dong2013magnetic}.

Thermalization can be controlled by varying the Hamiltonian, while the system is simultaneously coupled to the bath. 
We now consider the general case, where the Hamiltonian does not commute with itself at different times, $[\hat H(t),\hat H(t')] \ne 0$. A prerequisite for obtaining control, is to derive a dynamical description which is  accurate and consistent with thermodynamics. Such a dynamical description has been formulated in Ref. \cite{dann2018time}, where a Non-Adiabatic Master Equation (NAME) was developed incorporating the effect of the external driving. 

Within this framework we can address the issue of actively speeding up the thermalization. 
Typically, the rate of approaching equilibrium is proportional to deviation of the state from the fixed point of equilibrium. As we get closer to the target, the rate decreases. Broadly speaking, the strategy of speeding up the thermalization
is to first generate coherence which moves the system away from the instantaneous attractor. As a result, the relaxation is 
enhanced. At the final stage the system is rotated, converting the coherence to energy to reach the desired thermal state.
The speedup comes with a price since, in contrast to the unitary strokes, during the open strokes we cannot separate the unitary drive from
the dissipative loss.

\subsection{Shortcut to equilibrium protocols}
\label{sec:STE protocols}

Shortcut to equilibrium protocols (STE), are active control protocols, generating a rapid transition between two equilibrium states with different Hamiltonians, while the system is coupled to a bath of a fixed temperature. This control task requires modifying the system entropy, in contrast to the common scheme of unitary control. 
The shortcut protocols fall within the framework of control of open quantum systems. The rules of the game we consider restrict the active control to only the system Hamiltonian. The bath Hamiltonian is set and cannot be controlled, and the bath remains in a thermal state due to its enormous size and the negligible influence of the qubit.

In the presence of non-adiabatic driving, the external field dresses the system, which consequently effectively modifies the system-bath interaction. As a result, the external driving enables indirect coherent control of the system's dissipative dynamics. As previously stated, control of the qubit state  during the isothermal strokes
requires  prior knowledge of the open system dynamics. In turn, to describe the reduced dynamics of the open system, one first requires a closed form solution of the driven isolated system. To be specific, the derivation of the Non-Adiabatic Master Equation (NAME) requires, as an input, the
free dynamics of the driven system \cite{dann2018time}. The solution is non-trivial in the presence of non-adiabatic driving, when the Hamiltonian does not commute with itself at different times. For arbitrary driving, constructing an explicit solution of  the dynamical propagator requires a time-ordering procedure \cite{messiah1962quantum}. 

 We have developed an algebraic procedure to circumvent the time-ordering problem by employing a dynamical operator basis. This technique is closely related to the inertial theorem \cite{dann2018inertial}. The theorem implies that for a closed operator algebra, the dynamical propagator can be obtained for a family of non-adiabatic protocols, characterized by a slow `acceleration' of the drive. The associated solutions and driving protocols are termed  inertial solutions and protocols. These solutions are conveniently expressed as linear combinations of the eigenoperators of the propagator.

For the qubit working medium, represented by Eq. (\ref{eq:hamil}), the inertial protocol is characterized by a slowly varying adiabatic parameter, i.e., $\f{1}{\Omega}\f{d\mu}{dt}\ll 1$. Under this condition, the  dynamics approximately follows the inertial solution. This solution is conveniently expressed in terms of the dynamics of the basis of operators $\v v \b t =\{\hat{H},\hat{L},\hat{C}\}^T$ Eq. (\ref{eq:hlc})
\begin{equation}
    \v{v}\b t=\f{\Omega\b t}{\Omega\b 0}P\b{\mu\b t} e^{-i\int_{0}^{t}D\b{t'}\Omega\b{t'}dt'}P^{-1}\b{\mu\b t}\v v\b{0}~~,
    \label{eq:inertial_solution}
\end{equation}
where $P$ is a $3$ by $3$ matrix which is dependent on the instantaneous adiabatic parameter $\mu\b t$, see Appendix \ref{apsec:explicit_expressions}, and $D=\rm{diag}\b{0,\kappa,-\kappa}$ with $\kappa = \sqrt{1+\mu^2}$. The three operators obtained from $\b{\hbar\Omega\b t}^{-1} P^{-1}\v{v}\b t$ are eigenoperators of the propagator. We introduce a scaled version of these operators $\v g =\{\hat \chi,\hat \sigma,\hat \sigma^\dagger\}^T$, satisfying an eigenvalue type relation $\hat{\sigma}^H\b{t}=\hat{U}^\dagger\b{t}\hat{\sigma}\b{0}{\hat{U}\b{t}}=e^{-i\int_0^{t}dt'\kappa\b{t'}\Omega\b{t'}}\hat{\sigma}\b 0$  where $\hat{U}\b{t}$ is the propagator and superscript $H$ designates operators in the Heisenberg picture.
The operator $\hat{\chi}^H\b{t}=\hat{\chi}\b{0}$ is the inertial invariant, i.e., the eigenoperator with a vanishing eigenvalue.
Expressing the eigenoperators  in terms of the $\{ \hat H,\hat L, \hat C \}$ basis, we obtain
\begin{eqnarray}
\label{eq:chi-sigma}
    \hat \chi\b t=\f{\sqrt{2}}{\kappa\hbar\Omega}\b{\hat H+\mu \hat C}\\
    \hat \sigma\b t=\f 1{\kappa \hbar \Omega}\b{-\mu \hat{H}-i\kappa \hat{L}+\hat C}~~,
    \nonumber
\end{eqnarray}
where all the parameters may be time-dependent.
The eigenoperators in $\v g$ are orthonormal with respect to the inner product in Liouville space, $\b{\hat{A},\hat{B}}=\rm{tr}\{\hat{A}^\dagger\hat{B}\}$, and satisfy the $SU(2)$ commutation relations of the form $\sb{\hat{\sigma},\hat{\sigma}^{\dagger}}=-\sqrt{2}\hat{\chi}$, $\sb{\hat{\chi},\hat{\sigma}}=-\sqrt{2}\hat{\sigma}$.
Appendix \ref{appendixA} summarizes  the relation between the various basis sets of expansion operators $\v{s}=\{\hat{S}_x,\hat{S}_y,\hat{S}_z\}^T$, $\v v=\{\hat{H},\hat{L},\hat{C}\}^T$ and $\v g =\{\hat{\chi},\hat{\sigma},\hat{\sigma}^\dagger\}$, see also Fig. \ref{fig:1} for a geometric representation.

Combining the inertial solution, Eq. (\ref{eq:inertial_solution}), for the isolated system dynamics with the NAME leads to a master equation for a broad range of driving protocols. The master equation  is valid from first principles under the following conditions: (i) The bath dynamics are rapid relative to the  typical timescales of both the system and the driving, $\tau_s$ and $\tau_d$, i.e, $\tau_b\ll\tau_s, \tau_d$, where $\tau_b$ is the typical timescale of the decay of correlations in the bath (ii) The system-bath relaxation time $\tau_r$ is large relative to the system and bath timescales, i.e, $\tau_r\gg \tau_s,\tau_b$. (iii) The driving protocol satisfies the inertial condition, $\f{1}{\Omega}\f{d\mu}{dt}\ll 1$. Condition (i) is associated with Markovian dynamics and (ii) corresponds to a weak system-bath interaction, also known as the weak coupling limit \cite{davies1974markovian,breuer2002theory}.

In the interaction picture relative to the system-bath bare Hamiltonian, the qubit's open system dynamics  obtains the familiar GKLS form \cite{dann2020fast}
\begin{eqnarray}
\label{eq:open_sys_dynamics}
    \f d{dt}\tilde{\rho}=\tilde{\cal{L}}\b t\sb{\tilde{\rho}}=k_\downarrow\b{\alpha\b t}\b{\hat{\sigma}\tilde{\rho}\b{t}\hat{\sigma}^{\dagger}-\f{1}{2}\{\hat{\sigma}^{\dagger}\hat{\sigma},\tilde{\rho}\b t\}}\\~~~~~~~~~~~~~~~~~~~+k_{\uparrow}\b{\alpha\b t}\b{\hat{\sigma}^{\dagger}\tilde{\rho}\b t\hat{\sigma}-\f 12\{\hat{\sigma}\hat{\sigma}^{\dagger},\tilde{\rho}\b t\}}~~.
    \nonumber
\end{eqnarray}
Here, $\hat{\sigma}$ and $\hat \sigma^\dagger$ designate operators at initial time and overscript tilde denotes operators in the interaction picture. The kinetic coefficients of Eq. (\ref{eq:open_sys_dynamics}) depend on the spectral features of the bath and the effective time-dependent frequency $\alpha$. This frequency serves as an effective generalized Rabi frequency of the driven system
\begin{equation}
    \alpha\b t= \kappa\b{t} \Omega\b t~~ = \sqrt{1+\mu(t)^2} \,\, \Omega\b t~~.
\label{eq:alpha}
\end{equation}

 In the quantum adiabatic regime, $\mu\ra 0$ and $\alpha$ converges to  the instantaneous Rabi frequency, $\Omega\b t$. For $\mu>0$, the effective frequency $\alpha(t) > \Omega\b t$. This is the outcome of an effective dressing of the system by the driving. As a consequence of the rapid driving, the bath interacts with the dressed system, leading to deviations from the adiabatic dynamics. For the general case, there may be multiple effective frequencies  $\{\alpha\}$. Their exact form depends on a particular system-bath interaction and the defined spectral density \cite{dann2018time,dann2020fast}.

For the present analysis we assume a bosonic bath with an Ohmic spectral density. The system-bath interaction is taken as $\hat{H}_{sb}=i g\sum_k\sqrt{\f{2\pi \omega_k}{V \hbar }}\b{\hat{b}_k-\hat{b}_k^\dagger}\hat S_y$, where $\hat{b}_k^\dagger$ and $\hat{b}_k$ are the creation and annihilation operators of the $k$'th bath oscillator, and  $\omega_k$ is the oscillator frequency. The coupling strength is represented by $g$ and $V$ is the reservoir size. For a large reservoir in equilibrium, the kinetic coefficients become 
\begin{eqnarray}
    k_{\downarrow}\b{\alpha}=\f{g^2\alpha}{ \hbar c \kappa}\b{1+N\b{\alpha}}   
    \label{eq:kinetic_coefficients}\\
    k_{\uparrow}\b{\alpha}=\f{ g^2 \alpha}{\hbar c \kappa }N\b{\alpha}~~.
    \nonumber
\end{eqnarray}
where $c$ is the speed of event propagation in the bath and $N\b{\alpha}=1/\b{\exp\b{\hbar \alpha/k_B T}-1}$ is the Bose-Einstein distribution, characterizing the correlations between bath modes at frequency $\alpha$. It is simple to verify that these kinetic coefficients satisfy detailed balance with respect to $\alpha$, Eq. (\ref{eq:detailed balance}). This property is essential for a thermodynamically consistent dynamical description \cite{kosloff2013quantum,alicki2018}. In the adiabatic limit, the kinetic coefficients converge to adiabatic rates and the Lindbald jump operators to the creation annihilation operators of the two-level system. As expected, equation (\ref{eq:open_sys_dynamics}) then converges to the adiabatic master equation \cite{albash2012quantum}.

The NAME of the driven qubit, Eq. (\ref{eq:open_sys_dynamics}) propagates the qubit state in the direction of the instantaneous attractor. The attractor is defined by the relation ${\cal L}\b t \sb{\tilde{\rho}_{i.a}} = \tilde{\rho}_{i.a}  $, where $\tilde{\cal{L}}\b t$ is a superoperator which generates the dynamics in the interaction picture. 
For the qubit, the attractor is in the direction of $\hat \chi$, mixing energy and coherence. The attractor is  rotated by the angle $\xi=\arccos(1/\sqrt{1+\mu^2})$ from the energy axis.
The attractor can be expressed 
in the Gibbs form: 
\begin{equation}
    \tilde{\rho}_{i.a}\b t =Z^{-1} e^{-\f{\hbar \alpha\hat{\chi}}{\sqrt{2}k_B T}}~~, 
    \label{eq:TLS_instant_attractor}
\end{equation}
where $Z=\rm{tr}\b{e^{-\hbar\alpha\hat{\chi}/\sqrt{2}k_B T}}$ is the partition function. In the presence of driving, the attractor varies in time, and the system continuously aspires towards a changing target, but does not manage to reach it. Only at the initial and final times, is the driving stationary and the qubit reaches the attractor.

The qubit control is based on the master equation, Eq. (\ref{eq:open_sys_dynamics}). Our present control target is to speed up the thermalization while changing the qubit Hamiltonian. This step will be employed in the open strokes of Carnot-type engines in Sec. \ref{sec:local}.
Specifically, we desire a control protocol which transfers an initial Gibbs state, defined by $\Omega \b 0 =\Omega_i$ and temperature $T$,  to a final Gibbs state 
of the same temperature and final frequency $\Omega\b{t_f}=\Omega_f$.
Moreover, we assume that the system at initial and final times is stationary with no external driving.
The control agents are the parameters of the free Hamiltonian $\omega\b t$ and $\eps\b t$. Notice that these parameters only indirectly effect the master equation. To find a control we opt to employ a reverse engineering approach, in which, we propose a trajectory for the qubit state that forms a solution to the master equation. In turn, this solution determines the kinetic coefficients of the master equation, from which we can extract the direct control parameters.

Preforming the analysis in the interaction representation relative to the bare system Hamiltonian simplifies the control scheme.
In this frame the Lindblad jump operators, $\hat{\sigma}$ and $\hat \sigma^\dagger$, vary slowly with $\mu$.

The control trajectory, which is a dynamical solution of Eq. (\ref{eq:open_sys_dynamics}), is obtained by
representing the state $\tilde \rho$ in terms of the basis of eigenoperators $\v g$ of the free dynamics, Appendix \ref{appendixA}:
\begin{equation}
    \tilde{\rho}=\f 12\hat I+c_{\sigma}\hat \sigma+c_{\sigma^{\dagger}}\hat \sigma^{\dagger}+c_{\chi}\hat \chi~~,
    \label{eq:rho_open_sys}
\end{equation}
where $c_{r}=\rm{tr}\b{\hat{r}\tilde{\rho}}$, with
$r=\sigma, \sigma^\dagger,\chi$, are time-dependent coefficients. Substituting Eq. (\ref{eq:rho_open_sys}) into (\ref{eq:open_sys_dynamics}) and utilizing the orthogonality of the eigenoperators, leads to an equivalent representation of the dynamics
\begin{eqnarray}
\f d{dt}c_{\chi}=-\b{k_{\downarrow}\b t+k_{\uparrow}\b t}c_{\chi}-\f 1{\sqrt{2}}\b{k_{\downarrow}\b t-k_{\uparrow}\b t} \label{eq:chi_sigma1}\\
\f d{dt}c_{\sigma}=-\f 12\b{k_{\downarrow}\b t+k_{\uparrow}\b t}c_{\sigma}~~.
\label{eq:chi_sigma2}
\end{eqnarray}
and similarly for $c_{\sigma^{\dagger}}$. These equations completely determine the system dynamics and form the template for coherent control. What is missing is are the boundary condition. 
The choice of the initial and final Gibbs state along with the condition of stationarity at initial and final times  imposes boundary conditions on Eq. (\ref{eq:chi_sigma2}).

We can simplify the problem by eliminating Eq. (\ref{eq:chi_sigma2}). For the boundary conditions (and any initial  diagonal state in the energy representation) the coefficients $c_\sigma \b 0= c_{\sigma^\dagger}\b 0=0$ and $\mu=0$. These relations together with Eq. (\ref{eq:chi_sigma2}) imply that $c_\sigma$ and $c_{\sigma^\dagger}$ vanish at all times.

We can now focus on a single equation, Eq. (\ref{eq:chi_sigma1}), with boundary conditions:\\
$c_\chi\b{0}=-\frac{1}{\sqrt{2}}\rm{tanh}\b{\f{\hbar\Omega\b 0}{2k_B T}}$, $c_\chi \b{t_f}=-\frac{1}{\sqrt{2}}\rm{tanh}\b{\f{\hbar\Omega\b{t_f}}{2k_B T}}$, and $\mu\b 0=\mu\b{t_f}=0$. In addition, the initial and final states and  Eq. (\ref{eq:chi_sigma1}) imply that $\dot{c}_\chi\b 0 =\dot{c}_\chi\b{t_f}=0$.

To proceed, we determine the trajectory solution through the coefficient $c_{\chi}$.
We choose the most simple polynomial solution which is compatible with the boundary conditions. In this case, a third order polynomial is sufficient. In terms of a dimensionless parameter
$s=t/t_f$, the solution reads
\begin{equation}
    c_\chi\b{s}=c_\chi\b 0+3\Delta s^{2}-2\Delta s^{3}~~,
    \label{eq:poly_chi}
\end{equation}
where $\Delta = c_\chi\b{t_f}-c_\chi\b{0}$.
Next, we substitute the solution Eq. (\ref{eq:poly_chi}) into  Eq. (\ref{eq:chi_sigma1}) and obtain the kinetic coefficients, from which we can extract $\alpha\b t$, Eq. (\ref{eq:kinetic_coefficients}). These steps are achieved utilizing a common numerical solver.

The control function $\Omega\b t$ is now evaluated by solving the master equation $\Omega= \alpha/\kappa$ for a set of defined controlled parameters $\omega\b t$ and $\epsilon\b t$.
In practice the master equation  depends on the generalized Rabi frequency $\Omega \b t$ and $\dot \phi$ (through $\mu$), this means that we have an additional freedom in the control parameters of the Hamiltonian.

We chose to parameterize the control parameters in terms of the time-dependent frequency $\Omega$ and the phase $\phi$:
\begin{eqnarray}
    \omega\b t=\Omega\cos\b{\phi}
    \label{eq:control_protocols}\\
    \epsilon \b t = \Omega\sin\b{\phi}~~.
    \nonumber
\end{eqnarray}
In this parametrization the adiabatic parameter becomes $\mu=-\dot{\phi}/\Omega$ and the effective frequency can then be expressed as 
\begin{equation}
\alpha=\sqrt{1+\b{{\dot{\phi}}/{\Omega}}^2}\Omega~~.
\label{eq:alpha2}
\end{equation}
To set the angle $\phi$, we study two protocols which differ by their boundary conditions. The first is a quadratic function of time $\phi\b t=a\b{t-{2t^2}/{3t_f}}$, where is $a$ a dimensionless free parameter taken to be equal to the numerical value of $1/t_f^2$ in the model units. This protocol leads to a final value for the angle which scales with the duration time $\phi\b{t_f}\propto t_f$. The second protocol starts at $\phi\b 0 =0$ and ends at $\phi\b{t_f} =\pi/2$, where the direction of the final Hamiltonian is rotated by ninety degrees relative to the initial Hamiltonian. Introducing a polynomial which complies with the boundary conditions leads to $\phi\b t={\pi}t^2\b{3 t_f-2t}/\b{6t_f^3}$. Both protocols satisfy the required condition of stationarity at initial and final times: $\dot{\phi}\b 0=\dot{\phi}\b{t_f}=0$. Finally, solving Eq. (\ref{eq:alpha2}) for $\Omega\b t$ leads to the control protocol.

Overall, the constructed shortcut to equilibration (STE) protocol rapidly modifies the system entropy, transferring an initial thermal state with a Rabi frequency $\Omega_i$ to a thermal state of a frequency $\Omega_f$ at the same temperature. In Ref. \cite{dann2020fast}
a different STE protocol has been introduced, utilizing a product state consisting of exponentials, see Appendix \ref{appendixA}. In contrast, here we choose a linear combination of eigenoperators, Eq. (\ref{eq:rho_open_sys}), which is the natural approach for a system described by a compact algebra. This choice has the advantage of leading to a simpler analysis.

\subsection{Thermodynamic cost of finite-time thermalization}
\label{subsec:5.4}
Fast driving moves the system away from equilibrium, leading to enhanced dissipation. The thermodynamic cost can be characterized by the entropy production rate
\begin{eqnarray}
\label{eq:entropy_production_name}
    \Sigma^u\equiv -\f{d}{dt}{\cal D}\b{\hat{\rho} |\hat{\rho}_{i.a}}=-k_{B}\rm{tr}\b{{\tilde{\cal L}}\sb{\tilde \rho}\rm{ln}\tilde{\rho}}+k_{B}tr\b{\tilde{\cal L}\sb{\tilde \rho}\rm{ln}\tilde{\rho}_{i.a}} \\~~~~~~~~~~~~~~~~~~~~~~~ =-k_{B}\rm{tr}\b{{\tilde{\cal L}}\sb{\tilde \rho}\b{\rm{ln}\tilde{\rho}+\f{\hbar\alpha}{\sqrt{2}T}\hat \chi}}~~.
    \nonumber
\end{eqnarray}
In the infinitely long time limit, the state $\tilde{\rho}$ converges to $\tilde {\rho}_{i.a}$ and the entropy production rate vanishes. 
The entropy production in this limit has been studied recently \cite{scandi2020quantum} and related to fluctuation theorems.

During the shortcut to equilibrium protocols $c_{\sigma,\sigma^\dagger}=0$  and the state is completely characterized by the expectation value $c_{\chi}=\rm{tr}\b{\hat \chi\tilde{\rho}}$,  $\tilde{\rho}=\f{1}{2} I + c_{\chi}\hat{\chi}$. Alternatively, this state can be  represented in a Gibbs form $\tilde{\rho} =Z^{-1}e^{-\beta \hat{\chi}}$. 
The role of $\beta$ motivates introducing an effective temperature of the qubit in the interaction representation: $T' \equiv \hbar \alpha/\sqrt{2}k_B \beta$.
Such a form allows a straightforward interpretation of Eq. (\ref{eq:entropy_production_name}), this is achieved by the following derivation. We begin by utilizing the Gibbs form of $\tilde{\rho}$ and insert Eq. (\ref{eq:TLS_instant_attractor}) into  (\ref{eq:entropy_production_name}) to obtain
\begin{equation}
    \Sigma_{\chi}^u=k_{B}\b{\f1{T'}-\f1{T}}\b{\f{\hbar\alpha}{\sqrt{2}k_{B}}}tr\b{\dot{\tilde{\rho}}\chi}
    \label{eq:Sigma_interm}~~~~.
\end{equation}
This relation can be interpreted as the product of a thermodynamic force $\propto -\nabla \f 1{T}$ and heat current in units of the energy quanta $\hbar \alpha$.
Next, we express $\tilde{\rho}$ in terms of $\beta$ to obtain: $\dot{\tilde{\rho}}=-\b{\mean{\chi}+1}\dot{\beta}\tilde{\rho}$, 
with $\mean{\hat{\chi}}=-\f{1}{\sqrt{2}}\rm{tanh}\b{\f{\beta}{\sqrt{2}}}$, and 
\begin{equation}
    \dot{\beta}=\rm{tr}\b{\f{d\tilde{\rho}}{dt}\tilde{\rho}^{-1}\hat{\chi}}=-\f 1{\sqrt{2}}\sb{\b{1+e^{-\sqrt{2}\beta}}k_{\downarrow}-\b{1+e^{\sqrt{2}\beta}}k_{\uparrow}}~~.
    \label{eq:beta_dot}
\end{equation}
Substituting Eqs. (\ref{eq:beta_dot}), (\ref{eq:kinetic_coefficients}) into Eq. (\ref{eq:Sigma_interm}) leads to the final expression
\begin{equation}
    \Sigma_{\chi}^u=\b{\f1{T'}-\f1{T}}\b{\hbar\alpha}
    \f{k_\downarrow\b{\alpha}\mean{\hat{\chi}}\b{\mean{\hat{\chi}}+1}}{2\b{1+e^{\hbar\alpha/k_{B}T'}}}\b{e^{-\hbar\alpha/k_{B}T'}-e^{-\hbar\alpha/k_{B}T}}~.
    \label{eq:Sigma_final}
\end{equation}
As expected, we obtain a positive entropy production. The first and last terms in the brackets have opposite signs while the expectation value of $\hat{\chi}$ satisfies $-1<\mean{\chi}<0$  for a positive temperature.  This leads to a symmetric dependence on the temperature gap $\Delta T = T-T'$, i.e. the entropy production  depends only on the magnitude of the gap and is independent of whether the working medium effective temperature is hotter or colder relative to the bath temperature.

In the high temperature limit $\hbar \alpha/k_B\ll  T, T'$ the relation can be further simplified, leading to entropy generation that scales as the square difference between inverse temperatures
\begin{equation}
    \Sigma_{\chi}^u\approx- k_{B} k_{\downarrow}\mean{\chi}\b{\f{\hbar \alpha}{2 k_B}}^2
    \b{\f 1{T'}-\f 1{T}}^{2}~~.
    \label{eq:sigma-class}
\end{equation}

We next derive the entropy production rate for a general initial state which includes coherence, following the inertial solution, Eq. (\ref{eq:chi_sigma2}). We begin by expressing the qubit state as a maximum entropy state $\tilde{\rho}=\bar{Z}^{-1}\exp\b{-\b{\bar{\beta}\hat{\chi}+\bar{\gamma_x}\hat{\sigma}_x+\bar{\gamma}_y\hat{\sigma}_y}}$, where $\sigma_{x}=\f{1}{\sqrt{2}}\b{\sigma + \sigma^{\dagger}}$ and $\sigma_{y}=\f{i}{\sqrt{2}}(\sigma - \sigma^{\dagger})$ see Appendix \ref{appendixA} for further details. 
The existence of such a form is guaranteed from the closure property of the operator algebra and the Baker–Campbell–Hausdorff formula \cite{alhassid1978connection}. Defining the effective thermodynamic forces ${\cal F}_l$ and effective"temperature":
${\cal F}_{\chi} =\frac{1}{T_{\chi}} -\frac{1}{T}$, where $T_{\chi}=\f{\hbar\alpha}{\sqrt{2}k_{B}\bar{\beta}}$, ${\cal{F}}_{\sigma_x}=\f{ k_B \bar{\gamma}_x}{ \hbar\alpha}$ and   ${\cal{F}}_{\sigma_y}=\f{ k_B \bar{\gamma}_y}{ \hbar\alpha}$
leads to the entropy production rate:
\begin{equation}
    \Sigma^u=\sum_{l=\chi,\sigma_x,\sigma_y }{\cal{F}}_l{\cal J}_{l}~~,
    \label{eq:Sigma_w_coherence}
\end{equation}
where ${\cal J}_{l}=\f{\hbar \alpha}{\sqrt 2}\rm{tr}\b{\dot{\tilde{\rho}} \hat{l}}$.
We can further simplify the fluxes ${\cal{J}}_l$ by utilizing the linearity of the trace and the derivative operations and the dynamics of eigenoperators expectation values, Eq. (\ref{eq:chi_sigma2}). This leads to 
\begin{equation}
    {\cal J}_{\chi}=-\f{\hbar\alpha\Gamma}{\sqrt{2}}\b{\mean{\hat{\chi}}-\mean{\hat{\chi}}_{i.a}}\,\,\,\,;\,\,\,\,{\cal J}_{\sigma_x}=-\f{\hbar\alpha\Gamma}{2}\mean{\hat{\sigma}_x}\,\,\,\,;\,\,\,\,{\cal J}_{\sigma_y}=-\f{\hbar\alpha\Gamma}{2}\mean{\hat{\sigma}_y}~~,
    \label{eq:flux}
\end{equation}
with $\Gamma = k_\downarrow+k_\uparrow$ and $\mean{\hat{\chi}}_{i.a}=-\f{1}{\sqrt{2}}\rm{tanh}\b{\f{\hbar \alpha}{2k_B T}}$.
The form of the entropy production rate resembles the heat transfer entropy production law of classical non-equilibrium thermodynamics \cite{onsager1931reciprocal,onsager1931reciprocal-2,de2013non}, but with a nonlinear relation between flux and force.
The difference between the effective inverse temperature and the bath temperature constitutes the thermodynamic forces, while ${\cal J}_{l}$ are the associated thermodynamic fluxes. The expression obtain a similar form, however, a fundamental difference between Eq. (\ref{eq:Sigma_w_coherence}) and the classical expression exists. In the classical expression the relations between the thermodynamic fluxes and forces is strictly phenomenological. Commonly, only the first order is considered and the fluxes are taken to be linear functions of the forces. For example, Fick's law for diffusion of matter relates the diffusion flux to the gradient in concentration, or Fourier's law for heat conduction relates the heat transport to the gradient of inverse temperature. In contrast, the framework of open quantum systems, which we currently employ, allows deriving the relation between thermodynamic fluxes and forces from a  microscopic description.

In the high temperature limit when $\bar \beta$
and $\bar \gamma$ are small we recover the linear response relation between fluxes and forces: ${\cal J}_{\chi} \approx L {\cal F}_{\chi}$,  ${\cal J}_{\sigma_x} \approx L {\cal F}_{\sigma_x}$ and 
and ${\cal J}_{\sigma_y} \approx L {\cal F}_{\sigma_y}$, where $L=\f{\Gamma}{k_B} \left(\f{\hbar \alpha}{2} \right)^2$, see Appendix \ref{appendixA} for further details. As a result, the entropy production rate in the linear response region becomes
\begin{equation}
\Sigma^u=\sum_l L {\cal F}_l^2 ~~~.
\label{eq:onsager}
\end{equation}
It should be noted that the diagonal Onsager matrix is a consequence of the fact that in the interaction representation the dynamics of the coherence
is separated from $\chi$. Once we rotate to the $\{\hat H, \hat L ,\hat C \}$ basis we will get symmetric coupling elements between energy and coherence in the Onsager matrix.
We stress that the current derivation, leading to the linear response result, is not based on the adiabatic assumption of a perturbation with respect to the Gibbs state  \cite{brandner2020thermodynamic}.
 
Overall, in the general case we observe three independent forces and fluxes responsible for entropy production, a heat flux and two fluxes associated with loss of coherence.

\section{Local cycles}
\label{sec:local}

Closing the cycles requires concatenating the four strokes. We distinguish
two families of cycles which differ by the coherence operation: global or local. In local cycles the coherence vanishes on the four switching points between strokes. Global cycles on the other hand, maintain coherence throughout the cycle and will be treated in section \ref{sec:global}.

\subsection{Local Otto cycle}

\begin{figure}[htb!]
\centering
\includegraphics[width=10.2 cm]{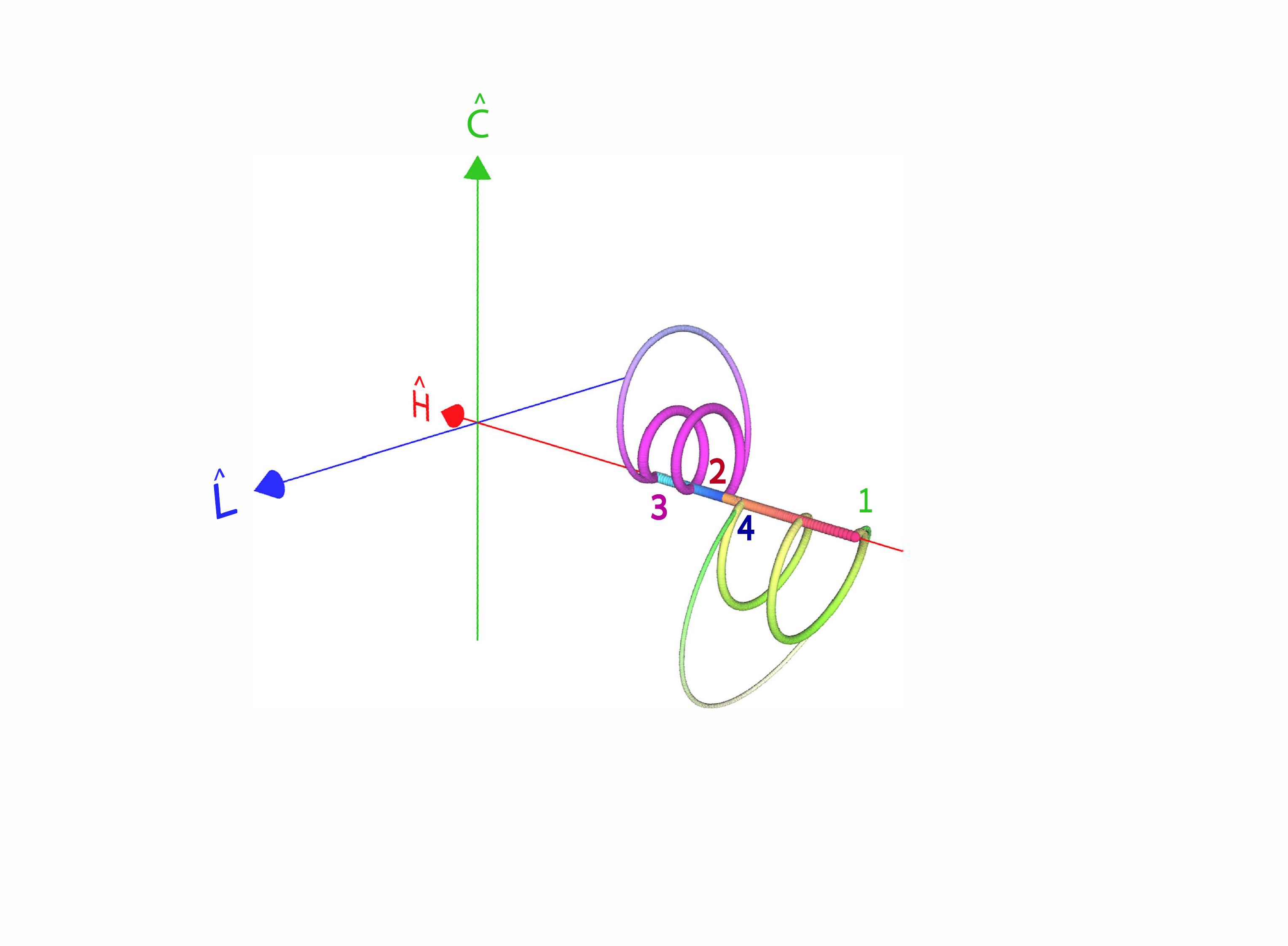}
\caption{Frictionless shortcut (local) Otto cycle plotted in the
$\{\mean{\hat{H}},\mean{\hat{L}},\mean{\hat{C}}\}$ space. The hot isochore $1 \rightarrow 2 $ is represented by red thick line, and the cold isochore $3 \rightarrow 4 $ is shown in blue. The expansion $2 \rightarrow 3 $ and compression $4 \rightarrow 1 $ unitary strokes begin and end on the energy axis, as the qubit exhibits no coherence between consecutive strokes. The fastest two solutions of shortcuts with constant $\mu$ are shown. The single large loop (thick line) corresponds to the fastest solution with $l=1$ in Eq. (\ref{eq:quantum}), corresponding to $\tau_{min}$ and the second solution includes two small loops, $l=2$.}
\label{fig:7}
\end{figure}

Local cycles are obtained by employing  shortcuts
to adiabaticity (STA) on the adiabats \cite{hoffmann2011time,del2014more}.
The chosen protocols are characterized by a minimum 
unitary stroke time $\tau_{adi}(l=1)$, which forces a finite optimum thermalization time. As a result, maximum 
power is obtained for a total finite cycle period \cite{feldmann2000performance}, Fig. \ref{fig:7} displays such a cycle. 

Optimizing the thermalization  period has been addressed in Ref. \cite{feldmann1996heat}.
The main variable influencing the power output is the polarization difference $\bar S _2' -\bar S_1'$
(Cf. Fig. \ref{fig:4}). Therefore, the gaps $|S_2 -S_2'|$ and $|S_1 -S_1'|$ are optimized to achieve finite heat transport. The described procedure leads to \cite{feldmann1996heat}
\begin{equation}
 \bar S _2' -\bar S_1' = (\bar S _2 -\bar S_1) F(x,y)~~,
 \label{eq:fxy}
\end{equation}
where $F(x,y)=\frac{(1-x)(1-y)}{1-xy}$, with $x=e^{-\Gamma_h \tau_h}$
and $y=e^{-\Gamma_c \tau_c}$. Here, $\tau_h$ and $\tau_c$ are the time allocation for thermalization. In addition, optimizing for $\Gamma_c=\Gamma_h$ leads to $\tau_h=\tau_c$.
Optimizing for power under the constraint of a finite-time allocation during
the unitary strokes, $\tau_{adi}$, leads to \cite{rezek2006irreversible}:
\begin{equation}
    x + \Gamma \tau_{adi} = \sinh(x)
    \label{eq:optx}
\end{equation}
For small $x$ Eq. (\ref{eq:optx}) can be solved to obtain 
$\tau_h=\tau_c=\frac{1}{\Gamma}(\Gamma \tau_{adi}/3)^{1/3} $. 
In this limit, the optimal power of local Otto becomes
\begin{equation}
{\cal P}_{Otto}^{L.O} =\frac{1}{(\Gamma \tau_{adi}/3)^{1/3}} \f{1}{4}\Gamma  \Delta \Omega \Delta \bar S  =\frac{1}{(\Gamma \tau_{adi}/3)^{1/3}}{\cal P}_{Otto}^{B.B}
\label{eq:maxpower}~~,
\end{equation}
which is smaller than the bang-bang power ${\cal P}_{Otto}^{B.B}$
Eq. (\ref{eq:opower}).

The power of the local Otto cycle as a function of cycle time is shown in Fig. \ref{fig:power_vs_cycletime_STE} displaying the typical maximum power. The efficiency of the engine $\eta_{Otto}^{L.O}$ Eq. (\ref{eq:e-otto}) is independent of cycle period (Cf. Fig. \ref{fig:eff_vs_cycletime_STE} ).
In practice, such a cycle can be analysed by means of stochastic thermodynamics \cite{seifert2012stochastic} since coherence has been eliminated from the analysis by employing shortcut protocols.
A similar result employing a different derivation can be found in \cite{abiuso2019non}.

\subsection{Local Carnot cycle}
\label{subsec:Carnot-STE}

\begin{figure}[htb!]
\centering
\includegraphics[width=5.5 cm]{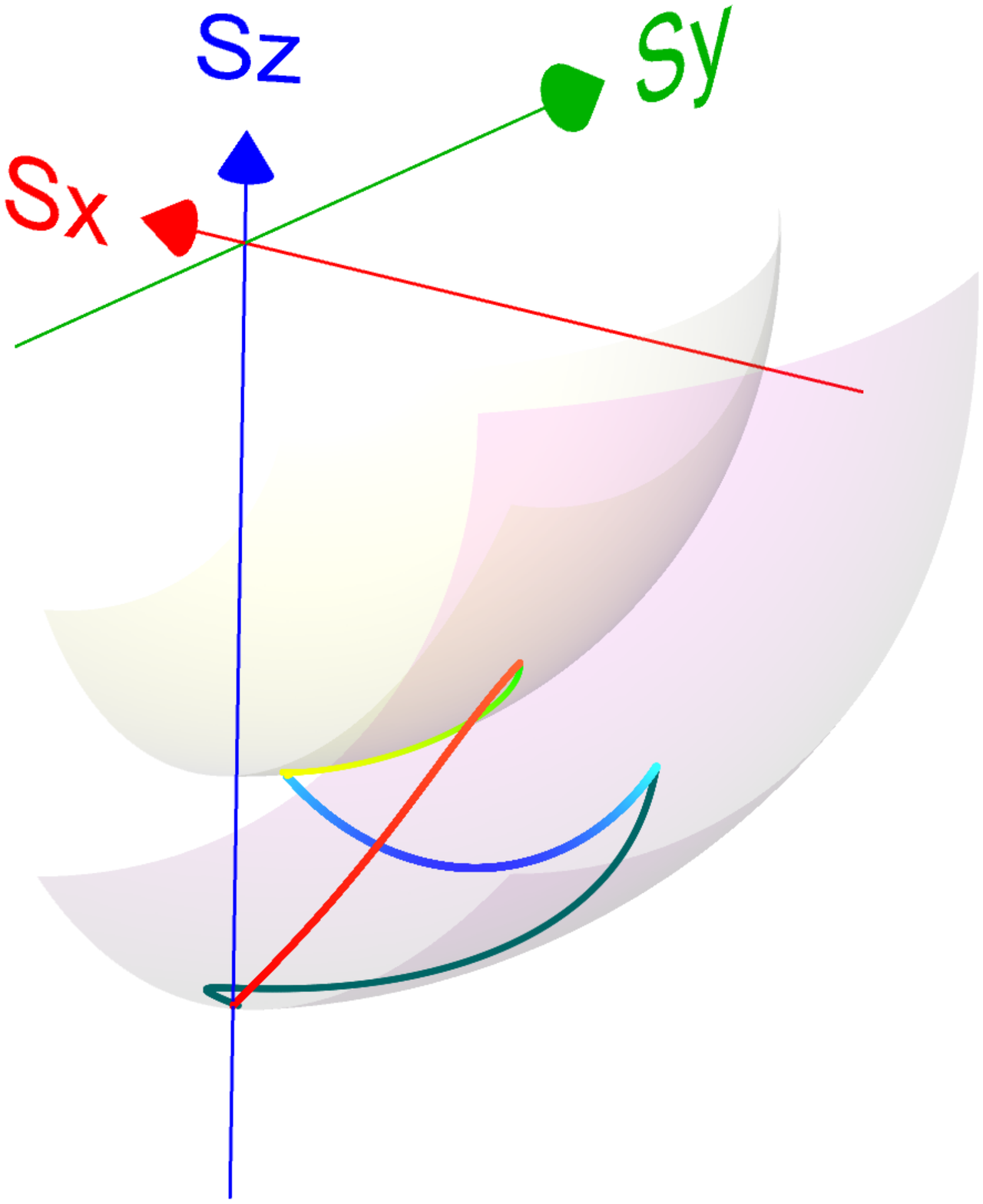}
\caption{Shortcut Carnot cycle: Polarization change during the four strokes. The two hemispheres represent constant polarization for the two unitary strokes.
The red part of the cycle trajectory is the hot isotherm connecting the two
constant polarization hemispheres. The blue section is the cold isotherm.
Green sections represent adiabats, where the the top curve corresponds to the expansion stroke and the bottom curve to the compression.}
\label{fig:Scomponents_STE}
\end{figure}

A local Carnot cycle, also called the `Shortcut Carnot' cycle, is constructed by combining two shortcut to equilibrium protocols (open-strokes) and two shortcuts to adiabaticity protocols (unitary strokes), see Fig. \ref{fig:Scomponents_STE}. It is characterized by the same cycle parameters as the Carnot cycle, while operating at finite speed, thus producing power. The rise in power does not come for free, as rapid driving increases dissipation, leading to a reduction in efficiency. Thus, the common tradeoff between efficiency and power is obtained from a first principle derivation, highlighting the quantum origins of the empirical phenomena associated with friction. 

\begin{figure}[htb!]
\centering
\includegraphics[width=5cm]{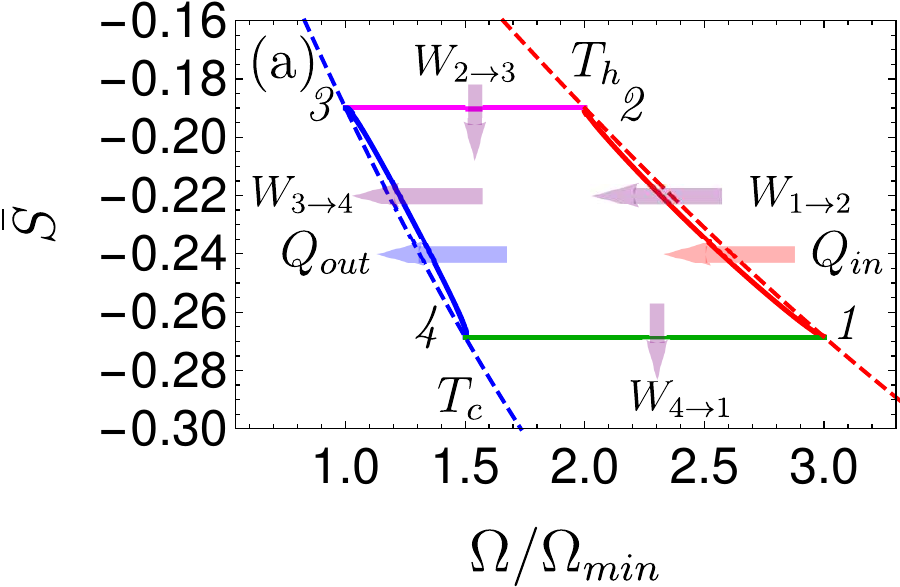}
\includegraphics[width=5cm]{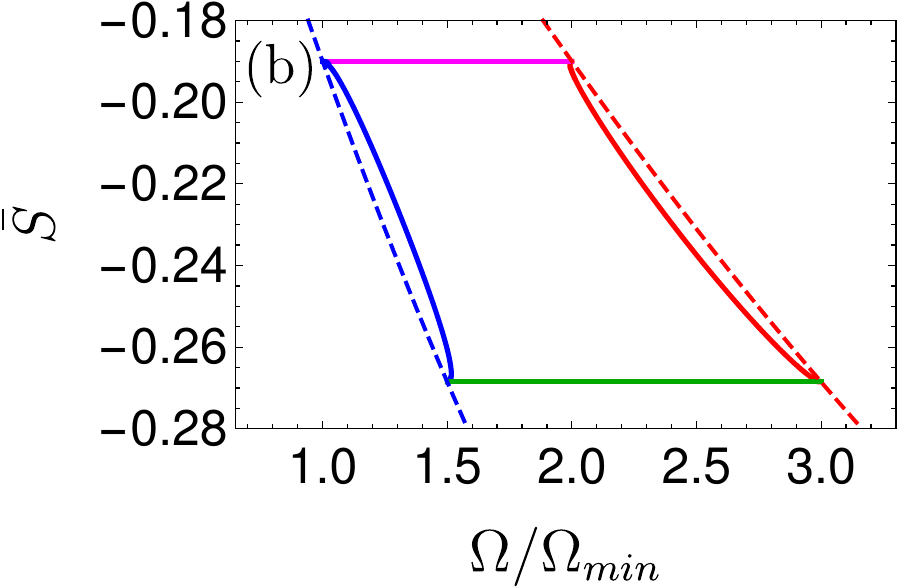}
\includegraphics[width=5cm]{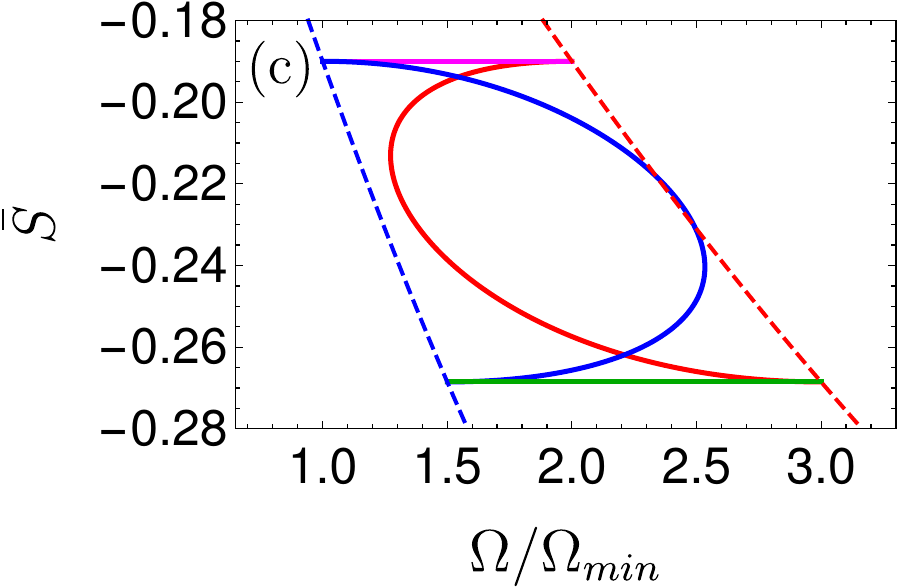}
\caption{Shortcut Carnot cycle: Polarization as a function of the Rabi frequency for the  four different cycle duration's: (a) $\tau_{cyc} = 192$ (b) $\tau_{cyc} = 108$ (c) $\tau_{cyc} = 9$, units of $\b{2\pi/\Omega_{min}}$, with $\Omega_{min}=\Omega_3\, \rm{m.u.}$ (model units $\hbar=k_B=c=1$). The hot and cold isotherms are represented by dashed red and blue lines and the cycle points are denoted by numbers. Incoming (outgoing) arrows designate consumption (extraction) of work or transfer of heat to (from) the qubit. For slow driving the cycle lies close to the reversible Carnot cycle, Panel (a). Increasing the driving speed leads to dissipation and deviations from reversible operation, Panel (b). Eventually, below a the transition cycle time $\tau_{trans} =12.7 \b{2\pi/\Omega_{min}} $, the cycle transitions to a dissipator operation mode, Panel (c). Cycle parameters are summarized in Table \ref{table:cycle_parameters}.  }
\label{fig:S_vs_Omega_STE_cycle}
\end{figure}

The shortcut cycle is constructed by setting the bath temperatures $T_h$ and $T_c$, the minimum Rabi frequency $\Omega_{min}=\Omega_3$ and compression ratio ${\cal{C}}=\Omega_1/\Omega_3$. The remainder of the cycle parameters are then determined by the condition that the working medium is at equilibrium with the bath at the four corners of the cycle, see Fig \ref{fig:S_vs_Omega_STE_cycle} Panel 1. This condition implies the relations $\Omega_4 T_h=\Omega_1 T_c$ and $\Omega_2 T_c =\Omega_3 T_h$, cycle parameters are given in Table \ref{table:cycle_parameters}. 
In contrast to the ideal Carnot cycle, the strokes including exchange of energy with the bath are denoted as open-expansion and open-compression. This change in nomenclature highlights the fact that during these strokes the qubit constitutes an open quantum system and at intermediate times along the strokes, the qubit is in a non-equilibrium state.

The adiabats (unitary strokes) are accelerated by employing shortcuts to adiabaticity (STA) protocols, characterized by a constant adiabatic parameter $\mu$, see Sec. \ref{subsec:STA}. The dynamics of these protocols are governed by the propagator $\Lambda_{adi}={\cal U}_1{\cal U}_2$ given in  Eqs. (\ref{eq:u1}) and (\ref{eq:prop}). Shortcuts to adiabaticity protocols are then achieved by setting the stroke duration $\tau$ such that ${\cal{U}}_2$ is proportional to the identity. The net effect for an initial state with no coherence is a total scaling of the energy. This is achieved for $\tau$, satisfying $\kappa\:\!\:\!\theta(\tau)=2\pi l$, with $l\in \mathbb{N}$ Eq. (\ref{eq:prop}). In the following analysis we choose $l=1$, Eq. (\ref{eq:tmin}).

Other STA protocols are possible, nevertheless 
the specific choice of an STA protocol only slightly affects the qualitative thermodynamic cycle performance. Different STA protocols lead to the same state-to-state transformation, while generating different transient dynamics and having different stroke durations. In principle, if the energy of the driving is not bounded, one can achieve the adiabats in vanishing time by utilizing the bang-bang protocols, Sec. \ref{subsec:STA}. The net effect of different stroke duration is therefore just an additional constant to the cycle time. Overall, the qualitative thermodynamic performance is determined by the isothermal protocols \footnote{We present the description of the chosen STA protocol for the sake of completeness.}. 

Acceleration of the open-strokes is obtained by employing STE protocols, which are described in detail in Sec. \ref{sec:STE protocols}.
The speedup relies on non-adiabatic dynamics and generation of coherence at intermediate times.  The STE protocols are engineered to incorporate both the unitary effect, which leads to rise in coherence, and the dissipative interaction that induces decay of coherence. These two contributions are combined to induce conversion of all the coherence of the working medium to energy at the final stage of the protocol. When the driving is slow, only a small amount of coherence is generated and the evolution is close to an isothermal process. 
The close proximity of the polarization during the  open-strokes
is observed in 
Fig. \ref{fig:S_vs_Omega_STE_cycle} Panel A,  (dashed lines).

Accelerating the driving generates larger coherence accompanied by a thermodynamic cost.
This link between coherence and thermodynamic cost follows from the properties of the  dynamical propagator at constant $\mu$,  Eq. (\ref{eq:prop}). 
During the open-strokes, the coupling to the bath leads to decay of coherence. This decay increases with the amount of coherence present. As a result, rapid driving leads to enhanced dissipation which reduces the cycle performance. 

\begin{figure}[htb!]
\centering
\includegraphics[width=7cm]{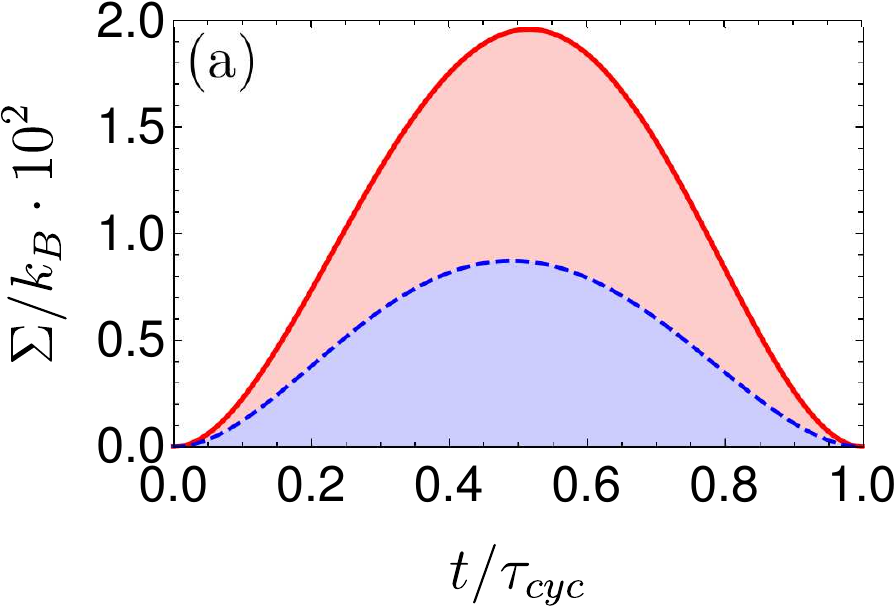}
\includegraphics[width=7cm]{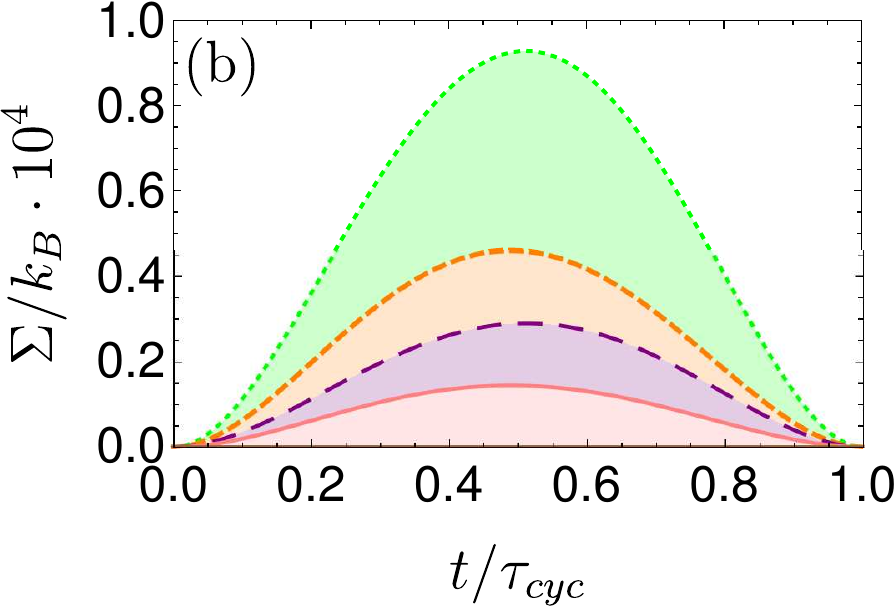}
\caption{Entropy production rate for the open strokes of the shortcut Carnot cycle as a function of normalized time for various cycle times. Panel (a): Open-compression (red) and open-expansion (dashed blue) for a short cycle time $\tau_{cyc}= 9 \b{2\pi/\Omega_{min}}$. 
Panel (b):  Open-compression (dotted green/ dashed purple) and open-expansion (dashed orange/ continuous pink) for a large cycle time $\tau_{cyc}= 108\,/\,192$  $\b{2\pi/\Omega_{min}}$. The three cycle times correspond to the cycles plotted in Fig. \ref{fig:S_vs_Omega_STE_cycle}. Decreasing the cycle time increases the dissipation and results in a greater entropy production. The compression strokes include cooling the qubit, which requires greater amounts of entropy production relative to the open-expansion strokes for the same stroke times.       }
\label{fig:EP_vs_normalized_time}
\end{figure}

A visual representation of this phenomenon is shown in Fig.
\ref{fig:S_vs_Omega_STE_cycle}, 
which compares three 
Carnot-type cycles with varying cycle times.
The amount of extracted work during a single cycle is related to the area enclosed by the $\bar{S}\b{\Omega}$ plot. 
As the cycle time decreases the open-strokes deviate further from the isotherms (Panel 2), consuming more work and dissipating larger amounts of energy and coherence. Fig. \ref{fig:EP_vs_normalized_time} shows the entropy production rate on the open strokes for these cycles. 
At the beginning and the end of the stoke the entropy production rate Eq. (\ref{eq:entropy_production_name}) is zero since 
the protocol is designed to reach equilibrium on the four corners of the cycle. The area under the lines is the total entropy production.
As expected, the entropy production increases for decreasing stroke duration.
Eventually, the cycle transitions to an accelerator operation mode, where work is consumed during both open-strokes (Panel 3 of Fig. \ref{fig:S_vs_Omega_STE_cycle}) which enhances the entropy production.

\begin{figure}[htb!]
\centering
\includegraphics[width=7.2cm]{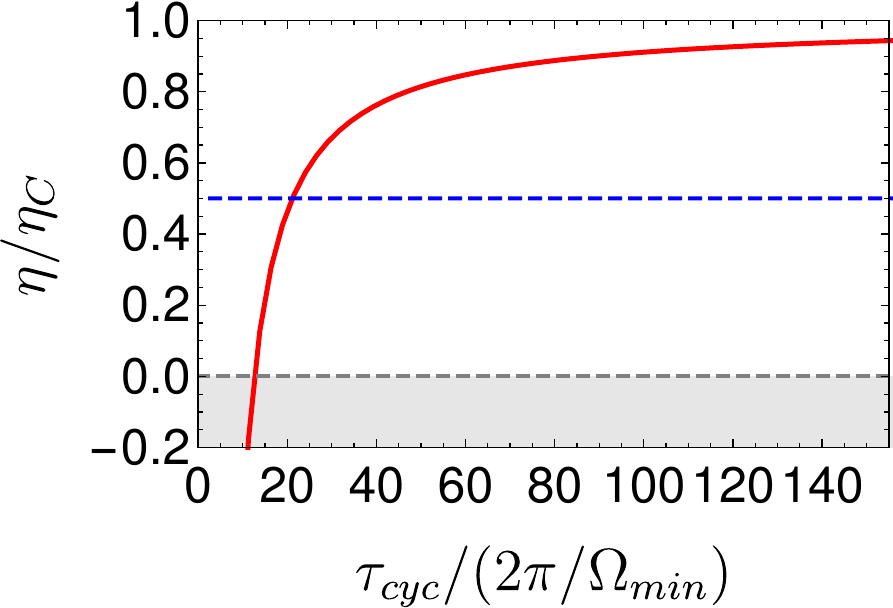}
\caption{Normalized efficiency as a function of the cycle time for a local Carnot cycle (thick red) and local Otto cycle (blue dashed). In the local Carnot cycle long cycle times lead to close to reversible dynamics, optimizing the efficiency towards the Carnot bound $\eta_C$. For short cycle times dissipation of energy and coherence leads to a degradation of efficiency. Eventually, resulting in a transition from an  engine operation mode ($\eta\equiv-{\cal W}/{\cal Q}_{h}>0$)  to an accelerator operation mode ($\eta<0$). The two studied protocols for the angle $\phi\b t$ (below Eq. (\ref{eq:alpha2}) cannot be distinguished in this graph. In the local Otto cycle the efficiency $\eta_{Otto} = 1-\Omega_c/\Omega_h$ is independent of the cycle time.}
\label{fig:eff_vs_cycletime_STE}
\end{figure}

\begin{figure}[htb!]
\centering
\includegraphics[width=7.5 cm]{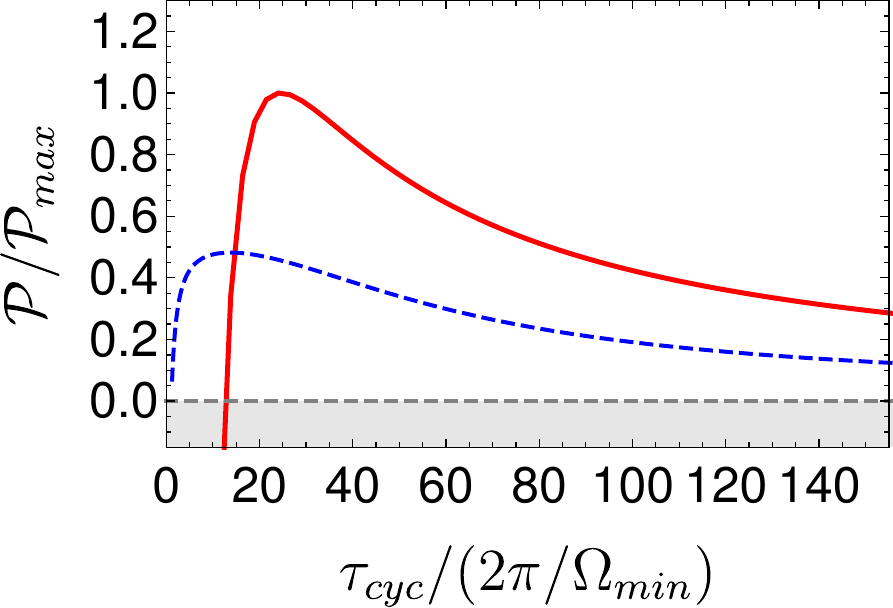}
\caption{Power as a function of cycle time for a local Carnot cycle (red thick line) and local Otto cycle (blued dashed line). Slow driving leads to a reduction in power ${\cal{P}}\equiv-{\cal W}/\tau_{cyc}$. Moreover, under rapid driving dissipation reduces the net extracted work, leading to an optimal power  of ${\cal{P}}_{max}=5.19\cdot10^{-3} \rm{m.u.}$ 
for $\tau_{cyc}\approx 24\, \b{2\pi/\Omega_{min}}$ for the local Carnot and ${\cal{P}}_{max}=2.5\cdot 10^{-3} \rm{m.u.}$ 
for $\tau_{cyc}\approx  \b{2\pi/\Omega_{min}}$ for the local Otto cycle. }
\label{fig:power_vs_cycletime_STE}
\end{figure}

\begin{figure}[htb!]
\centering
\includegraphics[width=7.5cm]{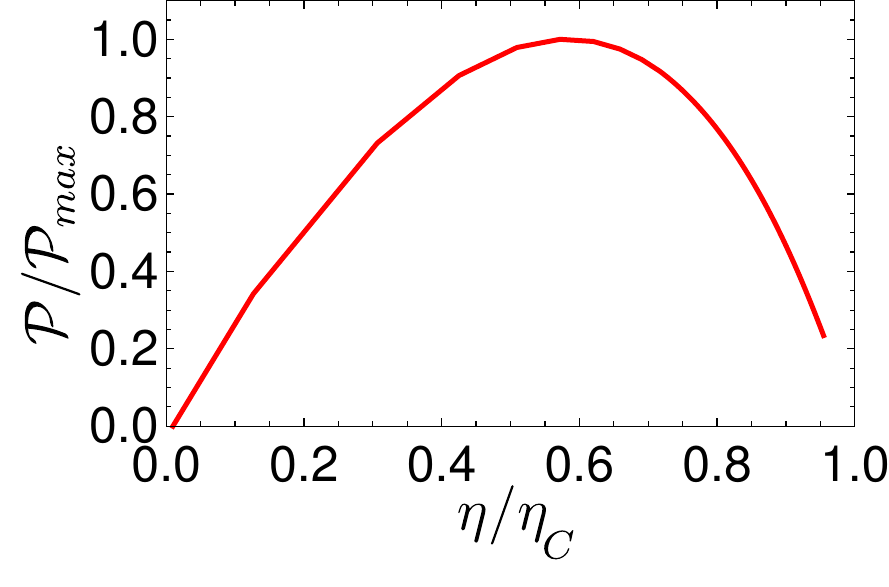}
\caption{ Power as a function of efficiency for the  local Carnot cycle. The typical behaviour is a manifestation of the tradeoff between efficiency and power. The efficiency at maximum power $\eta_{\rm{max}{\cal{P}}}\approx0.57$ exceeds the Curzon–Ahlborn efficiency $\eta_{CA}=1-\sqrt{T_c/T_h}\approx 0.3$. This result is not surprising, as the operation speed goes beyond the low dissipation regime \cite{esposito2010efficiency}. }
\label{fig:power_vs_eff_STE}
\end{figure}

In the opposing limit of long cycle times, the dynamics are adiabatic and the efficiency approaches the Carnot efficiency $\eta_C$, Fig. \ref{fig:eff_vs_cycletime_STE}. The improved efficiency is obtained on account of a reduction in power, see Fig. \ref{fig:power_vs_cycletime_STE}. Optimal power is obtained for relatively short cycle times $\tau_{cyc}\approx 24\, \b{2\pi/\Omega_{min}}$ for the Carnot type cycle and a shorter time
of $\tau_{cyc}\approx 11\, \b{2\pi/\Omega_{min}}$ for the Otto cycle.
Overall the power of the shortcut Carnot cycle exceeds the local Otto cycle for almost all cycle times.
Fig. \ref{fig:power_vs_eff_STE} shows the typical efficiency-power tradeoff for the shortcut Carnot  cycle.

Surprisingly, when comparing the two different protocols for the angle $\phi$ (see below Eq. (\ref{eq:alpha2}), the performance of the cycle is almost independent of the chosen protocol for $\phi$. The primary difference between the two cycles concerns the amount of coherence generated during the open-strokes. As expected the protocol which includes a rotation of $\pi/2$ in $\phi$ exhibits much larger coherence along the stroke.
We expect that for higher values of $\mu$, protocols which are characterized by a rotation of the Hamiltonian, will generate more coherence. In turn, this will shift the transition point of the engine to an accelerator to larger cycle times. 
 
\section{Global cycles}
\label{sec:global}

Closing globally coherent cycles requires more than just connecting the four strokes since the four corners of our cycle are no longer 
required to be Gibbs states. In general, we prescribe a periodic driving protocol and the qubit is thereby driven to a limit cycle \cite{kosloff2002discrete}.

\begin{figure}[htb!]
\centering
\includegraphics[width=9 cm]{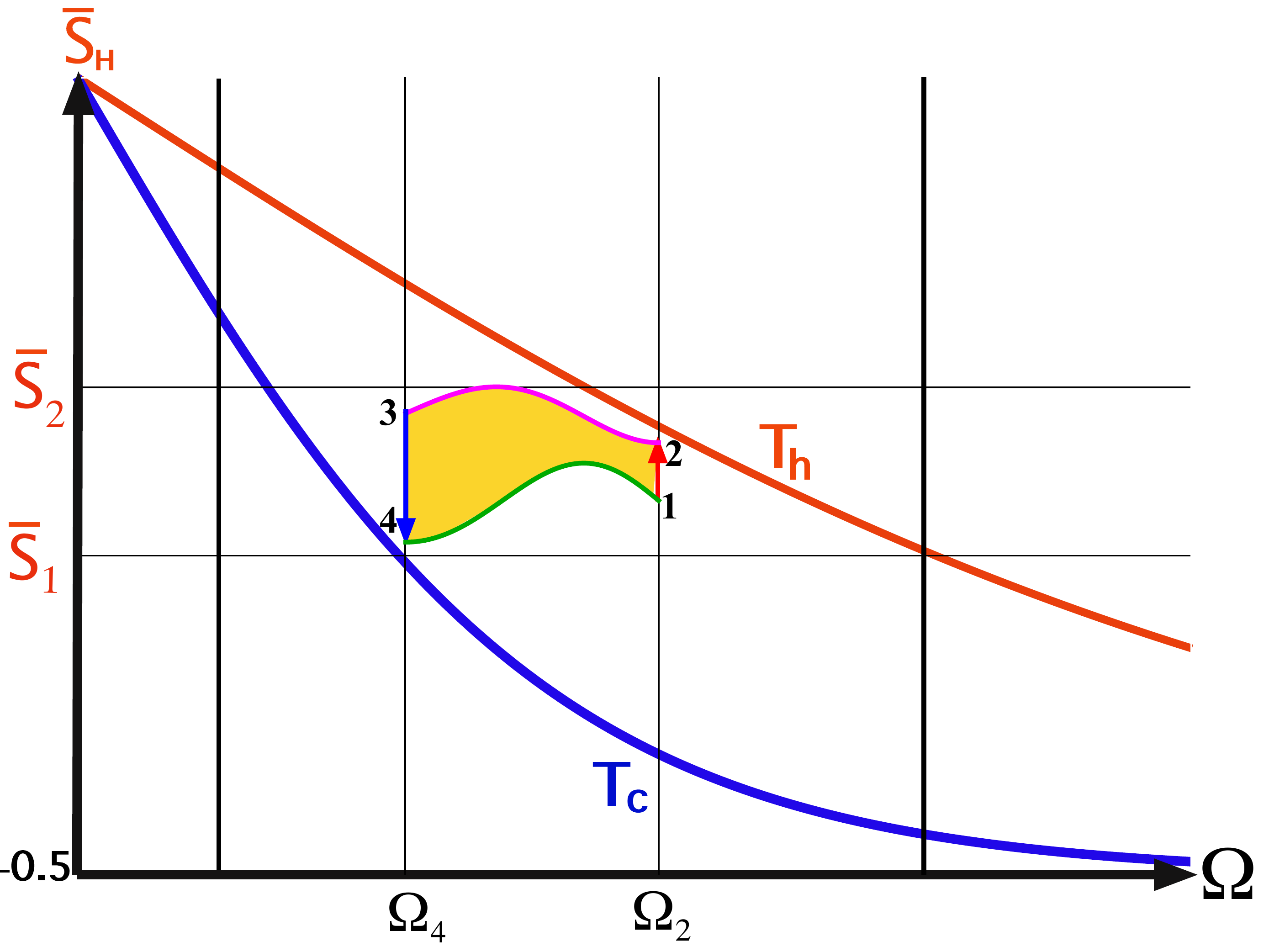}
\caption{Global Otto cycle with friction. Projection of the polarization on the energy axis as a function of the generalized Rabi frequency.  The coherence at the end of the adiabats dissipates during the isochores. Due to the decay of this coherence, the polarization along the adiabats always exceeds its initial value. In the presented cycle $\mu>\mu_{l=1}$.}
\label{fig:5}
\end{figure}

We will start by examining the Otto cycle which is easier to analyse.
Fig \ref{fig:5} shows an example of a global Otto cycle.
During the unitary strokes of the Otto cycle, $\Lambda_{h\rightarrow c}$ and $\Lambda_{c\rightarrow h}$, non-adiabatic dynamics generates coherence, which carries the
system away from the energy direction. 
\cite{solfanelli2020nonadiabatic,alecce2015quantum,francica2019role}.
This coherence subsequently decays during the isochoric strokes. Note that if STA protocols are used on the unitary strokes, no coherence ever dissipates, leading to no friction, and the discussion from the frictionless treatment in sections \ref{subsec:elemOtto} and \ref{subsec:otto-opt} applies. 
Since our goal is to understand the behavior including friction, we use constant $\mu$ protocols for the unitary strokes. These are frictionless only for quantized stroke durations. 
We present results for the power and the efficiency as a function of the cycle time. Since our STA protocols are frictionless only for quantized times, the behavior in Figs. \ref{fig:eff_vs_cycletime_Endo} and  \ref{fig:power_vs_cycletime_Endo}  shows oscillations for cycle times smaller than $\tau(l=1)$.
Interestingly, except for some wild oscillations for very small times, the power of the coherent Otto cycle is monotonically decreasing in the cycle time, reaching its maximum for the sudden cycle in the limit of $\tau \ra 0$ for small $\Phi$. We therefore begin with a closer look at this case.

\subsection{Global Otto cycle and the sudden limit}

We now analyse the influence of coherence in the sudden limit. In general, the working medium Hamiltonian $\hat{H}\b t$, Eq. (\ref{eq:hamil}), does not commute with itself at different times, generating substantial coherence under rapid driving. 
As will be demonstrated, this coherence has a direct effect on the thermodynamic performance and the cycle's operation mode. 

The sudden operation is characterized by only two types of strokes:  unitaries and isochores. When the Hamiltonian parameters are varied instantaneously, the working medium dynamics is dominated by the unitary part (adiabats). Any finite coupling with the bath (weak in our analysis) only negligibly affects the working medium state. During the isochores, the control parameters remain constant and a small amount of heat transfer occurs. Note that the initial portion of an isochore has the largest temperature difference between the bath and our qubit so the sudden cycle uses only this fastest heat exchange, explaining how the power can be maximum in the zero time limit.  

As in the general case, the evolution of the working medium during the sudden Otto cycle is constructed by combining the propagators of the adiabats $\Lambda^{sudd}_{i\ra f}$, Eq. (\ref{eq:sudden_adiabats_prop}), and the propagators for the isochores $\Lambda_{sudd}^{iso}$. 
The propagator for the isochores are obtained by substituting the basis operators  $\{\hat{H},\hat{L},\hat{C},\hat{I}\}$ into the 
Heisenberg form of the master equation, Eq. (\ref{eq:lind}), and expanding the solution up to first order in the stroke time $\tau$. This leads to 
\begin{equation}
    \Lambda^{sudd}_{i}=\sb{\begin{array}{cccc}
1-\Gamma_i \tau & 0 & 0 & \Gamma_i \tau \Omega_i\mean{ \bar{S}_{eq}\b{\Omega_i,T_i}}\\
0 & 1-{\Gamma_i \tau}/{2} & -\Omega_i \tau & 0\\
0 & \Omega_i \tau & 1-{\Gamma_i \tau}/{2}& 0 \\
0 & 0 & 0 & 1
\end{array}}~~,
\end{equation}
where $i=h,c$ indicates the frequency of the hot and cold baths. In the studied sudden Otto cycle $\Omega_{h/c}=\Omega_{2/4}$ and the kinetic rates are taken to be equal $\Gamma_i=\Gamma_c=\Gamma_h$.
Concatenating the stroke propagators in the suitable order generates the sudden cycle propagator \begin{equation}
\Lambda_{cyc}^{sudd}=\Lambda_{c \rightarrow h}^{sudd}\Lambda_{c}^{sudd}\Lambda_{h\rightarrow c}^{sudd}\Lambda_{h}^{sudd}~~.
\label{eq:prop_sudden}
\end{equation}
We next solve for the invariant of the limit cycle $\Lambda_{cyc}^{sudd}\v{v} = \v v$, where the elements of $\v{v}$
give the expectation values of the basis operators $\{\hat{H},\hat{L},\hat{C}\}$ at the beginning of the hot isochore. This information is sufficient to determine the qubits state throughout the cycle, and in turn, allows evaluating the thermodynamic quantities.

\begin{figure}[htb!]
\centering
\includegraphics[width=7 cm]{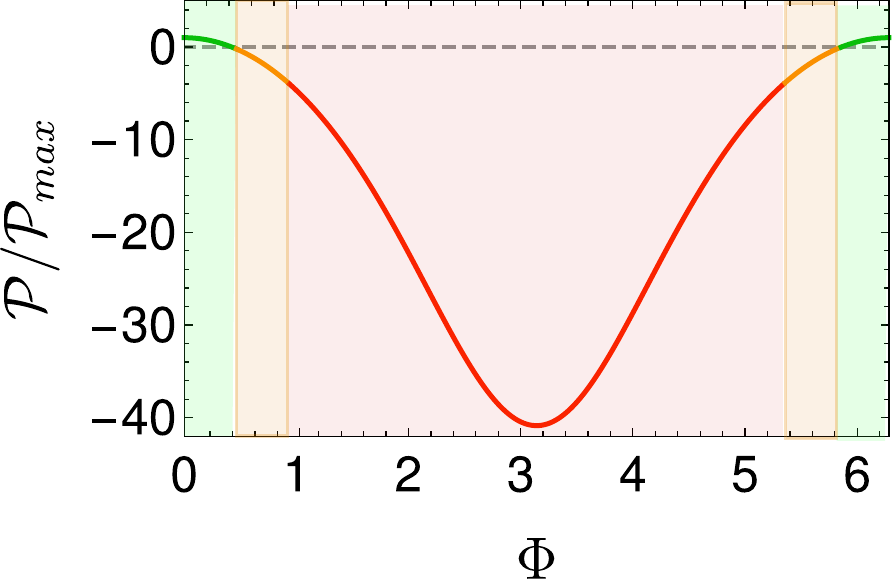}
\includegraphics[width=7 cm]{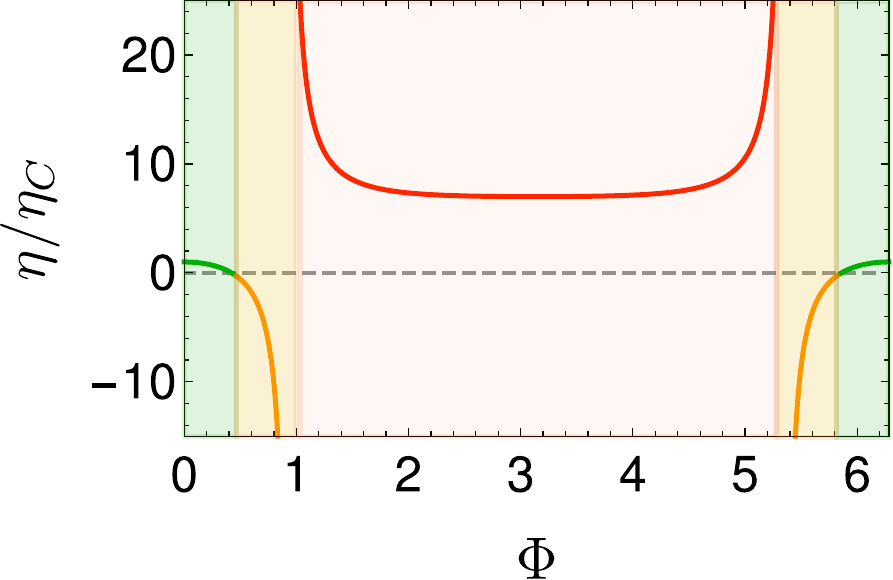}
 \caption{(a) Normalized power and (b) efficiency as a function of the relative phase $\Phi$. When the phase values are near $2\pi k$, $k\in\mathbb{Z}$, only small amounts of coherence are generated and the cycle operates as an engine (green lines), producing a positive power output. Once the phase deviates from the optimal values, the net work becomes positive and the cycle operates as an accelerator (orange lines, ${\cal W}>0$, ${\cal Q}_h>0$, ${\cal Q}_c<0$), accelerating the flow from hot to cold. When the coherence generated during the adiabats dissipates to both baths on the isochores, the cycle transitions to a dissipator $\eta>1$ (red line, ${\cal Q}_c,{\cal Q}_h<0$, ${\cal W}>0$). The model parameters are: $T_c=5$, $T_h=10$, $\Omega_c=6$, $\Omega_h=8$, $\Gamma_h\tau_h=\Gamma_c\tau_c=0.01$, where $\tau_h$ and $\tau_c$ are the stroke duration of the hot and cold isochores. } 
\label{fig:Sudden_power}
\end{figure}

We find that in the sudden limit, the cycle's performance is highly sensitive to the coherence generation along the adiabats. The amount of accumulated coherence is determined by the relative phase $\Phi$, see Eq. (\ref{eq:sudden_adiabats_prop}). For $\Phi= 2 \pi k$, $k\in\mathbb{Z}$, the Hamiltonian commutes with itself at different times and the state remains diagonal in the energy basis. In contrast, for intermediate values of $\Phi$, coherence builds up along the adiabats and dissipates during the isochores. The dissipation leads to a reduction in power and efficiency. We find that in the sudden limit, extraction of power is only obtained for small amounts of coherence. This regime corresponds to phase values close to $\Phi=2\pi k$, see Fig. \ref{fig:Sudden_power} Panel A. It is important to note that generated coherence on a unitary stroke is employed to
reduce the work against friction in the consecutive unitary stroke.
Eliminating this coherence on the isochores will transform the engine into a dissipator.

It is convenient to characterize the cycle performance in terms of the standard expression of efficiency: $\eta=-{\cal W}/{\cal Q}_{h}$. Under an engine operation mode, the efficiency remains within the range $0\leq\eta\leq\eta_C$.  With increasing coherence generation (increasing $\Phi$), the dissipated work exceeds the extracted work, leading to a net positive work. In this operation regime work is consumed (${\cal W}>0$), while heat keeps flowing from the hot bath to the cold bath (${\cal Q}_{h}>0,{\cal Q}_{c}<0$ ). Thus,  $\eta$ becomes negative, see Fig. \ref{fig:Sudden_power}  Panel B. When further increasing the coherence generation, the qubit starts dissipating energy to both baths (${\cal Q}_c,{\cal Q}_h<0$), this implies that $\eta$ changes its sign abruptly and becomes positive.
Maximum coherence generation is achieved for $\Phi=\pi k$, which corresponds to an equal magnitude of the $x$ and $z$ components of the Hamiltonian (Cf. \ref{subsec:otto-opt}.)
The isochores include off-diagonal terms ($\propto \Omega_i \tau$), which couple the coherence operators $\hat{L}$ and $\hat{C}$. This coupling originates from the unitary contribution to the open-system dynamics \footnote{The unitary term is of form $\f{i}{\hbar}\sb{\hat{H},\hat{X}}$ in the Heisenberg equation of motion for the operator $\hat{X}$.}, and tends to complicate the solution by coupling the dynamics of all three operators
along a complete cycle. In practice, we find that this coupling only slightly affects the results, improving the power by a very small amount ($\sim 10^{-2}{\cal{P}}_{max}$). This typical behaviour justifies discarding the coupling terms when evaluating the efficiency and power. Without these terms, their expressions in the sudden limit ($\tau_{cyc}\ra 0$) read 
\begin{eqnarray}
    \eta = \b{8c_{\Phi}\Omega_h\Omega_{c}\left({\bar S_{eq}^{c}}\Omega_{h}+{\bar S_{eq}^{h}}\Omega_{c}\right)-\Omega_{c}\Omega_{h}\b{\Omega_h{\bar S_{eq}^{h}}+\Omega_c{\bar S_{eq}^{c}}}\b{c_{2\Phi}+7}}/G
\end{eqnarray}
\begin{equation}
    {\cal P}= \frac{\Gamma\left(8c_{\Phi}\left(\bar{S}_{eq}^{c}\Omega_{h}+\bar{S}_{eq}^{h}\Omega_{c}\right)-\b{\Omega_{c}\bar{S}_{eq}^{c}+\Omega_{h}\bar{S}_{eq}^{h}}\b{c_{2\Phi}+7}\right)}{4\b{c_{2\Phi}-17}}~~,
\end{equation}
with  $G\equiv{\Omega_{h}\b{8\mean{S_{eq}^{c}} \Omega_c\Omega_{h}c_{\Phi}-\mean{S_{eq}^{h}}\Omega_h\Omega_{c}\b{c_{2\Phi}+7}}}$
and using the shorthand notation $c_{x}=\cos \b{x}$ and $s_{x}=\sin \b{x}$. In the evaluation of the power we assumed equal stroke duration.

\subsection{Global Carnot-type constant adiabatic parameter cycle}
\label{subsec:golobalc}

By definition, our Carnot-type cycles are constrained to be in equilibrium at switching points between two adjacent strokes. This requirement implies that coherence is only maintained "locally" within the strokes and defines what we mean by local coherence operation of our engines.
In the following analysis, we lift this restriction to study the properties of "global" coherence operation. 

A globally coherent cycle is constructed from two open-strokes and two adiabats. For our implementation, the value of $\mu$ is kept constant for the entire protocol. We study the performance of the limit-cycle, which maintains coherence throughout the cycle. 
In order to produce power we reduce the compression ratio ${\cal C}$ of the cycle while maintaining the same bath temperatures. Two globally coherent Carnot cycles are studied, which differ by the inner frequencies of the cycle, $\Omega_2$ and $\Omega_4$. The frequencies are chosen to fit an endoreversible Carnot cycle with a constant temperature gap, in the first cycle $\Delta T_h=\Delta T_c = 1 \,\rm{m.u.}\,$ and for the second cycle $\Delta T_h=\Delta T_c = 2 \,\rm{m.u.}\,$.
Cycle parameters are summarized in Table \ref{table:cycle_parameters}. These cycles maintain a non-vanishing heat flow in the desired direction on the open-strokes. 

The protocol choice of a constant adiabatic parameter allows an additional degree of freedom in the choice of the Hamiltonian controls $\omega \b t$ and $\epsilon\b t$. We choose to set them as in Eq. (\ref{eq:control_protocols}). We then obtain a relation between the phase and Rabi frequency: $\Omega\b t =-\dot{\phi}/\mu$.  For small $\dot{\phi}$ the protocols can be achieved very rapidly while keeping  $\mu$ small, thus, still maintaining quantum adiabatic evolution. However, in such regime a slow change in the phase implies that $\omega \b t$ and $\epsilon\b t$ are nearly proportionate to one another, resulting in a Hamiltonian which commutes with itself at different times. In order to study the influence of coherence on the thermodynamic performance, we require a substantial change in phase. For this reason, we determine the driving protocols by setting both initial and final Rabi frequencies, $\Omega_i$ and $\Omega_f$ and phases $\phi_i$ and $\phi_f$. In the quantitative analysis we choose $\Phi=\phi_f-\phi_i=\pi/2$, meaning that the Hamiltonian direction rotates  from the $z$ to the $x$ axis during the open-expansion stroke ($-\pi/2$ on the open-compression stroke).

\begin{figure}[htb!]
\centering
\includegraphics[width=8.5cm]{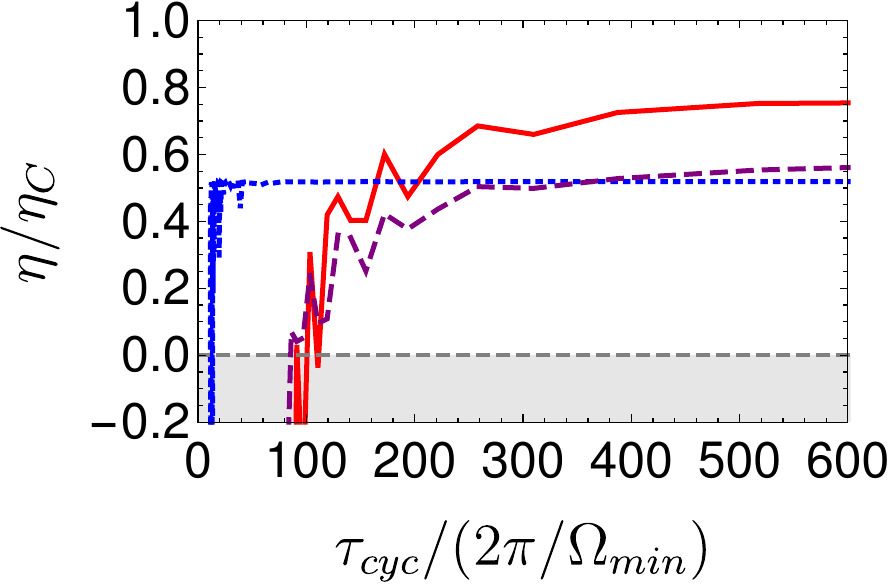}
\caption{ Normalized efficiency as a function of the cycle time for the Globally coherent Carnot (contineous red and dashed purple) and Otto cycles (dotted blue). In the slow driving regime, coherence only degrades the extracted work output and the efficiency. As the cycle time increases less coherence is generated and the efficiency increases monotonically. In the rapid driving regime, the cycle exhibits a quantum operation mode, where the cycle performance is dominated by coherence. In this driving regime, When coherence generation and consumption is coordinated with the stroke times, the cycle efficiently produces work. On the other hand, for stroke time leading to induced dissipation of coherence, the work extraction declines and the cycle may transfer to a dissipator operation mode ($\eta<0$). This sensitivity to coherence leads to an oscillatory dependence for short cycle times. The efficiency of the Carnot-global cycle exceeds the efficiency of the global Otto cycle at long cycle times. This result  stems from the reduced compression ratio of the global Otto cycle. This relative performance reverses for short cycle times. In this driving regime, the global Otto maintains a close to optimal efficiency where the Carnot cycle performance degrades and the cycle ceases to operate as an engine. For the chosen cycle parameters the Carnot efficiency obtains a value of  $\eta_C=0.75$. The two Globally Carnot cycle differ by their $\Omega_2$ and $\Omega_4$ frequencies. As a result, the effective temperature gap of the purple cycle is larger compared to the red cycle. Comparing to the power plot (Fig. \ref{fig:power_vs_cycletime_Endo}) the cycle with lower efficiency exhibits a larger maximum power. The Globally coherent Carnot and global Otto cycles parameters are  summarized in Tables \ref{table:cycle_parameters} and \ref{table:gen_parameters}. }
\label{fig:eff_vs_cycletime_Endo}
\end{figure}

\begin{figure}[htb!]
\centering
\includegraphics[width=8.5cm]{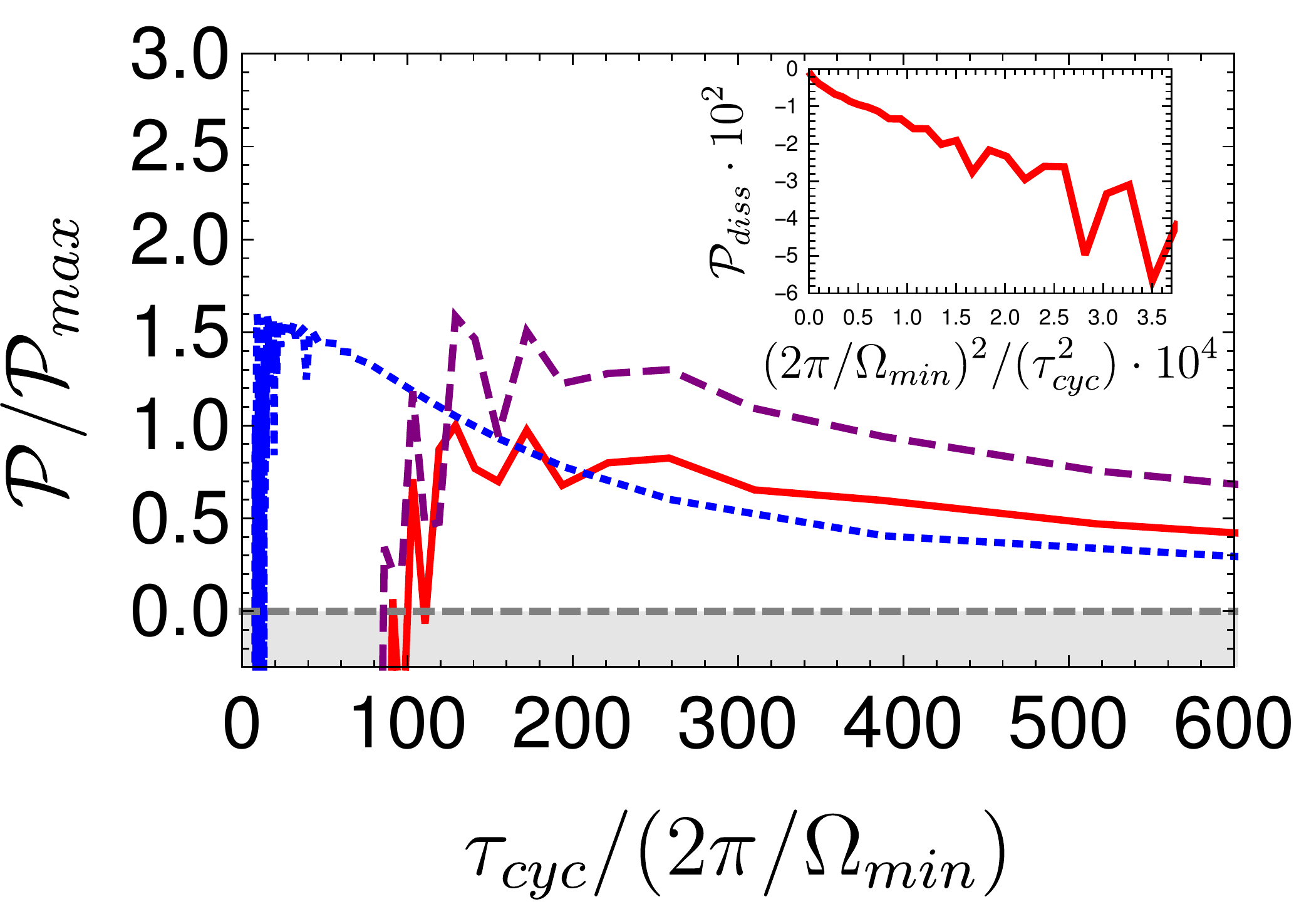}
\caption{ Power as a function of the cycle time for the Globally coherent Carnot (continuous red and dashed purple) and global Otto (dotted blue) cycles. The maximum power for the two Carnot cycle are ${\cal{P}}_{max}=1.8\cdot10^{-3} \rm{m.u.}$ for the purple ${\cal{P}}_{max}=1.14\cdot10^{-3} \rm{m.u.}$\, for the red, and for the Otto ${\cal{P}}_{max}=1.8\cdot10^{-3} \rm{m.u.}$\, . Inset: Dissipated power ${\cal{P}}_{diss}={\cal{P}}-|{\cal W}_{ideal}|/\tau_{cyc}$ as a function of a scaled scaled $1/\tau_{cyc}^2$. For large cycle times the dissipated work scales as $1/\tau_{cyc}$ and the dissipated power as  ${\cal{P}}_{diss}\propto 1/\tau_{cyc}^2$. This result is in accordance with a linear response analysis. Cycle parameters are presented in table \ref{table:cycle_parameters}.  }
\label{fig:power_vs_cycletime_Endo}
\end{figure}

Global coherence operation allows coherence, generated in one stroke, to be converted to energy and utilized during the adjacent strokes. Accelerating the driving enhances this phenomenon by generating greater coherence, which eventually dominates the cycle's performance. Using the coherence measure $\textfrak{C}$, Eq. (\ref{eq:coherence}), we observe that when ${\textfrak{C}}>0.01$, relative to a maximum value of $0.5$, strong interference takes place which are manifested in oscillations in power and efficiency. The coherence value should also be compared to the typical value of $|\bar{S}_H|$, which is of the order $\approx 0.1$. Figure \ref{fig:eff_vs_cycletime_Endo}  presents the scaled efficiency for varying cycle times. In the slow driving regime coherence only degrades the extracted work and efficiency increases monotonically  with the cycle time. In contrast, for sufficiently fast driving, the efficiency oscillates rapidly due to interference. If the generated coherence is utilized efficiently, the cycle extracts more work and the efficiency  improves. Moreover, optimal power is obtained in the fast driving regime, Fig. \ref{fig:power_vs_cycletime_Endo}. 

On the other hand, if generation and consumption of coherence is not coordinated with the stroke times (related to the cycle time) the dissipation increases, decreasing the efficiency. Overall, the oscillations in efficiency constitute a signature of a quantum operation mode \cite{dann2020quantum}, dominated by coherence.

Generally, the efficiency at long cycle durations surpasses the local optima seen at short cycle times. This is a consequence of strong dissipation of coherence under rapid driving. Even when the generation and consumption of coherence is fully coordinated with the stroke duration, still greater amount of coherence leads to greater dissipation on the open strokes, and a reduction in efficiency. Therefore, no quantum advantage is expected in this scenario.  

\begin{figure}[htb!]
\centering
\includegraphics[width=7.5cm]{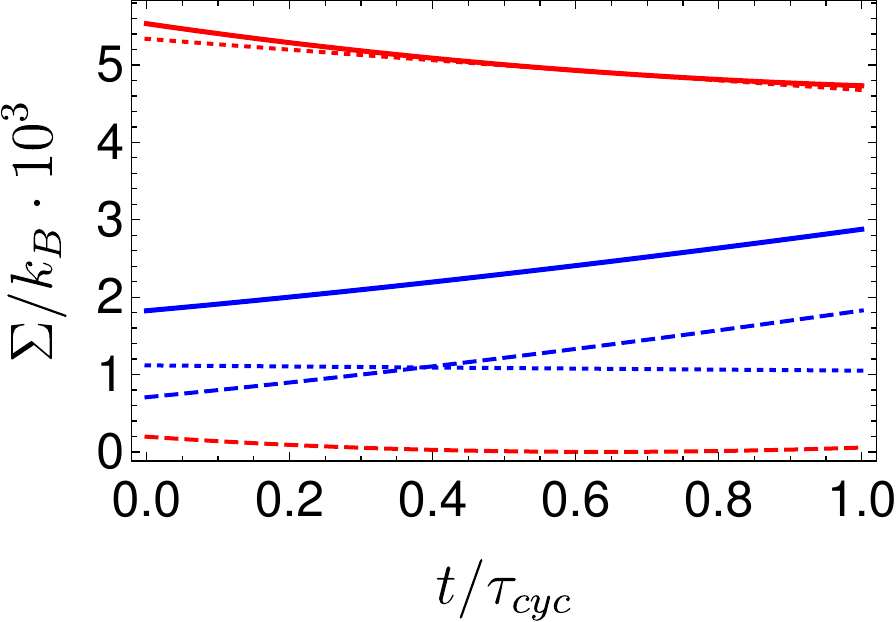}
\caption{Entropy production rate as a function of normalized time for the Globally coherent Carnot cycle with $|\mu|=0.3$. The total entropy production rate for the open-expansion and open-compression strokes are shown in thick red and blue lines, correspondingly. These are a sum of entropy production due to the flux of $\mean{\hat{\chi}}$, ${\cal F}_\chi {\cal J}_\chi$ (dashed lines),  and coherence-like terms ${\cal F}_{\sigma_x} {\cal J}_{\sigma_x}+{\cal F}_{\sigma_y} {\cal J}_{\sigma_y}$ (dotted lines). The various contributions are as expected positive. With decreasing $\mu$ the coherence-like terms decrease and the term ${\cal F}_\chi {\cal J}_\chi$ is the dominant contribution to the entropy production. The breakup of the entropy production to in the $\{\hat{H},\hat{L},\hat{C}\}$  basis will show a similar pattern. }
\label{fig:entropy_prod_const_mu}
\end{figure}

The entropy production rate of these cycles is almost constant throughout the cycle Fig. \ref{fig:entropy_prod_const_mu}.
This is a confirmation that the cycle is always far from the instantaneous attractor. Shorter cycle periods lead to larger entropy production. The breakup of entropy production to an energy like term $\Sigma_{\chi}={\cal F}_{\chi}{\cal J}_{\chi}$ and a coherent part $\Sigma_{\sigma_{x/y}}={\cal F}_{\sigma_{x/y}}{\cal J}_{\sigma_{x/y}}$, Sec. \ref{subsec:5.4}, show 
similar values for large $\mu$ which means that coherence dominates the cycle. For small $\mu$ (large cycle times) the entropy production is dominated by $\Sigma_{\chi}$ which can be attributed to 
irreversible heat transport.

In the asymptotic limit (large $\tau_{cyc}$), the working medium remains in the linear response regime during the open strokes. This regime is characterized by low dissipation and a typical $1/\tau_{cyc}$ scaling law of the dissipated energy. Similarly, one can introduce the dissipated power, defined as 
${\cal{P}}_{diss}={\cal{P}}-|{\cal W}_{ideal}|/\tau_{cycle}$, where the ideal work ${\cal W}_{ideal}$ is achieved in the large time limit. In the linear response regime the dissipated power is expected to scale asymptotically as $1/\tau_{cyc}^2$. Under small $\mu$ (slow driving) the globally coherent cycle exhibits such typical behaviour, as showcased in the inset of Fig.  \ref{fig:eff_vs_cycletime_Endo}.

We can compare the performance of the Globally coherent Carnot and Otto cycle. Both cycles maintain coherence throughout the cycle, where in the global Otto cycle coherence is generated only during the unitary strokes. 
The turnover to an operation mode which is strongly influenced by interference requires faster driving and larger value of coherence measure ${\textfrak{C}}>0.1$. As a result, the coherent affected operation mode at shorter cycle time. This characteristic behaviour can be witnessed in Fig. (\ref{fig:eff_vs_cycletime_Endo}). The shorter cycle times allow the global Otto cycle to posses comparable maximum power with respect to the Globally coherent Carnot cycle with higher efficiency.

\section{Quantum signature: constant  aiabatic parameter cycles maintaining global coherence.}
\label{sec:Q_signiture_global_coherence}

A quantum signature is defined as a measurable quantity of the system which affirms non-classical behaviour \cite{uzdin2015equivalence,lostaglio2020certifying}. In the present scenario, we search for thermodynamic properties which are susceptible. Unlike classical features, quantum properties are sensitive to any measurement that extracts information on the system state. This feature allows validating the quantum signature by analysing the affect of measurements on the cycle performance. Specifically, we compare the globally coherent Carnot cycle efficiency to the efficiency in the presence of weak quantum measurements of energy in the instantaneous energy basis, which are performed on the unitary strokes.

\begin{figure}[htb!]
\centering
\includegraphics[width=7.5cm]{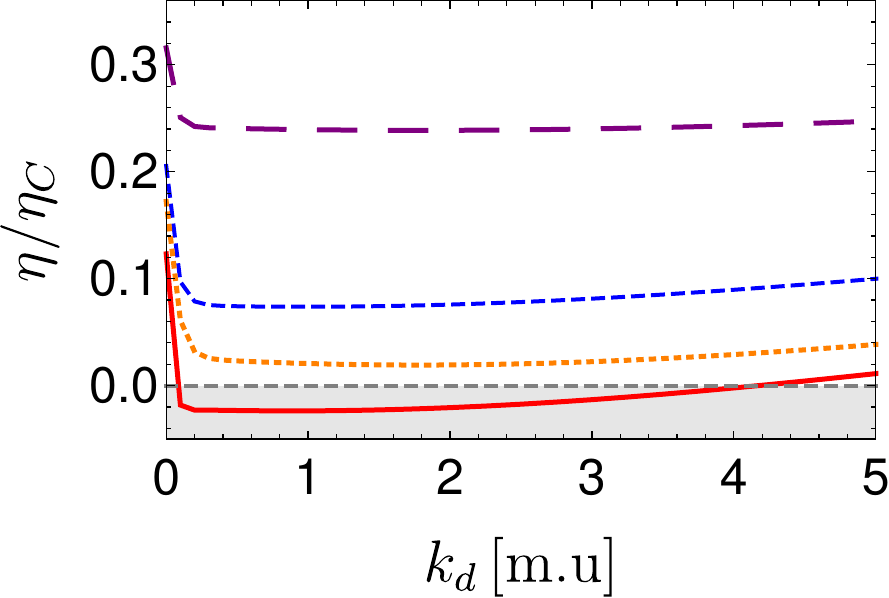}
\caption{Efficiency as a function of dephasing constant $k_d$ for varying cycle times for the globally coherent Carnot cycle: Red-continuous $\tau_{cyc}=102.5$, blue dashed $\tau_{cyc}=103$, orange dotted $\tau_{cyc}=105$ ,purple long dashed $\tau_{cyc}=129$, units of ($2\pi/\Omega_{min}$).  
\label{fig:eff_vs_k_dephasing}}
\end{figure}  

The weak measurement back action effectively leads to a double commutator term $-k_d\sb{\hat{H},\sb{\hat{H},\hat{X}}}$ in the master equation for the system operator $\hat {X}$ \cite{diosi2005weak}. Such a term leads to pure dephasing with a dephasing constant $k_d$. We compare the affect of dephasing for different cycle times for the global Carnot cycle. 
For a specific cycle time, the results shows a decrease in efficiency for small dephasing constant $k_d$, Fig. \ref{fig:eff_vs_k_dephasing}. This regime corresponds to weak measurements that only slightly influence the system dynamics and decrease the coherence. When the system is perturbed weakly, the dephasing only increases the dissipation, and therefore, reduces the efficiency. Beyond a critical value, stronger measurements (increasing the value of $k_d$) lead to an opposite effect and improve the efficiency. This result is related to the Zeno effect \cite{facchi2000quantum,uzdin2018markovian} and quantum lubrication \cite{feldmann2006quantum}, as continuously measuring the qubit forces it to remain on the energy shell. In return, this leads to less coherence generation and therefore, reduced dissipation. The measurement backaction and the present thermodynamic analysis, should be taken with certain care. Once the qubit state is being monitored it ceases to be an isolated system and the measurement may be accompanied by an additional heat transfer \cite{elouard2017role,elouard2018efficient}. The additional heat arising from the weak measurement is not taken into account in our analysis.

As expected, the influence of the measurement reduces with increasing cycle times, see  Fig. \ref{fig:eff_vs_k_dephasing}.
Slower driving reduces the amount of coherence throughout the cycle, thus, diminishing the affect of dephasing on the thermodynamic performance. In the quantum adiabatic limit, the system remains on the energy shell and the measurement does not disturb the system.

\begin{table}
\caption{Shortcut (local) and Globally coherent cycle parameters are given in the model units (m.u.), satisfying $\hbar=k_B=c=1$. The parameters for the Globally Carnot cycle correspond to the continuous red line in Figs. \ref{fig:eff_vs_cycletime_Endo}, \ref{fig:power_vs_cycletime_Endo}, while the parameters in brackets correspond to the purple dashed lines. }
\begin{center}
\begin{tabular}{|p{2.7cm}||p{2.5cm}|p{2.8cm}|p{2.1cm}|p{2.8cm}| }
 \hline
Parameters	& Local Carnot & Globally coherent Carnot & Local Otto & Globally coherent Otto  	\\
 \hline
 \hline
$\Omega_1$		& $12$		&	$10$ & $8$		&	$9$\\
$\Omega_2$		& $8$		&	$9$ ($6.857$) & $8$		&	$9$\\
$\Omega_3$		& $4$		&	$6$ & $6$		&	$6 \f{2}{3}$\\
$\Omega_4$		& $6$		&	$6  \f{2}{3}$ ($8.75$)& $6$		&	$6 \f{2}{3}$\\
Hot bath temperature		& $T_h=10$		& $T_h=10$	& $T_h=10$		& $T_h=10$	\\
Cold bath temperature		& $T_c=5$		& $T_c=5$ 	& $T_c=5$		& $T_c=5$\\
 \hline
\end{tabular}
\end{center}
\label{table:cycle_parameters}
\end{table}

\begin{table}
\caption{Stroke parameters are given in the model units (m.u.), satisfying $\hbar=k_B=c=1$.}
\begin{center}
\begin{tabular}{|p{7cm}||p{2cm}|}
\hline
Parameters	& Value	\\
\hline
\hline
Coupling constant $A\equiv g^2/2 \hbar c$		& $0.01$	\\
Integration step size 		& $10^{-3}$	\\
\hline
\end{tabular}
\end{center}
\label{table:gen_parameters}
\end{table}

\section{Discussion}

\subsection{What the qubit can and cannot do}

The qubit QM model can generate expressions for thermodynamic quantities, based on first-principle derivations under the paradigm of open quantum systems. It incorporates all the features we expect from finite-time thermodynamics: tradeoff between efficiency and power, irreversible process, finite heat transport, friction, heat leaks. A major advantage of the qubit model is its simplicity. 
Nevertheless, the model is able to elucidate the main issues of finite-time-thermodynamics, but not all of types of effects. It is important to stress what phenomena we omitted from this paper, either  since they deserve further study
or because the model is restricted.

The present qubit model by construction is limited in describing many body effects on engine performance. 
For example, entanglement in engines \cite{zhang2007four,wang2009thermal,ji2012entangled}, engines based on many body localization \cite{halpern2019quantum} as a working fluid,  
collective and critical quantum effects 
in engines \cite{hardal2015superradiant,campisi2016power,niedenzu2018cooperative,mukherjee2020universal,jaramillo2016quantum}, and  synchronization \cite{jaseem2020quantum}.

In the introduction we stated four possible sources of irreversibility, two sources which were not included in the present analysis are heat leaks and switching losses. In any realistic engine, there is always a residual system-bath coupling even during the unitary strokes \cite{correa2015internal}. As a result, additional heat currents from the hot to the cold reservoir occur. Moreover, such interaction causes additional dephasing. These effects are not counted in our models.
In addition, in all the four stroke cycles presented, we ignored the energetic cost of switching the coupling to the bath, $g$, on and off  \cite{barra2015thermodynamic,de2018reconciliation}. Such switching occurs as the cycle transitions between unitary and open strokes. If one chooses in Eq. (\ref{eq:htotal}) a system bath interaction which satisfies $\sb{\hat{H}_{s-h/c},\hat{H}_s+\hat{H}_{h/c}}=0$ the energetic cost of switching the coupling on and off vanishes.


The miniaturization of engines emphasizes the role of fluctuations.
Fluctuations add another twist to the tradeoff between power and efficiency \cite{shiraishi2016universal,pietzonka2018universal,funo2018quantum,denzler2020power}. 
It has recently been claimed that the possibility of heat engines to have finite power output, operate close to Carnot efficiency and  exhibit only small fluctuations is excluded \cite{pietzonka2018universal}. For steady-state heat engines, driven
by a constant temperature difference between the two heat baths, it has been claimed that out of these three
requirements only two are compatible. The present qubit model could be 
a unique platform for testing these ideas \cite{silaev2014lindblad,erdman2019maximum}.

\subsection{Further considerations}
There are infinitely many thermal cycles that can operate between given hot and cold baths, and produce power. These cycles differ by the externally controlled protocols and the switching points between the strokes.  Optimization can be applied to the control protocols to enhance power or to minimize entropy production. 

In the present study we considered only a restricted class of control strategies, and mostly emphasized control strategies that optimize individual strokes. Such control and optimization relies on the prior knowledge of the equations of motion of the working medium. 
The control of the qubit is based on the full $SU(2)$ algebra.
This reflects the physical intuition that in practice the control
operators do not commute in general with the system Hamiltonian.
As a result $[\hat H(t),\hat H(t')] \ne 0$.

In the unitary strokes we explored shortcuts to adiabaticity (STA) protocols, Sec. \ref{subsec:STA}. Without any restriction, employing STA protocols allows carrying out strokes with vanishing stroke duration. Restricting the energy or the coherence stored within the controller leads to a minimum stroke duration for frictionless operation.  When analysing the complete engine cycle, the time allocated to the unitary strokes was found to have no qualitative effect on the cycle performance. 

The thermalization process during the open-strokes can be controlled as well. In the study of the Carnot-type cycles, we employed protocols which speedup the thermalization process, with the cost of additional dissipated work and concomitant entropy production. The utilized protocols achieve the target thermal state rapidly, but are by no means optimal. 
Thermalization strokes are a much newer development and what features might make them optimal is not yet clear. For example, it is not at all clear that our STE protocols, which cash in all the coherence at the end of the stroke, are desirable. Cashing in this coherence before the end of the stroke may not be helpful as conversion during the following unitary stroke is easily handled. In fact, macroscopic optimizations of finite-time Carnot cycles \cite{salamon1981finite} would lead to maximum energy exchange for a given entropy change of the working fluid and suggests that better use of the heat exchange time would be to utilize it fully by keeping the coherence for conversion during the following unitary stroke. Our expectation for an optimal implementation would be one that keeps the entropy production rate constant \cite{tsirlin2020averaged,tondeur1987equipartition}, and examining these rates in Fig. \ref{fig:entropy_prod_const_mu} shows that our constant $\mu$ protocol comes reasonably close.

Global optimization was only carried out for the frictionless case, in the Otto (Sec. \ref{subsec:otto-opt} ) \cite{feldmann1996heat} and Carnot (Sec. \ref{subsec:endo_reversible}) \cite{geva1992quantum} cycles. These studies lead to the conclusion that a maximization of the power is accompanied by a maximum entropy production. For cycles with friction we conjecture that one can construct cycles that balance between the entropy production and produced power. In the general case, including friction, finding an optimal cycle is a subject of great interest, and remains to be explored in future efforts. 

Besides serving as a comparison to Carnot-type cycles, our treatment of Otto cycles shows off some interesting new features. The fact that both the power and the entropy production of the cycle are proportionate to the change in polarization gives this cycle a unique character. In particular, it implies that the point of maximum power is the point of maximum entropy production, i.e. the two objectives are diametrically opposed for this engine. Some light can be shed on this situation by realizing that the thermal losses are set by the temperature gap between the qubit and the bath at their highest values at the beginning of the open strokes. After that, this gap decays with the only control being the time spent on the stroke. This forces the heat exchange and the entropy production to be largest at the beginning of the stroke with rate decreasing with longer stroke duration. 

This line of reasoning is also what led us to the closer examination of the sudden cycle for which only the very initial segment of the open branches are used. By the above line of reasoning, this initial segment is the fastest heat exchange. Using instantaneous counter-diabatic driving for the unitary strokes leads to overall frictionless operation of the sort discussed in reference \cite{feldmann1996heat}. The interesting feature of sudden cycles, without STAs for the unitary branches, is that instantaneous driving produces significant coherence which is actually very useful for the cycles' performance. The unitary jumps are reversible, hence there is no cost to going forwards and backwards and at the end we get all the invested work back \cite{korzekwa2016extraction}. The same coherence in the forward jump is used to power the backwards jump. The only difference between a forward and backward jump and sudden engine operation is the very brief stops in contact with the baths, during which some coherence decays. This cost in coherence however is not enough to kill all the power and the sudden cycles give an important example of an engine in which coherence helps. This is contrary to conjectures in the literature that coherence is always an undesirable in heat engine operation \cite{brandner2020thermodynamic}. Our findings show that this conjecture, while valid for slow operation, does not appear to be true of all types of operation; there exist valid benefits of coherence. 

Increasing the driving increases coherence generation. On the open branches, this coherence results in rather significant frictional losses which quickly brings us to the turnover point where the friction dominates the cycle performance and the engine no longer produces work. This turnover point occurs for in the Carnot cycle for much smaller values of the coherence than for the Otto cycle, presumably because in the Carnot cycle the open branches generate additional coherence. It is also the reason our graphs of the efficiency and the power for the Carnot cycle cannot reach lower cycle times, cf. Figs \ref{fig:eff_vs_cycletime_STE}, \ref{fig:power_vs_cycletime_STE}, \ref{fig:eff_vs_cycletime_Endo}, and  \ref{fig:power_vs_cycletime_Endo}. 
The smooth behavior of both the efficiency and the power as a function of the cycle time for local cycles gives way to oscillations at short times for global cycles (Figs. \ref{fig:eff_vs_cycletime_Endo}, and  \ref{fig:power_vs_cycletime_Endo}). Coherence by nature oscillates, and these oscillations result in effectively constructive and destructive interference with the oscillation during the following stroke in our global cycles.
Note that this feature also shows up for the global Otto but at much faster cycle times. In general, the global Otto cycle is less sensitive to coherence (the coherence related operators and energy are on the same scale). While the sensitivity to coherence depends on the temperature gap between the system and bath during the open strokes, this lower sensitivity to coherence for the Otto cycle holds for any comparable gaps. 

\subsection{Comparing to the Harmonic working fluid}

   Engine models with the qubit and harmonic oscillator 
   working medium have been the most popular quantum systems in the study of quantum heat devices \cite{geva1992classical,feldmann1996heat,rezek2006irreversible,kosloff2017quantum,rossnagel2016single,abah2019shortcut,insinga2016thermodynamical,insinga2018quantum,deffner2018efficiency}. These models share many common features, including the tradeoff between power and efficiency, and obtain the Carnot bound in the limit of large cycle time.  
  Moreover, in the limit of low temperatures the harmonic oscillator converges to the qubit model, and the thermodynamic performance should be equivalent. Despite the similarities there are qualitative differences in the thermodynamic performance. The major differences between the two models can be traced to the dynamical algebra of the two, $SU(2)$ and the Heisenberg-Weyl group $H_3$. The  former algebra is compact, while the latter is non-compact. A direct consequence is that the heat capacity of the harmonic oscillator increases with the temperature, saturating for high temperatures. In contrast, the capacity of the qubit reaches a maximum value and then asymptotically decreases as $T^{-2}$ in the high temperature regime. 
   
   The different algebra influences the dynamics as well, for example, for a constant adiabatic parameter protocol (non-adiabatic driving) the effective frequency of the qubit increases, while in the harmonic case the effective frequency decreases. In turn, the effective frequency determines the relaxation rate towards the instantaneous attractor. In both models this rate increases monotonically with the effective frequency, and the relaxation rate will be influenced in an opposite manner. In addition, the detailed balance condition is also modified, which means the internal temperature of the qubit is reduced in the presence of non-adiabatic driving.
   Comparing the present global cycle to an analogous harmonic cycle \cite{dann2020quantum}, we find  that the qubit cycle has a greater sensitivity to the presence of coherence (short cycle times). A possible explanation of this result is the shift to a lower internal temperature and a higher relaxation rate, which destroys the coherence and nulls the extracted work.
    
In the operation of the engines, both working mediums allow performing shortcut protocols \cite{del2014more} on the unitary and open strokes.
For frictionless and shortcut cycles, the harmonic Otto cycle exhibits a maximum efficiency when optimizing the compression ratio, which corresponds to the classical endoreversible result $\eta_{CA}$, Eq. (\ref{eq:eta_ca}) \cite{rezek2006irreversible,kosloff2017quantum}. This result is independent of the power of the engine. In contrast,  the qubit model reaches the Curzon-Ahlborn efficiency for an endoreversible  cycle in the high temperature limit, Sec. \ref{subsec:endo_reversible} \cite{geva1992quantum}. \\

\subsection{High temperature limit} 

With the motto of learning from example, we can  employ the qubit model to elucidate the path from the quantum first principle derivation to the classical FTT results. The key is the high temperature limit. This means 
that the polarization $|\bar S| $ is small and can be used 
to expand the thermodynamical expressions to first order.
In the elementary Carnot-type cycle this expansion leads to the Curzon-Ahlborn efficiency $\eta_{CA}$, Eq. (\ref{eq:ca}), at maximum power
without referring to the linear Newtonian  heat transfer law
or the low dissipation limit \cite{abiuso2019non,abiuso2020optimal,brandner2020thermodynamic}. If we consider the cost of driving (Subsection \ref{subsec:5.4}), we find that the entropy production rate $\Sigma^u$ at the high temperature limit can be cast into the template of the Onsager relations, Eq. (\ref{eq:onsager}).

The qubit engine model operates in the low dissipation limit
when the cycle period is very large. We observe the expected limit
(insert of Fig. \ref{fig:power_vs_cycletime_Endo} ), as the dissipated power scales inversely with the square of the cycle period ${\cal P}_{diss} \propto 1/\tau_{cyc}^2$. This is generically true, as discussed in the next subsection.\\

\subsection{Dissipation}

The problem of operating a heat engine while trying to minimize dissipation has a simple general answer: turn off the engine so nothing happens. This of course has a dissipation cost of zero; you cannot do better. In order to get an interesting answer to the minimum entropy production question, we have to require something to happen. For a heat engine, one natural choice is to carry entropy $\Delta {\cal S}$ from the hot bath at temperature $T_h$ to the cold bath at temperature $T_c$. Once such a constraint is specified, the interesting optimizations for a finite-time heat engine range from minimum entropy production to maximum power, with maximum efficiency as merely an intermediate point \cite{salamon1981finite}.

The low dissipation limit has been recently employed to bound the dissipation in a heat engine using the notion of thermodynamic distance \cite{brandner2020thermodynamic,Abiuso2020geometric,salamon1983thermodynamic}. This distance, defined on thermodynamic states using the second derivative of the entropy, bounds the finite-time cost of driving a system along a given trajectory in the linear response regime, i.e. the slow process limit. In fact the minimum cost of driving a system along a path of length $\cal L$ in time $\tau$ is ${\cal L}^2/\tau$. Applied to our engine, when the qubit traces the cycle of length $\cal L$, the dissipated power for slow processes must be at least ${\cal L}^2/\tau^2$ as matching our observations, cf. Fig \ref{fig:power_vs_eff_STE}. How to geometrically bound the dissipation in the context of our non-adiabatic master equation formalism for faster driving is not clear. A major difference is the coupling of energy and coherence. This issue
is left for future efforts.  

\subsection{Experimental connections}


Realization of engines with a working medium composed of an ensemble of spins is an expected development, such as in an NMR experiment \cite{peterson2018experimental}. 
The surprise is the ability to operate an engine with a single spin.

Experimental realization of single qubit engines and refrigerators  is in the process of rapid development.
In part this progress is part of a larger effort in developing quantum technology. This breakthrough is due to the ability to
cool the ambient environment to temperatures in the range, or colder than, the qubit energy gap \cite{pekola2019thermodynamics}. Moreover, the rapid progress in manipulations, designed for quantum information processing, can be employed for the unitary strokes of quantum heat engines.
Recently, experimental realizations of four-stroke cycles were demonstrated \cite{von2019spin,ono2020analog} as well as a two-stroke engine  \cite{klatzow2019experimental}. 
Another application for quantum engines is in
quantum sensing and in particular, thermometery. Suggestions  based on
the transition point between a quantum engine cycle and a refrigerator have recently been proposed \cite{bhattacharjee2020quantum,levy2020single}.

A more immediate goal is quantum refrigeration \cite{ronzani2018tunable,pekola2019thermodynamics}. 
Since, any quantum device operates in ultracold temperatures, 
the drive for miniaturization will require an on-the-chip
quantum refrigerator replacing the cumbersome dilution 
refrigerators of today.


\section{Conclusions}

The qubit thermal engine has been a source of insight concerning finite-time thermodynamics for 30 years, with its origins dating back to a time when qubits were still two-level-systems. Among the lessons from the model was the role of coherence in friction like phenomena on the unitary strokes. 
Further analysis revealed that the generation of coherence occurs on  the unitary strokes and is separated from its dissipation, which occurs when the qubit is in contact with the thermal bath. 
This insight led first to the notion of quantum friction and later to the exploration of shortcuts to adiabaticity (STA). The analysis of the qubit engine 
generated a unifying overview of these finite-time thermodynamic phenomena. We have tried to present such an overview alongside our new findings. 

Our study explores a new chapter in the behavior of the qubit: the study of a driven isothermal process. 
Recent progress in open system dynamics, \cite{dann2018time,dann2018inertial,dann2020fast} 
allowed the treatment of thermalization processes, driven processes with time dependent Hamiltonians in contact with a heat bath. This allowed the Carnot cycle to be analysed for shorter times than the previous linear response treatments.

This breakthrough was achieved by using a basis of eigenvectors of the instantaneous propagator of the system dynamics in the interaction representation while exploiting the dynamical $SU(2)$ algebra of the qubit. 
The key to the breakthrough was the non-adiabatic master equation (NAME) \cite{dann2018time} that could correctly describe the dynamics of thermalization in a thermodynamically consistent way and eliminate issues of time reordering \cite{dann2018inertial}. In the qubit model, the instantaneous attractor is rotated from the energy direction by an amount that depends on the speed of the driving and reveals the details of the coupling between energy and coherence.

In the present manuscript we have analysed local cycles where the coherence is required to vanish at the switching points between strokes as well as global cycles where coherence is set only by the driving protocols and is carried from one stroke to the next. For local cycles we designed and implemented an STE protocol that mimics an isotherm. For the global cycles we used only constant adiabatic parameter $\mu$ trajectories. Both these types of cycles are analysed and compared. As a result, for the qubit engine we are now able
to assess the role of coherence along the isothermal strokes. The global cycles exhibit oscillations in their efficiency and power once we reach small cycle times (fast driving). These oscillations are due to the oscillations in coherence and how the timing of the switching between strokes happens to catch the coherence oscillation of the previous stroke. The Carnot-like cycles show enhanced sensitivity to coherence. This coherence can be reduced by weak measurement 
of the energy causing pure dephasing. Dephasing is damaging for short cycles periods but can be beneficial for intermediate cycle times.  Another coherence related finding concerns the global Otto cycle in the sudden limit near which the coherence of the engine acts as a useful flywheel.

Our analysis calculates the entropy production rate from first principles. Using inertial coordinates for the qubit, we generally find that this entropy production naturally takes a flux-times-force form. In the high temperature limit we find a linear relation between these fluxes and forces and an associated Onsager relation.

The present manuscript managed to compare the known behavior of local and global Otto and Carnot-like cycles. That is a lot to compare. We tried to focus on the new emergent phenomena in a coherent Carnot-type cycle. We found that to discuss these phenomena we needed a backdrop of related results to compare to. The above attempted synthesis is the outcome.

\vspace{6pt} 




\section*{Historical Background}
This study is a chapter in a 25 year collaboration between Peter Salamon and Ronnie Kosloff, initiated by the study of an Otto spin engine with Tova Feldmann and Eitan Geva \cite{feldmann1996heat}. The focus on the Carnot cycle was inspired by a comment by Peter during a visit to Jerusalem, pointing out a discrepancy in the quantum heat engine studies between the many studies of Otto cycles and the few studies of Carnot and other cycles. The reason was the lack of an adequate thermodynamically consistent master equation for the isothermal
strokes. It took four years to address the issue. The first step is the development of the non adiabatic master equation by  Roie Dann and Amikam Levy \cite{dann2018time}. Roie Dann continued the development setting it on firm theoretical grounds with the inertial theorem. Returning to the original objective, a shortcut to equilibration was developed \cite{dann2020fast}. The present study incorporates a broad perspective as well as many new results on the role of coherence in entropy production in the Carnot and Otto cycles.


\ack
{We would like to thank our
many contributors to this study: Tova Feldmann, Eitan Geva, Lajos Diosi, Jose P Palao, Yair Rezek, Karl Heinz Hoffmann, Amikam Levy, Raam Uzdin, Ander Tobalina, and Andrea Insinga.\\
This research was supported by the Adams Fellowship  Program of the Israel Academy of Sciences and Humanities and the Israel Science Foundation, grant number 2244/14.
}



%

\section*{Abbreviations}
The following abbreviations are used in this manuscript:\\
\\
\begin{tabular}{@{}ll}
GKLS & Gorini, Kossakowski, Linblad, Sudarshan; Master equation\\
CPTP & Completely Positive Trace Preserving map\\
NAME & Non-Adiabatic Master Equation\\
FEAT & Fastest Effectively Adiabatic transition\\
STA & Shortcut To Adiabticity\\
STE & Shortcut To Equilibrium\\
\end{tabular}

\appendix
\section{Representations of the qubit state.}
\label{appendixA}
\unskip

The qubit state can be described in many alternative ways. Each representation highlights a certain aspect of the engine.
We will now summarize the different approaches and the relation between them.
The basic construction relies on a set of orthogonal operators that form a closed Lie Algebra
$\{ \hat A \}$
\begin{equation}
[\hat A_i,\hat A_j] =\sum_k C_k^{ij} \hat A_k
\label{eq:lie}
\end{equation}
where $C_k^{ij}$ is the structure tensor of the algebra.
The orthogonality relation:
\begin{equation}
 \rm{tr}\{\hat A_i^\dagger\hat A_j \}=\delta_{ij}~~, 
 \label{eq:ortho}
\end{equation}
where the identity $\hat I$ is part of the set and all other operators are therefore traceless.
Under these conditions the state $\hat \rho $ can be expanded as a linear combination of the set
$\{ \hat A \}$
\begin{equation}
 \hat \rho = \frac{1}{N} \hat I
 + \sum_j \alpha_j \hat A_j~~,
 \label{eq:linear-rho}
\end{equation}
where $\alpha_j = \langle \hat A_j \rangle$ and $N$ is the size of Hilbert space.
An alternative formulation includes representing the state in terms of a
generalized Gibbs state \cite{alhassid1978connection}:
\begin{equation}
\label{eq:g-gibbs}
\hat \rho = \frac{1}{Z}\exp 
\left( \sum_j \lambda_j \hat A_j \right)~~.
\end{equation}
The generalized Gibbs state is 
the maximum entropy state subject to the constrains of
the expectation value 
$\langle \hat A_j \rangle = 
\rm{tr}\{\hat \rho \hat A_j \}$, this leads to a set of non linear equations which determines the Lagrange multipliers $\lambda_j$.
The forms (\ref{eq:linear-rho})
and (\ref{eq:g-gibbs}) are unique once the expectation values $\langle \hat A_j \rangle$ are known.

It is convenient to express the generalized Gibbs state as a product form \cite{wei1964global}
\begin{equation}
\label{eq:g-gibbs2}
\hat \rho = \frac{1}{Z}\prod_k \exp( 
  \gamma_j \hat A_j )~~.
\end{equation}
This form is not unique since it depends on the order
of operators. Once the order is set the coefficients $\{\gamma_j\}$ are determined from the expectation values $\{\langle \hat A_j \rangle\}$.

Specifically, for the qubit we employ the $SU(2)$ algebra and three sets of orthogonal bases.
The first  basis set $\v{s}=\{{\hat{S}_x},{\hat{S}_y},{{\hat{S}}_z}\}^T$ represents the static polarization,
where $\hat S_j$
are the spin operators with the commutation relation of the $SU(2)$ algebra $ [\hat S_i,\hat S_j] =i \hbar \epsilon_{ijk} \hat S_k$ and $\hat{\sigma_j}$. In terms of the Pauli operators $\hat{\sigma}_j$ they are expressed as $\hat{S}_j=\frac{\hbar}{2}\hat{\sigma}_j$. An arbitrary state is expressed as a linear combination of these operators in Eq. (\ref{eq:statew}). A geometric interpretation uses this set 
as a Cartesian basis in 3D Fig \ref{fig:1}. A time-dependent rotation around the $\hat S_y$ axis leads, up to a scaling, to the dynamical basis set
$\v v \b t =\{\hat{H},\hat{L},\hat{C}\}^T$,  Eq.(\ref{eq:hlc}). The $SU(2)$ algebra in the polarization basis defines a rotation in Liouville space
\begin{equation}
   {\cal R}_y(\phi)=\exp \left( \frac{i}{\hbar} [\hat S_y, \bullet ] \phi \right)  = e^{ \frac{i}{\hbar}\hat S_y \phi} \bullet e^{ -\frac{i}{\hbar}\hat S_y \phi}~~,
   \label{eq:rot}
\end{equation}
and the relation between the two basis sets
\begin{equation}
    \v v = \Omega(t) {\cal R}_y(\phi(t))  \v s
\end{equation}
where $\phi=\arccos(\omega/\Omega)$.
The explicit dependence of $\Omega(t)$ and $\phi(t)$ means that the dynamical basis set $\v v$ is time-dependent. 

In terms of the dynamical basis set,
the linear form of the state Eq. (\ref{eq:hlc-state}) is defined.
Since the Hamiltonian $\hat H$ is part of $\v v$, it is natural to define the generalized Gibbs state
\begin{equation}
\hat \rho = \f{1}{Z} \exp\left(-( \beta \hat H + \lambda \hat L + \gamma \hat C) \right)
\label{eq:hlc-gibbs}
\end{equation}
The standard Gibbs state is obtained when $\gamma=\lambda=0$ then
$\beta=\f{2}{\hbar \Omega} \tanh^{-1} \left( \f{2 \langle \hat H \rangle }{\hbar \Omega}\right)$.

The third basis set is obtained from the eigenoperators of the free propagator $\{\hat \chi,\hat \sigma,\hat \sigma^\dagger\}^T$, Eq. (\ref{eq:chi-sigma}). In terms of this set
the linear form of the state in the interaction representation is given by  Eq. (\ref{eq:rho_open_sys}).
In addition, the state can be expressed in a generalized Gibbs state form
\begin{equation}
\tilde{\rho}=\bar{Z}^{-1}\exp\b{-\b{\bar{\beta}\hat{\chi}+\bar{\gamma}\hat{\sigma}+\bar{\gamma}^*\hat{\sigma}^\dagger}}~~.
\end{equation}
An equivalent product form has been previously been employed in Ref. \cite{dann2020fast}
\begin{equation}
    \tilde{\rho}=Z^{-1}e^{\tilde \gamma \hat{\sigma}}e^{\tilde \beta \hat{\chi}}e^{\tilde \gamma* \hat{\sigma}^\dagger}~~,
    \label{eq:rho_TLS}
\end{equation}
where $\tilde{Z}\equiv \tilde{Z}\b t=\rm{tr}\b{\tilde{\rho}_S\b t}$  is the partition function, with time-dependent parameters $\tilde \gamma\b t$ and $\tilde \beta$.

In Sec. \ref{subsec:5.4} we modify the basis set $\v g=\{\hat \chi,\hat \sigma,\hat \sigma^\dagger\}^T$ to a hemitian basis: $\v g'=\{\hat \chi,\hat \sigma_x,\hat \sigma_y\}^T$ where $\hat \sigma_x=\f{1}{\sqrt{2}}(\hat \sigma+\hat \sigma^{\dagger})$
and $\hat \sigma_y=\f{i}{\sqrt{2}}(\hat \sigma-\hat \sigma^{\dagger})$. The new representation allows relating the basis
$\v v$ to $\v g$ by a scaling and rotation around the $\hat L$ axis
(Cf. Fig. \ref{fig:1})
\begin{equation}
   \v g' =\f{\sqrt{2}}{\hbar \Omega} {\cal R}_L (\xi) \v v
   \label{eq:rotL}~~,
\end{equation}
where $\xi = \arccos(1/\sqrt{1+\mu^2})$ and 
${\cal R}_L (\xi) = \exp \left(  \frac{i}{\hbar \Omega} [\hat L,\bullet] \xi\right)$.

When studying the entropy production rate in Sec. \ref{subsec:5.4} we take advantage of the form 
\begin{equation}
\tilde{\rho}=\bar{Z}^{-1}\exp\b{-\b{\bar{\beta}\hat{\chi}+\bar{\gamma}_x\hat{\sigma}_x+\bar{\gamma}_y\hat{\sigma}_y}}~~.
\end{equation}
The $\{\bar{\beta},\bar{\gamma}_x,\bar{\gamma}_y\}$ time-dependent parameters are defined  by the eigenoperators expectation values. The relations are given by 
\begin{equation}
    \mean{\hat{\chi}}_{int}=f\b{r}\bar{\beta}~~~;~~~\mean{\hat{\sigma}_x}_{int}=f\b{r}\bar{\gamma}_x~~~;~~~\mean{\hat{\sigma}_y}_{int}=f\b{r}\bar{\gamma}_y~~,
\end{equation}
where $\mean{\bullet}=\rm{tr}\b{\bullet \tilde \rho}$, $f\b{r}=-\f 1{\sqrt{2}r}\rm{tanh}\b{\f r{\sqrt{2}}}$ with $r=\sqrt{\bar{\beta}^{2}+\bar{\gamma}_{x}^{2}+\bar{\gamma}_{y}^{2}}$. In the large temperature limit $f\b{r}\approx -\f{1}{2}$, which leads to simple relations between the thermodynamic fluxes and forces in the classical regime.
Alternatively, the parameters can be expressed in terms of the expectation values 
\begin{equation}
    \bar{\beta}=s\b{k}\mean{\hat \chi}_{int}~~~;~~~ \tilde{\gamma}_{x}=s\b{k}\mean{{\hat \sigma}_{x}}_{int}~~~;~~~
    \tilde{\gamma}_{y}=s\b{k}\mean{{\hat \sigma}_{y}}_{int}~~,
\end{equation}
where $s\b{k}=\log\left(\frac{1-\sqrt{2}{k}}{\sqrt{2}{k}+1}\right)/\b{\sqrt{2}{k}}$ with $k=\sqrt{\mean{{\hat \sigma}_{x}}^{2}_{int}+\mean{\hat{\sigma}_{y}}^{2}_{int}+\mean{\hat \chi}^{2}_{int}}$



\section{Explicit expressions}
\label{apsec:explicit_expressions}
The transition matrix from the basis  operators $\v{v}$  to the basis operators $\v{g}$. The matrix appears in the inertial solution of the qubit, Eq. (\ref{eq:inertial_solution}): 
\begin{equation}
P=\sb{\begin{array}{ccc}
1 & -\mu & -\mu\\
0 & i\kappa & -i\kappa\\
\mu & 1 & 1
\end{array}}~~.
\end{equation}

\begin{table}[htb!]
\caption{Definitions and notations summary.}
\centering
\begin{tabular}{cc}
\hline
\textbf{Parameters}	& \textbf{Description} 	\\
\hline
 $\{\hat{H},\hat{L},\hat{C}\}$ ; $\v v$	& dynamical operator basis ; associated vector in Liouville space 	\\
 $\{\hat{S}_x,\hat{S}_y,\hat{S}_z\}$ ; $\v s$	& polarization operator basis ; vector in Liouville space	\\
 $\{\hat{\chi},\hat{\sigma},\hat{\sigma}^\dagger\}$ ; $\v g$	& eigenoperator basis ; vector in Liouville space 	\\
 $\{\hat{\chi},\hat{\sigma}_x,\hat{\sigma}_y\}$ ; $\v g'$	& eigenoperator basis ; vector in Liouville space 	\\
 $X^H$ & Heisenberg picture \\
 $\tilde{X}$ & interaction picture \\
 $\omega$ and $\epsilon$    & control parameters	\\
 $\Omega$	& generalized Rabi frequency	\\
$\v S$    &  polarization vector	\\
$\bar S$    &  polarization	\\
 $\bar S_H$	& projection of the polarization vector on the energy axis 	\\
 $\bar S_{eq}$	& thermal polarization 	\\
$T$    & bath temperature 	\\
$\{\Lambda\}$ 	& dynamical propagators 	\\
 ${\cal S}_{v.n}$   & von-Neumann entropy 	\\
 ${\cal S}_{H}$   & energy entropy 	\\
 $\sigma^u_{cyc}$ &  entropy production per cycle \\
 $\Sigma^u$ & entropy production rate \\
 $\Gamma$   & decay rate 	\\
 $\eta_C$ and ${\cal W}_C$	& efficiency and work of the Carnot cycle	\\
 $\eta_i$, ${\cal W}_i$, ${\cal P}_i$ and ${\cal Q}_i$	& efficiency, work, power and heat of the $i$'th cycle\\
 $\mu$ 	& adiabatic parameter 	\\
 $\kappa$ & Inertial scaling factor\\
 $\alpha$ 	& effective frequency 	\\
 $\textfrak C$ & coherence\\
 ${\cal W}_{fric}$ & work to counter friction\\
 $\phi$ &$\arccos (\omega/\Omega)$ \\ 
 $\Phi$ & $\phi_a-\phi_b$\\
  $P$ & transformation matrix between $\v v$ and $\v g$\\
  $D$ & eigenvalue matrix of the eigenoperators \\
   $T_i$ & effective temperatures \\
 ${\cal J}_y$ & thermodynamic fluxes \\
 ${\cal F}_y$ & thermodynamic force \\
\hline
\end{tabular}
\label{table:def_parameters}
\end{table}

\clearpage
\section*{References}
\bibliographystyle{unsrt}






\end{document}